\documentclass[prd,preprint,showpacs,eqsecnum,superscriptaddress]{revtex4}

\usepackage{amsfonts}
\usepackage{amssymb}
\usepackage{graphicx}
\usepackage{pstricks}
\usepackage{slashbox}

\newcommand{\be}{\begin{equation}}
\newcommand{\bea}{\begin{eqnarray}}
\newcommand{\ee}{\end{equation}}
\newcommand{\eea}{\end{eqnarray}}
\newcommand{\bpi}{\begin{picture}}
\newcommand{\bce}{\begin{center}}
\newcommand{\epi}{\end{picture}}
\newcommand{\ece}{\end{center}}

\newcommand{\ksm}{k\hspace{-0.24cm}\slash}

\newcommand{\psm}{p\hspace{-0.18cm}\slash}
\newcommand{\psmp}{p\hspace{-0.14cm}\slash}

\def\g{\widehat{\Gamma}}

\newcommand{\ie}{{\it i.e.}, }
\newcommand{\eg}{{\it e.g.}, }

\renewcommand{\theequation}{\arabic{section}.\arabic{equation}}

\begin{document}

\begin{flushright}
ECT*-08-05
\end{flushright}

\title{New Schwinger-Dyson equations for non-Abelian gauge theories}
\date{May 26, 2008}

\author{D. Binosi}
\email{binosi@ect.it}
\affiliation{European Centre for Theoretical Studies in Nuclear
  Physics and Related Areas (ECT*), Villa Tambosi, Strada delle
  Tabarelle 286, 
 I-38050 Villazzano (TN)  Italy}
 
\author{J. Papavassiliou}
\email{Joannis.Papavassiliou@uv.es}
\affiliation{\mbox{Departamento de F\'\i sica Te\'orica and IFIC, Centro Mixto, 
Universidad de Valencia-CSIC,}
E-46100, Burjassot, Valencia, Spain}

\begin{abstract}

We  show  that   the  application  of  the  pinch   technique  to  the
conventional  Schwinger-Dyson  equations  for  the  gluon  propagator,
gluon-quark  vertex,  and  three-gluon   vertex,  gives  rise  to  new
equations endowed  with special properties.  The  new series coincides
with the  one obtained  in the Feynman  gauge of the  background field
method, thus  capturing the extensive  gauge cancellations implemented
by the pinch  technique at the level of  individual Green's functions.
Its  building blocks  are the  fully dressed  pinch  technique Green's
functions  obeying Abelian  all-order Ward  identities instead  of the
Slavnov-Taylor identites satisfied by their conventional counterparts.
As a result, and contrary to  the standard case, the new equation for the 
gluon self-energy can be truncated  gauge  invariantly  at   any  order  
in  the  dressed  loop expansion.   The  construction  is  streamlined 
by  resorting  to  the
Batalin-Vilkovisky formalism  which allows for a  concise treatment of
all  the   quantities  appearing  in  the   intermediate  steps.   The
theoretical   and   phenomenological   implications  of   this   novel
non-perturbative framework are discussed in detail.

\end{abstract}


\maketitle
 
\section{Introduction} 

The                      Schwinger-Dyson                     equations
(SDE)~\cite{Dyson:1949ha,Schwinger:1951ex} provide  a formal framework
for tackling physics  problems requiring a non-perturbative treatment.
The SDE  constitute an infinite system of  coupled non-linear integral
equations for all Green's functions of the theory, and can be used, at
least  in principle, to  address questions  related to  chiral symmetry
breaking, dynamical  mass generation,  formation of bound  states, and
other non-perturbative effects \cite{Cornwall:1974vz,Marciano:1977su}.  In
practice, their usefulness hinges crucially on one's ability to devise
a  self-consistent truncation  scheme  that would  select a  tractable
subset of these equations,  without compromising the physics one hopes
to describe.  Inventing such a scheme for the SDE of gauge theories is
a highly non-trivial proposition. The problem originates from the fact
that the SDEs are built out of unphysical Green's functions; thus, the
extraction  of  reliable  physical  information depends  critically  on
delicate all-order cancellations,  which may be inadvertently distorted
in the  process of the truncation.   Several of the  issues related to
the truncation of  the SDEs of QED have been addressed  in a series of
articles~\cite{Curtis:1990zs,
Bashir:1994az,
Kondo:1988md,Sauli:2002tk}.  

The  situation  becomes even  more
complicated for  strongly coupled non-Abelian gauge  theories, such as
QCD~\cite{Mandelstam:1979xd}, mainly for the following reasons.

\begin{itemize}

\item[{\it i}.]  The  complications  caused  by  the  dependence  of  the  Green's
functions on the gauge-fixing  parameter are more acute in non-Abelian
gauge-theories,  as can  be  seen already  at  the level  of the  most
fundamental   Green's   function,   namely  the   two-point   function
(propagator)  of the  corresponding gauge  bosons.  In  QED  the photon
self-energy (vacuum  polarization) is independent of the 
gauge-fixing  parameter, both perturbatively (to all orders)
and  non-perturbatively; when 
multiplied by $e^2$ it forms  a physical observable,
the QED effective charge.  In  contradistinction, the gluon self-energy is
gauge-dependent  already at  one loop;  depending on  the gauge-fixing
scheme employed,  this dependence may  be more or less  virulent.  This
difference is of little practical importance when computing $S$-matrix
elements  at a fixed order in  perturbation theory,  given  that  the
gauge-dependence  of  the  gluon  self-energy is guaranteed to 
cancel against similar contributions from  other graphs, but has far-reaching
consequences  when  attempting  to  truncate  the  corresponding  SDEs,
written in some gauge. Contrary to what happens in the perturbative calculation,
even if one were to put together 
the non-perturbative expressions from these truncated SDEs to form
a physical observable, the gauge-cancellations may 
not go through completely, because the process of the truncation 
might have distorted them. Thus, there is a high probability of ending up
with a residual gauge-dependence infesting one's non-perturbative
prediction for a physical observable.

\item[{\it ii}.] In Abelian gauge theories the Green's functions satisfy naive 
Ward Identities (WIs): given the tree-level WI, the all-order 
generalization is obtained by simply replacing 
the tree-level expressions by the all-order ones. In general, 
this is not true for 
the Green's functions of non-Abelian gauge theories, where the WIs 
are modified beyond tree-level, and are 
replaced by more complicated expressions known as       
Slavnov-Taylor identities  (STIs)~\cite{Slavnov:1972fg,Taylor:1971ff}; 
in addition to the basic Green's functions of the theory 
they involve various composite ghost operators. 
In order to appreciate how the fact that the Green's functions 
satisfy STIs may complicate the truncation procedure of the SDEs,
let us consider the simplest STI (and WI in this case)
satisfied by the photon and gluon self-energies alike, namely
\be
q^{\alpha} \Pi_{\alpha\beta}(q) = 0 \,.
\label{fundtrans}
\ee
Eq.~(\ref{fundtrans}) is without a doubt the most fundamental 
statement at the level of Green's functions that one can obtain
from the BRST symmetry~\cite{Becchi:1974md
}; it affirms the transversality of the 
gauge-boson self-energy, be it a photon or a gluon, and is valid both  
perturbatively to all orders as well as non-perturbatively. 
The problem stems from the fact that 
in the SDE of $\Pi_{\alpha\beta}(q)$ enter higher order 
Green's functions, 
namely the fully-dressed fundamental vertices of the theory.  
 It is these latter Green's functions that in the Abelian context 
satisfy WIs whereas in the non-Abelian context satisfy STIs.
Thus, whereas in QED the validity of Eq.~(\ref{fundtrans}) 
can be easily seen at  
the level of the SDE, simply because 
$q^{\mu}\Gamma_{\mu}(p,p+q) =e\left[ S^{-1}(p+q)-S^{-1}(p)\right]$, in QCD 
proving Eq.~(\ref{fundtrans}) is very difficult, and requires the 
conspiracy of all full vertices appearing in the SDE.
Truncating the SDE naively usually amounts to leaving out some of these vertices, 
and, as a result, Eq.~(\ref{fundtrans}) is compromised.

\end{itemize}

The complications stemming from the 
two points points mentioned above are often compounded by 
additional problems related to the loss 
of multiplicative renormalizability and the inability to 
form renormalization-group invariant quantities.
 
Recently,  a  truncation scheme  for  the  SDEs  of non-Abelian  gauge
theories has been  proposed~\cite{Binosi:2007pi} that is based on  
the pinch technique (PT)~\cite{Cornwall:1981zr,Cornwall:1989gv} and  its  connection with  the  
background  field  method (BFM)~\cite{Dewitt:1967ub, Abbott:1980hw
} (see below).  
The way the PT resolves the difficulties related with 
points ({\it i}) and ({\it ii}) mentioned above is by imposing   a    drastic
modification already  at the  level of the  building blocks of  the SD
series, namely the off-shell  Green's functions themselves.  
The PT  is a  well-defined
algorithm  that  exploits  systematically  the symmetries  built  into
physical observables, such as  $S$-matrix elements or Wilson loops, in
order to construct new,  effective Green's functions, endowed with very
special  properties.
The basic observation, which essentially defines the PT, is that there
exists  a  fundamental  cancellation  between sets  of  diagrams  with
different kinematic  properties, such as  self-energies, vertices, and
boxes.  This  cancellation is driven  by the underlying  BRST symmetry
\cite{Becchi:1974md}, and  is triggered when a  very particular subset
of the longitudinal momenta circulating inside vertex and box diagrams
generate   out  of   them   (by  ``pinching''  internal   lines)
propagator-like  terms.   The latter  are  reassigned to  conventional
self-energy  graphs,  in order  to  give  rise  to the  aforementioned
effective  Green's   functions.   These  new   Green's  functions  are
independent of the  gauge-fixing 
parameter~\cite{Cornwall:1981zr,Cornwall:1989gv,Papavassiliou:1989zd,
Papavassiliou:1994pr,Binosi:2001hy},
satisfy ghost-free, QED-like  WIs instead  
of the  complicated STIs~\cite{Cornwall:1989gv,Papavassiliou:1989zd}, display only  physical  thresholds~\cite{Papavassiliou:1995fq,Papavassiliou:1996fn},  
have correct analyticity properties~\cite{Papavassiliou:1996zn}, 
and are  well-behaved at high energies~\cite{Papavassiliou:1997fn}.

Of central importance for the ensuing analysis is the  connection between the  PT and the
BFM.  The latter is  a special gauge-fixing
procedure  that preserves the  symmetry of  the action  under ordinary
gauge transformations with respect to the background (classical) gauge
field   $\widehat{A}^a_{\mu}$,   while   the  quantum   gauge   fields,
$A^a_{\mu}$, appearing  inside the loops, transform  homogeneously under the
gauge  group~\cite{Weinberg:1996kr}. As a result, 
the  background $n$-point  functions (\ie those involving $\widehat{A}^a_{\mu}$ fields)
satisfy QED-like WIs to all orders. 
The BFM 
gives rise to special Feynman rules (see Appendix~\ref{Frules}); most notably $(a)$
the tree-level vertices involving $\widehat{A}^a_{\mu}$ fields depend in general 
on the quantum gauge-fixing parameter $\xi_{Q}$, used to gauge-fix the 
quantum  gauge   fields, and $(b)$ the ghost sector 
is modified, containing symmetric gluon-ghost vertices, as well as 
two-gluon--two-ghost vertices. 
Notice an important point: the background $n$-point  functions
are gauge-invariant, in the sense that they satisfy (by construction) QED-like WIs, 
but are {\it not} gauge-independent, \ie they depend explicitly 
on  $\xi_{Q}$. The connection
between PT and  BFM~\cite{Denner:1994nn
,Papavassiliou:1994yi},  
demonstrated to be valid to all  orders~\cite{Binosi:2002ft
}, 
affirms that the (gauge-independent) PT $n$-point functions {\it coincide} with the
BFM $n$-point functions when the latter are
computed at the special value $\xi_{Q}=1$, also known as the background Feynman gauge (BFG).

Let us  now return to  points ({\it i}) and  ({\it ii}) and analyze them  from the
perspective of the PT.  In a  nutshell, the way point ({\it i}) is resolved,
for the  prototype case  of the gluon  self-energy, is  the following.
The  BFG  is  a privileged gauge,  
in  the  sense that  it  is  selected  {\it
dynamically} when  the gluon self-energy  is embedded into  a physical
observable (such as an on-shell test-amplitude).   Specifically, the BFG
captures the net propagator-like subamplitude emerging after QED-like
conditions have been replicated inside the test-amplitude, by means of
the PT  procedure. Thus, once  the PT rearrangements have  taken place,
the  propagator  is removed  from  the  amplitude  and is  studied  in
isolation: one considers the SDE for the background gluon self-energy,
$\widehat\Pi_{\alpha\beta}(q)$,  at $\xi_{Q}=1$.   Solving the  SDE in
the BFG  eliminates any gauge-related exchanges  between the solutions
obtained   for   $\widehat\Pi_{\alpha\beta}(q)$   and  other   Green's
functions, when put together  to form observables; thus, the solutions
are free of gauge artifacts.
Regarding point ({\it ii}), the key ingredient is that now all full vertices appearing in the new SDE  
satisfy {\it Abelian} WIs; as a result, 
gluonic   and  ghost   contributions   are  {\it
separately}   transverse,    within   {\it   each}    order   in   the
``dressed-loop'' expansion. Thus, it is much easier to devise truncation schemes that manifestly preserve the validity of
Eq.~(\ref{fundtrans}). 

The main results presented in this article are the following.
We provide a detailed and complete demonstration of  
how the application of the PT algorithm 
at the level of the conventional SD series leads 
to a new, modified SD series, with the special properties 
mentioned above. A preliminary discussion of this issue 
has already appeared in brief communication, dedicated to the 
SDE of the gluon self-energy~ \cite{Binosi:2007pi}. 
From the technical point of view, here we present a
significantly more concise and direct proof, by virtue  
of a crucial STI, that we employ for the first time. 
In addition, we extend the 
analysis to include the SDE for the quark-gluon and three-gluon 
vertices, which are important ingredients for 
obtaining a self-contained picture.
We emphasize that the three-gluon vertex relevant in this analysis is  
the one that would correspond, in the BFM language, to
$\Gamma_{\widehat{A}AA}$, i.e. one background gluon and two quantum ones merging,  
and {\it not} the fully 
Bose-symmetric vertex 
$\Gamma_{\widehat{A}\widehat{A}\widehat{A}}$ considered in~\cite{Cornwall:1989gv,Binger:2006sj}.
The reason is that it is the former vertex that appears in the SDE of  
the gluon self-energy,  in complete accordance with both the PT unitarity 
arguments~\cite{Papavassiliou:1999az} (see also Appendix \ref{Appendix:howtopinch}) and the independent diagrammatic rules of the BFM~\cite{Abbott:1980hw}.   

Furthermore, we address an important conceptual and practical issue,
related to the fact that,
qualitatively speaking, 
the new SD series expresses the BFG  Green's functions 
in terms of integrals involving the conventional ones. 
This is already evident at the two-loop level: 
the  two-loop BFM gluon self-energy is written in terms of 
integrals involving the conventional one-loop gluon self-energy.
This example might suggest, at first sight, that one cannot arrive at a 
genuine SDE involving the same unknown quantity on both sides, 
but there is a way around it. Specifically, 
the use of a set of crucial identities,  
relating the conventional and the BFM Green's functions, allow
one to convert the 
new SD series into a dynamical equation involving 
either the conventional or the BFM gluon self-energy only. 
 
The  paper  is  organized  as follows.   In  Section~\ref{General}  we
briefly review  the most  salient features of  the PT  methodology and
explain qualitatively how  the new SD series is  obtained and what are
its main  advantages compared to the conventional  SD series, focusing
on  the truncation  possibilities it  offers.  In  Section~\ref{fm} we
introduce the notation  and the formal machinery that  will be used in
the actual derivation of the  PT Schwinger-Dyson equations of QCD.  We
focus  particularly  on   the  Batalin-Vilkovisky  formalism  and  the
plethora of  relations that it  furnishes for the  various fundamental
and  auxiliary Green's  functions appearing  in our  construction.  In
Section~\ref{SDE} we  present the central result of  this work, namely
the detailed derivation  of the new set of SDEs.  There are three main
subsection,  dedicated to  the construction  of the  new SDEs  for the
quark-gluon  vertex, the  three-gluon  vertex, and  finally the  gluon
propagator.  In  Section~\ref{Concl} we discuss  some of the  main practical 
implications of  the new  SD series, and  present our  conclusions and
outlook.  

The      paper contains five      appendices.       In
Appendix~\ref{Appendix:howtopinch} we  discuss some subtleties  of the
extension of the  PT algorithm beyond one loop,  and in particular how 
to identify unambiguously the subset of three-gluon vertices that 
must undergo the PT decomposition. In  Appendix~\ref{Appendix:REn} we
describe the general strategy for carrying out  
the renormalization procedure to the new SD series obtained within the PT.  
Finally, the last  four appendices furnish
the   derivation  of   several  instrumental   formulas   employed  in
Section~\ref{SDE}, together with a complete set of Feynman rules.

\section{The new SDE series: General philosophy and main results\label{General}}

In this section we present the general ideas and outline the basic philosophy
of our approach, before diving into the complexities of the full SDE construction.
The style of this section is rather
qualitative; thus the reader who would like to skip the technicalities 
can get an overview of the theoretical and practical 
advantages offered by the new SD series, compared to the conventional 
formalism. 

\subsection{The difficulties with the conventional formulation\label{convform}}

Let us first focus on the conventional SD series for the gluon self-energy. 
Defining the transverse projector
\be
P_{\alpha\beta}(q)= g_{\alpha\beta} - \frac{q_\alpha
q_\beta}{q^2},
\label{projector}
\ee
we have for the full gluon propagator in the Feynman gauge ($\xi=1$)
\begin{equation}
i\Delta_{\alpha\beta}(q)= -i\left[P_{\alpha\beta}(q)\Delta(q^2) + 
\frac{q_{\alpha}q_{\beta}}{q^4}\right],
\label{prop_cov}
\end{equation}
with $\Delta_{\alpha\beta}^{ab}(q)=\delta^{ab}\Delta_{\alpha\beta}(q)$ (in what follows color factors will be omitted whenever possible).
The scalar function $\Delta(q^2)$ is related to the 
all-order gluon self-energy 
\be
\Pi_{\alpha\beta}(q)=P_{\alpha\beta}(q)\Pi(q^2),
\ee
through
\be
\Delta(q^2) = \frac{1}{q^2 + i\Pi(q^2)}.
\label{fprog}
\ee
Since $\Pi_{\alpha\beta}(q)$ has been defined in (\ref{fprog}) 
with the imaginary  factor $i$ factored out in front, it is simply 
given by the corresponding Feynman diagrams in Minkowski space.
The inverse of $\Delta_{\alpha\beta}(q)$ can be found by requiring that
\begin{equation}
i\Delta^{am}_{\alpha\mu}(q)(\Delta^{-1})^{\mu\beta}_{mb}(q)=\delta^{ab}g_\alpha^\beta,
\end{equation}
and is given by
\begin{equation}
\Delta^{-1}_{\alpha\beta}(q)=
 iP_{\alpha\beta}(q) \Delta^{-1}(q^2) + i q_{\alpha}q_{\beta},
\label{inv_prog}
\end{equation}
or, equivalently,
\begin{equation}
\Delta^{-1}_{\alpha\beta}(q)= ig_{\alpha\beta}q^2 - \Pi_{\alpha\beta}(q).
\label{inv_prop_pi}
\end{equation}
\begin{figure}[!t]
\includegraphics[width=14cm]{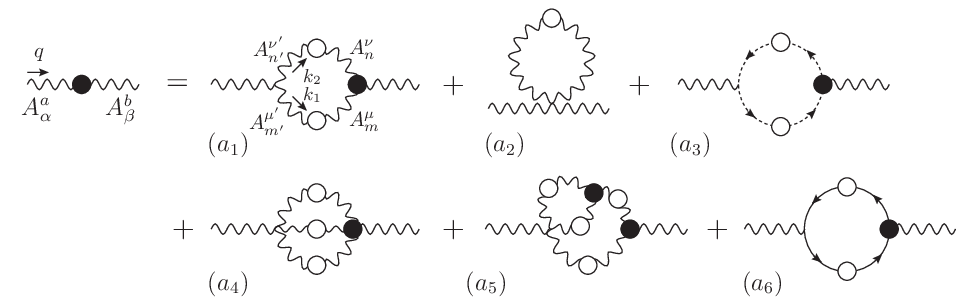}
\caption{Schwinger-Dyson equation satisfied by the gluon self-energy. 
The symmetry factors of the diagrams are $s(a_1,a_2,a_5)=1/2$, $s(a_3,a_6)=-1$, $s(a_4)=1/6$. 
White blobs represent connected Green's functions while black blobs represent one particle irreducible ones.}
\label{fig:gg_SDE-1}
\end{figure}
In Fig.~\ref{fig:gg_SDE-1} we show the SDE satisfied by the gluon self-energy. It reads
\be
\Delta^{-1}(q^2) P_{\alpha\beta}(q) = q^2 P_{\alpha\beta}(q) +  
i \sum_{i=1}^{6}(a_i)_{\alpha\beta}.
\label{convSD}
\ee
Of course,  in addition to the SDE for the gluon propagator, 
one must also include the corresponding SDE for the vertices; they
are normally expressed as  skeleton expansions 
in terms of the corresponding connected multi-particle kernels
(the proper treatment of the vertices is presented in Section~\ref{SDE}).

\begin{figure}[!t]
\includegraphics[width=8.5cm]{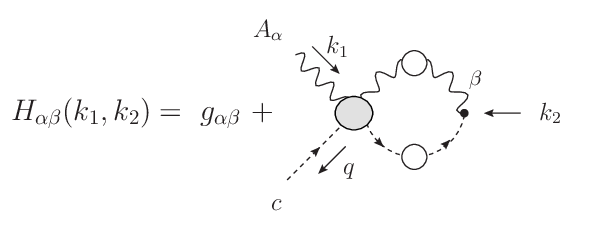}
\caption{The auxiliary function $H$ appearing in the three-gluon vertex STI. Gray blobs represent (connected) Schwinger-Dyson kernels (in this specific case the ghost-gluon kernel ${\cal K}$ appearing in the usual QCD skeleton expansion). }
\label{fig:H_aux}
\end{figure}

The main theoretical problem one encounters when 
dealing with the SDE given above is the fact that 
it cannot be truncated in a physically meaningful way.
The most obvious manifestation of this drawback is 
the following: after the truncation the fundamental 
Eq.~(\ref{fundtrans}) is violated.
To recognize the origin of this difficulty, 
note that Eq.~(\ref{fundtrans}) translates at the level of the SDE to
the statement
\be
q^{\alpha} \sum_{i=1}^{6}(a_i)_{\alpha\beta} =0.
\label{abcde}
\ee
The diagrammatic verification 
of (\ref{abcde}), \ie through contraction of the individual graphs by $q^{\alpha}$, 
is practically very difficult, essentially due to the 
complicated STIs satisfied by the vertices involved. 
The most typical example of such an STI is that of the 
conventional three-gluon vertex $\Gamma_{\alpha\mu\nu}(q,k_1,k_2)$ (all momenta entering), given by~\cite{Ball:1980ax}
\be
q^\alpha \Gamma_{\alpha\mu\nu}(q,k_1,k_2) =
\left[q^2D(q)\right]\left\{
\Delta^{-1}(k_2^2) P^{\gamma}_{\nu}(k_2) H_{\mu\gamma}(k_1,k_2)
-\Delta^{-1}(k_1^2) P^{\gamma}_{\mu}(k_1) H_{\nu \gamma}(k_2,k_1)\right\},
\label{sti3gv}
\ee
where  the auxiliary function $H_{\alpha\beta}$ is defined in Fig.~\ref{fig:H_aux}. The kernel ${\cal K}$ appearing in this function is the conventional connected
ghost-ghost-gluon-gluon kernel appearing in the usual
QCD skeleton expansion~\cite{Marciano:1977su,BarGadda:1979cz}. 
Notice also that $H_{\alpha\beta}(k,q)$ is related to the conventional gluon-ghost
vertex ${\Gamma}_{\beta}(k,q)$ (with $k$ the gluon and $q$ the anti-ghost momentum) 
by~ $q^{\alpha} H_{\alpha\beta}(k,q) = {\Gamma}_{\beta}(k,q)$
~\cite{Marciano:1977su,BarGadda:1979cz,Ball:1980ax,Pascual:1984zb}.

In addition, some of the pertinent STIs are either too complicated,
such as that of the conventional four-gluon vertex, or they 
cannot be cast in a particularly convenient  form.
For instance, in the case of the conventional gluon-ghost vertex, $\Gamma_{\mu}(q,p)$, 
the STI that one may obtain formally for  $q^{\mu}\Gamma_{\mu}(q,p)$ 
is the sum of two terms one of which is $p^{\mu}\Gamma_{\mu}(q,p)$; this
clearly limits its usefulness in applications.

The main practical consequence of these complicated STIs  
is that one cannot truncate (\ref{convSD}) in any obvious way 
without violating the transversality of the resulting $\Pi_{\alpha\beta}(q)$.
For example, keeping only graphs $(a_1)$ and $(a_2)$ is not correct even perturbatively,
since the ghost loop is crucial for the transversality of  $\Pi_{\alpha\beta}$
already at one-loop;
adding $(a_3)$ is still not sufficient for a SD analysis, because
(beyond one-loop) $q^{\alpha}[(a_1)+(a_2) + (a_3)]_{\alpha\beta} \neq 0$.

\subsection{The pinch technique \label{ptgeneral}}

The PT ~\cite{Cornwall:1981zr,Cornwall:1989gv} is a particular algorithm 
for rearranging the perturbative series 
 in such a way as to 
obtain  new Green's  functions 
that are independent  of the gauge-fixing 
parameter,  and satisfy to all orders 
ghost-free, QED-like WIs, 
instead of the usual STIs. The original motivation for resorting to it 
was precisely to devise a truncation scheme for the SDE that would 
preserve manifestly the gauge-invariance of the answer at every step.

Let us emphasize from the beginning  that, to date, there is no formal
definition of the PT procedure at the level of the functional integral
defining  the theory.  In  particular,  let us  assume  that the  path
integral has  been defined  using an arbitrary  gauge-fixing procedure
({\it i.e.}  linear covariant  gauges); then,  there  is no  known a  priori
procedure (such  as, {\it e.g.}, functional differentiation  with respect to
some combination of appropriately  defined sources) that would furnish
directly the  gauge-independent PT Green's  functions.  The definition
of the  PT procedure is operational,  and is intimately  linked to the
diagrammatic expansion of the theory  ({\it i.e.}  one must know the Feynman
rules).  In fact, the starting point of the PT construction can be any
gauge-fixing scheme that furnishes a set of well-defined Feynman rules
and   gauge-independent  physical   observables.    Specifically,  one
operates  at  a  certain  well-defined  subset of  diagrams,  and  the
subsequent  rearrangements give  rise to  the same  gfp-independent PT
answer, regardless of the  gauge-fixing scheme chosen for deriving the
Feynman  rules.   Note  however  that,  as  the  present  paper  amply
demonstrates (and as has already been emphasized in some of the earlier
literature), the PT is not diagrammatic, in the sense that
one does  not need  to operate  on individual graphs  but rather  on a
handful  of classes  of  diagrams (each  one  containing an  infinite
number  of  individual graphs).  This  is  the  enormous advantage  of
formulating the PT at the SD level.

The aforementioned  rearrangements of the PT are collectively implemented 
through the  systematic use of  the STIs satisfied by  certain Green's
functions  and  kernels; the latter constitute standard  ingredients  in  the
ordinary  perturbative  expansion  or  the SDE  of  the
various $n$-point Green's functions. In the Feynman gauge, which is 
by far the most convenient choice, 
the relevant STIs  are triggered 
by the action of a 
very special set of longitudinal  momenta. Specifically,
consider the subset of Feynman diagrams 
that have at least  one
{\it external} three-gluon vertex  
\be
\Gamma^{amn}_{\alpha\mu\nu}(q,k_1,k_2)=
-igf^{amn}\Gamma_{\alpha\mu\nu}(q,k_1,k_2).
\ee
By ``external''
we mean a vertex that has one of its legs irrigated by a physical momentum, to be denoted by $q$,
as opposed to a virtual momentum, 
that is being integrated over in the Feynman graph~\cite{Papavassiliou:1999az} 
(a detailed discussion of why only external vertices can pinch while all other three-gluon vertices inside the loops should remain unchanged is provided in Appendix~\ref{Appendix:howtopinch}).
Then  $\Gamma_{\alpha\mu\nu}(q,k_1,k_2)$ 
is decomposed as \cite{Cornwall:1981zr}
\bea
\Gamma_{\alpha\mu\nu}(q,k_1,k_2)&=&  \Gamma^{{\rm
F}}_{\alpha\mu\nu}(q,k_1,k_2)+ \Gamma^{{\rm
P}}_{\alpha\mu\nu}(q,k_1,k_2),     \nonumber  \\ 
\Gamma^{{\rm
F}}_{\alpha\mu\nu}(q,k_1,k_2)&=&(k_1-k_2)_\alpha  g_{\mu\nu}+2q_\nu
g_{\alpha\mu}-2q_\mu g_{\alpha\nu},    \nonumber \\    
\Gamma^{{\rm
P}}_{\alpha\mu\nu}(q,k_1,k_2)&=& k_{2\nu}g_{\alpha\mu}-
k_{1\mu}g_{\alpha\nu}.
\label{PTDEC}
\eea 
Evidently the above decomposition assigns a special role 
to the leg carrying the physical momentum $q$,  
and allows $\Gamma_{\alpha \mu \nu}^{{\rm F}}(q,k_1,k_2)$ to satisfy the tree-level WI
\be 
q^{\alpha} \Gamma_{\alpha \mu \nu}^{{\rm F}}(q,k_2,k_1) = 
(k_2^2 - k_1^2)g_{\mu\nu}, 
\label{WI2B}
\ee
where the rhs is the difference of two (tree-level) inverse 
propagators in the Feynman gauge.
Note that $\Gamma_{\alpha \mu \nu}^{{\rm F}}(q,k_1,k_2)$ ({\it i}) 
is Bose-symmetric only with respect to the
$\mu$ and $\nu$ legs, and ({\it ii}) it {\it coincides} with the BFM
three-gluon vertex 
involving a background gluon, ${\widehat A}_{\alpha}(q)$,
and two quantum gluons, $A_{\mu}(k_1)$ and $A_{\nu}(k_2)$, in the 
Feynman gauge (\ie when the quantum gauge-fixing parameter, $\xi_Q$, is chosen to be $\xi_Q=1$).
The term $\Gamma_{\alpha \mu \nu}^{{\rm P}}(q,k_1,k_2)$, which 
in configuration space corresponds to a pure divergence, plays the central role in the  
PT construction;  indeed, the main thrust of most PT demonstrations 
(in this article and many others before) is to essentially 
track down the precise action of the momenta contained inside $\Gamma_{\alpha \mu \nu}^{{\rm P}}$.
Specifically, $\Gamma_{\alpha \mu \nu}^{{\rm P}}$ 
contains the longitudinal  ``pinching'' momenta, which will get contracted 
with the kernels and Green's functions nested inside the remaining part of the diagram, triggering
the corresponding STIs; this, in turn, will produce 
the highly non-trivial rearrangements 
of the various terms characteristic of the PT. Quite remarkably, all these rearrangements 
finally amount to the modification of the ghost sector of the theory, 
reproducing dynamically the corresponding ghost sector of the BFM, leaving no residual terms behind.

The simplest example that demonstrates the action of the pinching momenta 
is the one-loop construction of the PT gluon self-energy
(see subsection~\ref{1lpt}): the STI triggered inside the conventional 
one-loop diagram $(a)$ in Fig.\ref {convse} is simply the tree-level version of Eq.~(\ref{sti3gv}), namely
\bea
k_1^{\mu} \Gamma_{\alpha \mu \nu}(q,k_1,k_2) &=& 
q^2 P_{\alpha\nu}(q) - k_2^2 P_{\alpha\nu}(k_2),
\label{sti3gvtree-1}\\
k_2^{\nu} \Gamma_{\alpha \mu \nu}(q,k_1,k_2) &=& k_1^2 P_{\alpha\mu}(k_1)
-q^2 P_{\alpha\mu}(q).
\label{sti3gvtree-2}
\eea
The terms proportional to an inverse propagator of the {\it external} leg ({\it i.e.}, to $q^2$), 
will cancel (when embedded into a physical process!) 
against similar contributions from other graphs (\eg vertex-graphs), 
also produced by the corresponding action of the pinching momenta inside them; in this case the 
 pinching momenta literary ``pinch out'' internal quark lines (hence the name of the technique).
Thus, {\it effectively}, inside a physical process,  
this particular subset of pinching contributions can be discarded altogether; this is the 
shortcut introduced in the ``intrinsic''~\cite{Cornwall:1989gv}. 
The remaining  pinching terms, namely those proportional to  the {\it internal} leg, 
are {\it instrumental} for obtaining the PT answer; 
in particular, they symmetrize the original $R_\xi$ 
ghost sector, so that it will finally coincide with the BFM ghost sector at $\xi_Q=1$ (see again subsection~\ref{1lpt}).

\subsection{Pinch technique and background field method: some conceptual issues}

As mentioned in the Introduction, the (gauge-independent) PT $n$-point functions {\it coincide} with the
BFM $n$-point functions when the latter are
computed at the special value $\xi_{Q}=1$ (BFG).
Even though this correspondence 
(and its correct interpretation) has been addressed in various places in the literature,
it may be useful to present a brief overview of some of the 
main subtleties associated with it. 

\begin{itemize}

\item[{\it i}.]  The objective of  the PT
construction is not to derive  diagrammatically the BFG, but 
rather to exploit the underlying BRST symmetry in order to 
expose a large number of cancellations, and eventually define 
gauge-independent Green's functions satisfying abelian WIs.
In fact, it was after more than a decade 
of independent PT  activity (when practically all one loop 
calculations had been carried out both 
in QCD and the electroweak sector) when the aforementioned correspondence was 
discovered (i.e. the PT results already existed, 
and then it was realized that they coincide with the results of the BFG).
Thus, while it is a remarkable and extraordinarily useful result
that the PT Green's functions can also be calculated in the
BFG, this needs a very extensive demonstration.  
Therefore, the correspondence must be  
verified at the end of the PT construction and should not be  
assumed beforehand. 

\item[{\it ii}.] It is well known that, at any order, the $S$-matrix $\widetilde{S}$ of the BFM,  
is equal to that in the conventional linear ($R_\xi$) gauges, {\it i.e.}, $\tilde{S}=S$
There is no way, however, to deduce from this equality 
the PT-BFG correspondence. Writing $S= \Gamma \Delta \Gamma + B$ and 
$\widetilde{S}= \widetilde{\Gamma} \widetilde{\Delta} \widetilde{\Gamma} + \widetilde{B}$, using that the 
box diagrams are equal in both schemes, {\it i.e.},  $B= \tilde{B}$, and, finally, 
observing that the PT does not change the unique $S$-matrix, one 
can deduce that  
$\Gamma \Delta \Gamma= \widehat{\Gamma} \widehat{\Delta} \widehat{\Gamma}$, and hence that 
$\widehat{\Gamma} \widehat{\Delta} \widehat{\Gamma}=\widetilde{\Gamma} \widetilde{\Delta} \widetilde{\Gamma}$.
But from this does {\it not} follow that $\widehat{\Delta}=\widetilde{\Delta}$ 
nor that $\widehat{\Gamma}=\widetilde{\Gamma}$; one must {\it prove  explicitly} the 
equality for {\it individual} Green's functions.

\item[{\it iii}.] We emphasize that the PT is a way of enforcing
gauge independence (and several other physical properties) for off-shell
Green's functions; the BFM, in a general gauge, is
not.  This is reflected in the fact that 
the BFM $n$-point  functions
are gauge-invariant, in the sense that they satisfy (by construction) QED-like WIs, 
but are {\it not} gauge-independent, \ie they depend explicitly 
on  $\xi_{Q}$. 
Had the BFM $n$-point functions been $\xi_{Q}$-independent, 
in addition to  being  gauge-invariant, there would be no need for introducing  
independently the PT.

\item[{\it iv}.] Notice that the  $\xi_{Q}$-dependent BFM Green's functions are 
not physically equivalent. This is best seen  
in theories with 
spontaneous symmetry breaking: the dependence of the 
BFM Green's functions on $\xi_Q$ gives rise 
to {\it unphysical} thresholds inside these Green's functions 
for $\xi_Q \neq 1$, a fact
which limits their usefulness for resummation purposes 
\cite{Papavassiliou:1995fq}.
Only the case of the BFG
is free from unphysical poles; that's because  
then (and only then) the BFM results collapse to the physical 
PT Green's functions.

\item[{\it v}.] The PT procedure has no a-priori knowledge of the BFM
built into it, despite the fact that the 
splitting of the regular three gluon vertex given in (\ref{PTDEC})
suggests such a preference.
Indeed, while  $\Gamma^{{\rm F}}$
 {\it coincides} with the BFG vertex, it only furnishes one 
piece of the final answer.
As already mentioned, the non-trivial part of the PT construction  
resides in what happens when the $\Gamma^{{\rm P}}$ part of the vertex
gets contracted with the Green's functions and kernels nested inside 
the corresponding SD diagram.
The correspondence with the BFG works finally 
only because
the WI triggered by $\Gamma^{{\rm P}}$ conspire in such a 
way as to reproduce dynamically the BFM ghost sector at $\xi_Q=1$, 
and nothing more. There is no a-priori way of knowing that this will 
indeed happen, and hence the need for the detailed demonstration 
presented in the next sections.

\item[{\it vi}.] Amplifying the previous point, notice that the  
PT works perfectly well in the context of  
non-covariant gauges (in fact it was first carried out in such a gauge), 
where the ghosts are decoupled from the $S$-matrix. Spectacularly enough, the 
PT procedure produces completely dynamically the necessary ghost sector,
from the STIs that are triggered. 

\item[{\it vii}.] Perhaps the most compelling fact 
that demonstrates that the PT and the BFM 
are intrinsically two completely disparate  methods is 
the following: one can apply the PT within the BFM.   
For example, 
the PT can be used to combine pieces of Feynman graphs in the
background Landau gauge, just as in any other gauge, and the usual 
PT results (those of the BFG) emerge. Operationally this is 
easy to understand: away from $\xi_Q =1$ even in the BFM there 
are longitudinal (pinching momenta) that will initiate the 
pinching procedure.  Ultimately, the BFG is singled out because of 
the total absence, in this particular gauge, of any such longitudinal 
momenta.

\item[{\it viii}.] We emphasize that  the PT construction  goes through
unaltered under  circumstances  where  the BFM
Feynman rules cannot even be  applied.  Specifically, if instead of an
$S$-matrix element  one were to consider a  different observable, such
as a current-current correlation function  or a Wilson loop (as was in
fact    done    by    Cornwall    in    the    original    formulation
\cite{Cornwall:1981zr}, and more recently in \cite{Binosi:2001hy}) one
could not start  out using the background Feynman  rules, because {\it
all} fields  appearing inside the  first non-trivial loop  are quantum
ones. Instead, by following the PT rearrangement inside these physical
amplitudes the unique PT answer emerges again.

\end{itemize}

\subsection{The pinch technique as a gauge-invariant truncation scheme\label{1lpt}}

Let us now see how the standard one-loop PT  construction 
contains the seed of a gauge-invariant truncation scheme for the SDE
of the gluon self-energy.
This exercise may seem trivial at first,
in the sense that 
no truncation is really needed,  given that the two diagrams 
comprising the full answer are elementary to calculate. However, it  
illustrates exactly   
how the PT rearrangement furnishes 
a transverse one-loop approximation for the conventional 
gluon self-energy, even if the (modified) ghost loop is omitted. 

\begin{figure}[t]
\includegraphics[width=13.5cm]{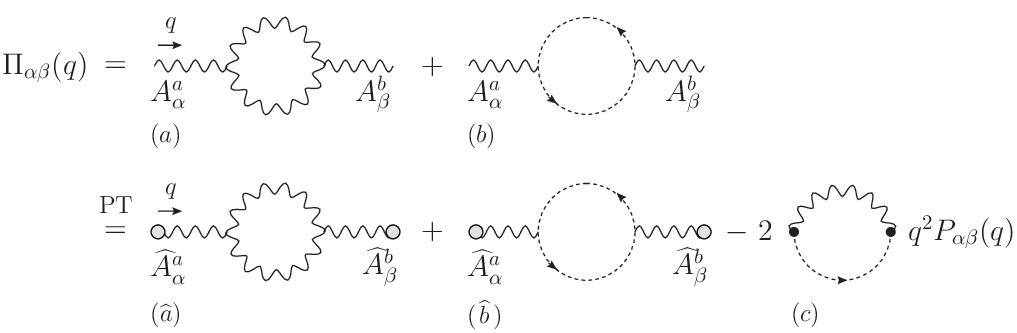}
\caption{The conventional one-loop gluon self-energy before (first line) and after (second line) the PT rearrangement.
A gray circle at the end of an external gluon line denotes that 
the corresponding gluon behaves as if it were a background gluon.}
\label{convse}
\end{figure}

In what follows we will use dimensional regularization, and will employ the short-hand notation $\int_k=\int d^d k/(2\pi)^d$, where $d=4-\epsilon$ is the space-time dimension. The conventional one-loop self-energy in the Feynman gauge, 
to be denoted by $\Pi_{\alpha\beta}^{(1)}(q)$, 
is given by 
the diagrams ($a$) and ($b$) in Fig.~\ref{convse}
(we set the ``seagull''-type contributions directly to zero, using the standard result $\int_k k^{-2} =0$).   
As is well known, neither ($a$) nor ($b$) is  transverse, and it is only their sum 
that furnishes a transverse answer for  $\Pi_{\alpha\beta}^{(1)}(q)$.
Specifically, setting 
\begin{equation}
f(q^2)=iC_A \frac{g^2}{48 \pi^2} \Gamma\left(\frac\epsilon2\right) \left(\frac{q^2}{\mu^2}\right)^{-\frac\epsilon2},
\end{equation}
with $C_A$ the Casimir eigenvalue of the adjoint representation ($C_A=N$ for $SU(N)$),
and dropping irrelevant constants, we have
\bea 
(a)_{\alpha\beta} &=& 
 \frac{1}{4}\, f(q^2) ( 19 q^2 g_{\alpha\beta} - 22  q_{\alpha}q_{\beta}),
\nonumber\\
(b)_{\alpha\beta} &=&
  \frac{1}{4}\, f(q^2) \,(q^2 g_{\alpha\beta} + 2   q_{\alpha}q_{\beta}),
\nonumber\\
\Pi_{\alpha\beta}^{(1)}(q) &=& 
5 q^2 f(q^2) P_{\alpha\beta}(q).
\label{abconv}
\eea

\begin{figure}[t]
\includegraphics[width=8.5cm]{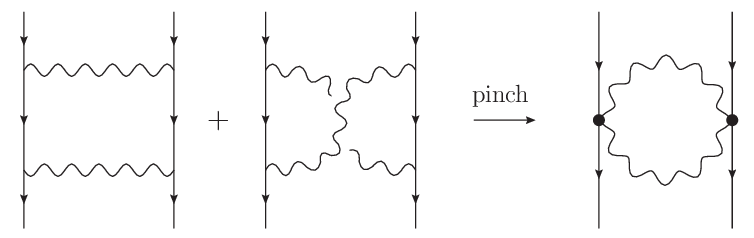}\\
\includegraphics[width=8.5cm]{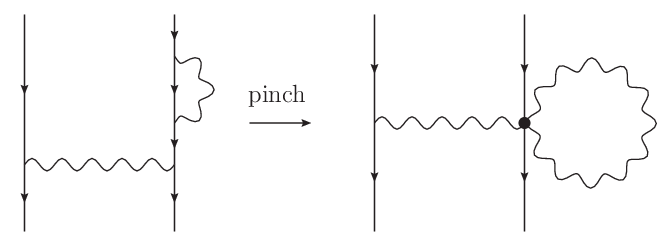}
\caption{\label{1l_pinch_contrib}Schematic representation of the pinching contributions one needs to consider away from the Feynman gauge $\xi=1$.}
\end{figure}

The application of the PT amounts to carrying out the following  
rearrangement of the two elementary three-gluon vertices
\bea
\Gamma_{\alpha \mu \nu}\Gamma^{\,\mu \nu}_{\beta}  &=&
\left[\Gamma^{{\rm F}}_{\alpha \mu \nu} + \Gamma^{{\rm P}}_{\alpha \mu \nu}\right]\,
\left[\Gamma^{{\rm F}\,\mu \nu}_{\beta} + \Gamma^{{\rm P}\,\mu\nu}_{\beta}\right]
\nonumber\\
&=&\Gamma^{{\rm F}}_{\alpha \mu \nu}\Gamma^{{\rm F}\,\mu \nu}_{\beta}
+\Gamma^{{\rm P}}_{\alpha \mu \nu}\Gamma^{\mu \nu}_{\beta}
+\Gamma_{\alpha \mu \nu}\Gamma^{{\rm P}\,\mu \nu}_{\beta}
-\Gamma^{{\rm P}}_{\alpha \mu \nu} \Gamma^{{\rm P}\,\mu\nu}_{\beta}.
\label{INPTDEC1}
\eea
Then, using the elementary WIs of Eq.s (\ref{sti3gvtree-1}) and~(\ref{sti3gvtree-2}) we have that 
\bea
\Gamma^{{\rm P}}_{\alpha \mu \nu}\Gamma_{\beta}^{\mu \nu}+
\Gamma_{\alpha\mu\nu}\Gamma^{{\rm P}\,\mu\nu}_{\beta}
&=& - 4 q^2 P_{\alpha\beta}(q) - 
2 k_{\alpha} k_{\beta} - 2 (k+q)_{\alpha}(k+q)_{\beta}, \label{GPG+GGP}\\
\Gamma^{{\rm P}}_{\alpha \mu \nu}\Gamma^{{\rm P}\,\mu\nu}_{\beta}  
&=& 2 k_{\alpha}k_{\beta}+ (k_{\alpha}q_{\beta}+q_{\alpha}k_{\beta}),
\label{GPGP}
\eea
where some terms have been set to zero by virtue of the dimensional regularization result $\int_k k^{-2} =0$.
Thus, one can cast $\Pi_{\alpha\beta}^{(1)}(q)$ in the following form:
\be
\Pi_{\alpha\beta}^{(1)}(q) = \frac{C_{A}g^{2}}{2}\left[ 
\int_k \frac{\Gamma^{{\rm F}}_{\alpha \mu \nu}\Gamma^{{\rm F}\,\mu \nu}_{\beta}}{k^2 (k+q)^2}
-2 \int_k \frac{(2k+q)_{\alpha}(2k+q)_{\beta}}{k^2 (k+q)^2}\right]
- 2  C_{A}g^{2} \int_k \frac{ q^2 P_{\alpha\beta}(q) }{k^2 (k+q)^2}. 
\label{propexp}
\ee
It is elementary to verify that each of the two terms in the square bracket on the rhs of 
(\ref{propexp})
are transverse; thus  the PT rearrangement has created 
three manifestly transverse structures. That in itself might not be so important, 
if it were not for the fact that these structures 
admit a special diagrammatic representation and a unique field-theoretic interpretation. 
Specifically, the two terms in the square bracket correspond precisely  
to diagrams $(\widehat{a})$ and $(\hspace{.05cm}\widehat{b}\hspace{.05cm})$ 
defining the one-loop gluon self-energy in the BFG,
to be denoted by $\widetilde\Pi_{\alpha\beta}^{(1)}(q)|_{\xi_Q=1}$;
note in particular the symmetrized gluon-ghost coupling. 
Thus, {\it it is as if the external gluons in $(a)$ and $(b)$ 
had been converted dynamically into background ones}.
As explained in \cite{Binosi:2007pi}, the third term on the rhs of (\ref{propexp}) 
is the one-loop 
expression of a special auxiliary Green's function, to be defined shortly;
it corresponds to diagram $(c)$ in  Fig.~\ref{convse}, and is 
generated from the first term on the rhs of (\ref{sti3gvtree-1}). 
The one-loop PT self-energy, to be denoted by $\widehat\Pi_{\alpha\beta}^{(1)}(q)$, 
is obtained by simply dropping this last term from the rhs of (\ref{propexp});
this defines the ``intrinsic'' PT~\cite{Cornwall:1989gv}. 
The completely equivalent way of 
saying this, corresponding to the ``$S$-matrix'' PT~\cite{Cornwall:1981zr}, is that 
the term corresponding to graph $(c)$ cancels exactly against a propagator-like
contribution extracted from the vertex graphs contributing to the full 
$S$-matrix element that one considers. 
Notice that this is true only in the Feynman gauge: away from $\xi=1$ additional pinching 
contributions need to be considered, {\it e.g.}, the ones coming from box and self-energy correction diagrams (Fig.~\ref{1l_pinch_contrib}).

\begin{figure}[t]
\includegraphics[width=8cm]{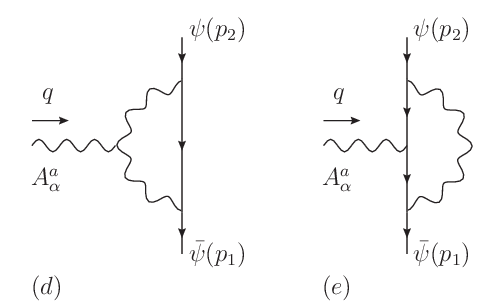}
\caption{The one-loop quark-gluon vertex appearing in the quark-quark elastic scattering  process.}
\label{convv}
\end{figure}

To see how the aforementioned cancellation comes about, 
let us now imagine
that the gluon self-energy $\Pi_{\alpha\beta}^{(1)}(q)$
is embedded into a physical 
process, such as the  $S$-matrix
element  for the  quark-quark elastic scattering  process 
$\bar\psi (r_1)\psi (r_2) \to \bar\psi(p_1)\psi (p_2) $,  with $q= r_1-r_2=p_2-p_1$ 
being the momentum transfer.
The one-loop quark-gluon vertex consists of the two graphs shown in Fig.~\ref{convv}.
Let us concentrate on the non-Abelian diagram $(d)$, and carry out the 
vertex decomposition of Eq.~(\ref{PTDEC})~\cite{Cornwall:1989gv}:
\bea
(d)
&=& \frac{1}{2} g^3 C_A t^{a} \int_k  
\frac{\Gamma_{\alpha \mu \nu}
\gamma^{\nu}S^{(0)}(\ksm + \psm_2)\gamma^{\mu}}{k^2(k+q)^2}
\nonumber\\
&=&\frac{1}{2} g^3 C_A  t^{a} \left[ \int_k 
\frac{\Gamma_{\alpha \mu \nu}^{{\rm F}}
\gamma^{\nu}
S^{(0)}(\ksm + \psm_2)\gamma^{\mu}}{k^2(k+q)^2}
+ 
\int_k 
\frac{\Gamma_{\alpha \mu \nu}^{{\rm P}}\gamma^{\nu}
S^{(0)}(\ksm + \psm_2)\gamma^{\mu}}{k^2(k+q)^2}
\right].
\eea
The first term on the rhs of the second line is the pure vertex-like part of $(d)$, while
the second term is purely propagator-like, as can be easily established  using
the elementary WI   
\be
 k_{\nu}\gamma^{\nu} = (\ksm + \psm - m) -  (\psm -m) 
\label{BasicWI}
\ee
The first term on the rhs of (\ref{BasicWI}) removes (pinches out) 
 the internal bare quark propagator \mbox{$S_0(\ksm + \psm)$},
whereas the second vanishes on shell, since 
$\bar{u}(p_2)(\not\! p_2 - m) = 0$ 
and  
$(\not\! p_1 - m)u(p_1)=0$. 
 Thus,
\be
\int_k 
\frac{\Gamma_{\alpha \mu \nu}^{{\rm P}}\gamma^{\nu}
S^{(0)}(\ksm + \psm_2)\gamma^{\mu}}{k^2(k+q)^2} 
\stackrel{\stackrel{{\rm PT}}{\rm Dirac\, Eq.}}{\longrightarrow}
2i \int_k \frac{1}{k^2(k+q)^2}\gamma_\alpha.
\ee
The self-energy-like 
contribution from the two vertex graphs (mirror graph included), 
to be denoted by $\Pi_{\alpha\beta}^{(1){\rm P}}(q)$, is given by
(longitudinal pieces may be added for free, due to current conservation)
\be
\Pi_{\alpha\beta}^{(1){\rm P}}(q) = 2 C_A g^2 
\int_k \frac{q^2\, P_{\alpha\beta}(q)}{k^2(k+q)^2}. 
\ee 

The PT one-loop quark-gluon vertex, to be denoted by
$\widehat\Gamma_{\alpha}^{a}(p_1,p_2)$, is given by~\cite{Papavassiliou:1989zd}
\bea
i\widehat\Gamma_{\alpha}^{a}(p_1,p_2) &=& g^2 t^{a} \left[
\frac{C_A }{2}\!  \int_k  
\frac{\Gamma_{\alpha \mu \nu}^{\rm F}
\gamma^{\nu} S^{(0)}(\ksm + \psm_2)\gamma^{\mu}}{k^2(k+q)^2}\right.\nonumber \\
&-&\left.\left(\!\frac{C_A}{2} -  C_f\!\right)\!
\int_k  
\frac{\gamma^{\mu} S^{(0)}(\ksm + \psm_2)\gamma^{\alpha}  S^{(0)}(\ksm + \psm_1)
\gamma_{\mu}}{k^2}
\right],
\label{PTvert}
\eea
where $C_f$ is the Casimir eigenvalue  of the fundamental representation [\mbox{$C_f=(N^2-1)/2N$} for $SU(N)$].
Now it is easy to derive the QED-like WI that the 
$\widehat\Gamma_{\alpha}^{a}(p_1,p_2)$ satisfies. Using (\ref{WI2B}),
we have that 
\bea
q^{\alpha} \widehat\Gamma_{\alpha}^{a}(p_1,p_2) &=&  -ig^3 t^{a}  C_f 
\left[
\int_k \frac{\gamma^{\mu} S^{(0)}(\ksm + \psm_2)\gamma_{\mu} }{k^2} -  
\int_k \frac{\gamma^{\mu} S^{(0)}(\ksm + \psm_1)\gamma_{\mu} }{k^2} 
\right]
\nonumber\\
&=& igt^{a}\left[ \Sigma (\psm_1) -  \Sigma (\psm_2)\right],
\label{WIvert}
\eea 
where $\Sigma(\psm)$ is the one-loop quark self-energy in the 
Feynman gauge~\cite{Papavassiliou:1994yi,Binosi:2001hy}.
 

Returning to the gluon self-energy, $\widehat{\Pi}^{(1)}_{\alpha\beta}(q)$ is
defined as 
\be
\widehat{\Pi}^{(1)}_{\alpha\beta}(q)=
\Pi^{(1)}_{\alpha\beta}(q)+ \Pi^{{\rm P}(1)}_{\alpha\beta}(q).
\ee
After carrying out the integrals one obtains
\bea 
({\widehat a})_{\alpha\beta} &=& 10 q^2 f(q^2)  P_{\alpha\beta}(q),
\nonumber\\
(\hspace{.05cm}{\widehat b}\hspace{.05cm})_{\alpha\beta} &=&  q^2 f(q^2) P_{\alpha\beta}(q),
\nonumber\\
(c)_{\alpha\beta}  &=& -6q^2  f(q^2) P_{\alpha\beta}(q),
\label{abcbfm}
\eea
and thus~\cite{Cornwall:1989gv}
\bea 
\widehat\Pi_{\alpha\beta}^{(1)}(q) &=& 
({\widehat a})_{\alpha\beta} + (\hspace{.05cm}{\widehat b}\hspace{.05cm})_{\alpha\beta}
\nonumber\\
&=&11  q^2  f(q^2) P_{\alpha\beta}(q)
= \widetilde\Pi_{\alpha\beta}^{(1)}(q)|_{\xi_Q=1}.
\label{ptbfm}
\eea
Note that the second line of (\ref{ptbfm}) expresses the PT-BFG correspondence 
at one loop~\cite{Denner:1994nn
}.

Then, Eq.~(\ref{propexp}) assumes the alternative form 
\be
\Pi_{\alpha\beta}^{(1)}(q) = \widehat\Pi_{\alpha\beta}^{(1)}(q) + (c)_{\alpha\beta}.
\label{BQIoneloop}
\ee
As has been explained in detail in~\cite{Binosi:2002ez}, and as we will see in the following sections, 
Eq.~(\ref{BQIoneloop})
is the one-loop version of a general identity~\cite{Grassi:1999tp
}, 
which we will call ``Background-Quantum'' identity (BQI)~\cite{Binosi:2002ez}, 
relating the conventional and the BFM self-energies in terms of an auxiliary Green's function, 
corresponding to graph $(c)$. This identity is valid 
to all orders in perturbation theory,
as well as non-perturbatively, and may be obtained either formally, by resorting to 
the Batalin-Vilkovisky (BV) formalism, or diagrammatically, as a by-product of the PT rearrangement of the 
conventional SD series; as we will see, in this latter case no reference to the BV formalism is necessary. 

Let us now focus on $\Pi_{\alpha\beta}^{(1)}(q)$ and  
imagine for a moment that no ghost loops may be considered when computing it, 
\ie the graphs $({\widehat b})_{\alpha\beta}$ must be omitted; 
in a SDE context  this ``omission'' would amount to a ``truncation'' of the series.
One may still obtain a {\it transverse} approximation  for 
$\Pi_{\alpha\beta}^{(1)}(q)$ with no ghost-loop, given by
\be
\Pi_{\alpha\beta}^{(1)} (q) = ({\widehat a})_{\alpha\beta} + (c)_{\alpha\beta} 
= 4  q^2 f (q^2) P_{\alpha\beta}(q).
\ee
Interestingly enough, the PT rearrangement offers already at one-loop   
the ability to truncate gauge-invariantly, \ie preserving  
the transversality of the truncated answer. 

\begin{figure}[!t]
\includegraphics[width=16cm]{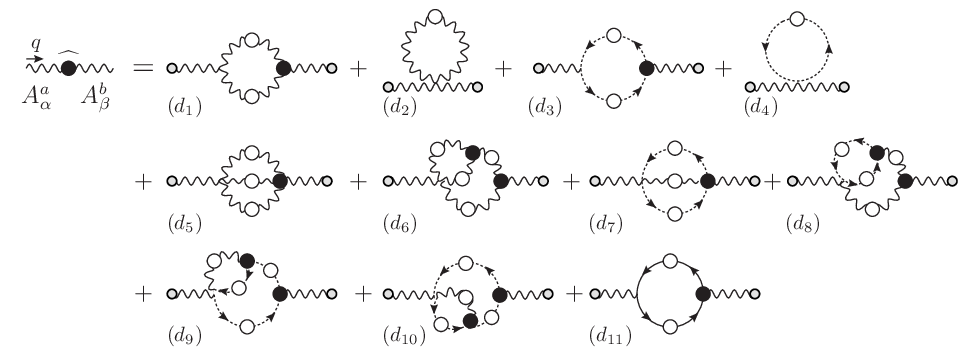}
\caption{The new Schwinger-Dyson series projected out dynamically by the PT algorithm. The symmetry factors are in this case $s(d_1,d_2,d_6)=1/2$, $s(d_5)=1/6$, and all the remaining diagrams have $s=-1$.}
\label{fig:PT_newSDE}
\end{figure}

\subsection{The new Schwinger-Dyson series\label{nsds}}

\begin{figure}[!t]
\includegraphics[width=9.5cm]{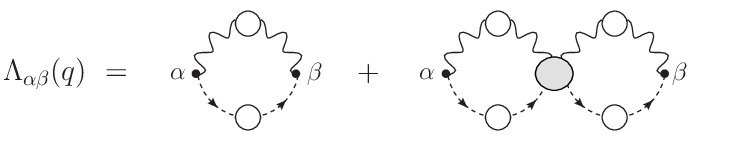}
\caption{Diagrammatic representation of the auxiliary functions $H$ and $\Lambda$.}
\label{fig:Lambda_aux}
\end{figure}

The implementation of the PT at the level of the SDE 
has been studied first in the context of scalar QED~\cite{Binosi:2006da}.
The corresponding construction in the case of quarkless
QCD has been recently carried out in a short communication~\cite{Binosi:2007pi},
where we restricted ourselves to the SDE of the 
gluon propagator. As has been explained there, the PT rearrangement
gives rise {\it dynamically} to a new SD series (see Fig. \ref{fig:PT_newSDE}), with the
following characteristics: on the rhs we have graphs that 
are made out of new vertices, but contain inside them the same gluon propagator as 
before, namely $\Delta_{\alpha\beta}(q)$. The new vertices, 
to be denoted by ${\g}_{\alpha\mu\nu}^{amn}$, ${\g}_{\alpha}^{anm}$, ${\g}_{\alpha\mu\nu\rho}^{amnr}$,
${\g}_{\alpha\mu}^{amnr}$, correspond 
precisely to the Feynman rules of the BFM in the Feynman gauge, \ie as already seen explicitly in the one-loop case,   
it is as if the external gluon had been converted dynamically into a background gluon.   
The lhs is composed from the sum of three terms:  
in addition to the term  $\Delta^{-1}(q^2) P_{\alpha\beta}(q)$, present there from the 
beginning, we have two additional contributions, 
$ 2 G(q^2) \Delta^{-1}(q^2) P_{\alpha\beta}(q)$ and $G^2(q^2) \Delta^{-1}(q^2) P_{\alpha\beta}(q)$,
which appear during the PT rearrangement of the rhs (and are subsequently 
carried to the lhs). 
The quantity  $G(q^2)$ is 
a special function, defined in terms of the gluon and ghost propagators as well as the auxiliary function 
$H_{\alpha\beta}$ of Fig.~\ref{fig:H_aux}.
Specifically, define
the following two-point function $\Lambda_{\alpha \beta}(q)$,   
(we suppress color indices)
\be
\Lambda_{\alpha \beta}(q) = C_{\rm {A}}
\int_k
H^{(0)}_{\mu\alpha}
D(k)\Delta^{\mu\nu}(q-k)\, H_{\nu\beta}(q-k,-q),
\label{gpert2}
\ee
with the diagrammatic representation shown in Fig.~\ref{fig:Lambda_aux}.
Then, $G(q^2)$ is defined as $i$ times the component of $\Lambda_{\alpha \beta}(q)$
multiplying $g_{\alpha\beta}$, namely
\be
\Lambda_{\alpha \beta}(q)= i g_{\alpha\beta} G(q^2) + \dots , 
\label{defG}
\ee
where the omitted terms are proportional to $q_{\alpha}q_{\beta}$.
Thus, the term appearing on the lhs of the new SDE is 
$\Delta^{-1}(q^2)[1+G(q^2)]^2 P_{\alpha\beta}(q)$.
So, one may write schematically
\be
\Delta^{-1}(q^2)[1+G(q^2)]^2 P_{\alpha\beta}(q) = q^2 P_{\alpha\beta}(q) + 
i \sum_{1=1}^{11}(d_i)_{\alpha\beta}, 
\label{newSDa}
\ee
or, equivalently, casting it into a more conventional form with the 
inverse of the unknown quantity isolated on the lhs, as 
\be
\Delta^{-1}(q^2)P_{\alpha\beta}(q)= 
\frac{q^2 P_{\alpha\beta}(q) + i \sum_{1=1}^{11}(d_i)_{\alpha\beta}}{[1+G(q^2)]^2}. 
\label{newSDb}
\ee

This new SD series has a very special structure. 
Let us first separate the diagrams on the rhs into four obvious categories:
one-loop (dressed) gluonic contributions [$(d_1)$ and $(d_2)$], 
one-loop ghost contributions [$(d_3)$ and $(d_4)$],
two-loop gluonic contributions [$(d_5)$ and $(d_6)$], and 
two-loop ghost contributions  [$(d_7)$, $(d_8)$, $(d_9)$ and $(d_{10})$]. 
It turns out that, 
by virtue of  the all-order WI
satisfied  by the  full vertices 
${\g}_{\alpha\mu\nu}^{amn}$, ${\g}_{\alpha}^{anm}$, ${\g}_{\alpha\mu\nu\rho}^{amnr}$,
${\g}_{\alpha\mu}^{amnr}$
appearing in the various diagrams, the contribution of each of the four subgroups  
is {\it individually} transverse.
Specifically, the four fundamental all-order WIs are given by
\bea
q^{\alpha}{\g}_{\alpha\mu\nu}^{amn}(q,k_1,k_2) &=&
gf^{amn}
\left[\Delta^{-1}_{\mu\nu}(k_1)
- \Delta^{-1}_{\mu\nu}(k_2)\right],
\label{3gl}\\
q^{\alpha}{\g}_{\alpha}^{anm}(q,k_1,k_2) &=&  igf^{amn}
\left[D^{-1}(k_1)- D^{-1}(k_2)\right],
\label{3gh}\\ 
q^{\alpha}{\g}_{\alpha\mu\nu\rho}^{amnr}(q,k_1,k_2,k_3) &=&
g f^{adr} {\Gamma}_{\nu\rho\mu}^{drm}(q+k_2,k_3,k_1)
+g f^{adn} {\Gamma}_{\nu\mu\rho}^{dmr}(q+k_3,k_1,k_2)
\nonumber\\
 &+&
g f^{adm} {\Gamma}_{\mu\nu\rho}^{dnr}(q+k_1,k_2,k_3),
\label{4gl}\\
q^{\alpha}
{\g}_{\alpha\mu}^{amnr}(q,k_1,k_2,k_3) &=&
-g f^{ame} \Gamma_{\mu}^{enr}(q+k_1,k_2,k_3)
-g f^{ane} \Gamma_{\mu}^{mer}(k_1,q+k_2,k_3)
\nonumber\\
&-& gf^{are}\Gamma_{\mu}^{mne}(k_1,k_2,q+k_3).
\label{4gh}
\eea
Using these WIs  one may show after some elementary operations that~\cite{Aguilar:2006gr} 
\bea
q^{\alpha}\left[ (d_1) + (d_2) \right]_{\alpha\beta} &=& 0,
\nonumber\\
q^{\alpha}\left[(d_3) + (d_4)\right]_{\alpha\beta} &=& 0,
\nonumber\\
q^{\alpha}\left[
(d_5) + (d_6)\right]_{\alpha\beta} &=& 0,
\nonumber\\
q^{\alpha}\left[(d_7) + (d_8) + (d_9) + (d_{10})\right]_{\alpha\beta} &=& 0.
\label{4tr}
\eea
[Notice that the one-loop dressed fermionic contributions $(d_{11})$ trivially satisfy this transversality property.] 

As has been pointed out in \cite{Binosi:2007pi},
this special property has far-reaching practical consequences 
for the  treatment of the SD series. Specifically, it furnishes a systematic truncation scheme
that preserves the transversality of the answer. For example, keeping only the 
diagrams in the first group, we obtain the truncated SDE
\be
\Delta^{-1}(q^2) P_{\alpha\beta}(q) = 
\frac{q^2 P_{\alpha\beta}(q) + i[(d_1)+(d_2)]_{\alpha\beta}}{[1+G(q^2)]^2}, 
\label{trua}
\ee
and from the first equation of (\ref{4tr}) we know that 
$[(d_1)+(d_2)]_{\alpha\beta}$ is transverse, \ie 
\mbox{$[(d_1)+(d_2)]_{\alpha\beta}= {(d-1)}^{-1} [(d_1)+(d_2)]^{\mu}_{\mu}P_{\alpha\beta}(q)$}.
Thus,  the transverse projector $P_{\alpha\beta}(q)$ appears 
{\it exactly}
on both sides of (\ref{trua}); one may subsequently 
isolate the scalar cofactors on both sides
obtaining a scalar equation of the form
\be
\Delta^{-1}(q^2) = 
\frac{q^2 + \frac{i}{(d-1)}[(d_1)+(d_2)]^{\mu}_{\mu}}{[1+G(q^2)]^2}. 
\ee
A truncated equation similar to (\ref{trua}) may be written for any other 
of the four groups, or for sums of these groups, 
without compromising the transversality of the answer. 
The price one has to pay for this advantageous situation 
is that one must consider in  addition the    
equation determining $G(q^2)$, \ie the $g_{\alpha\beta}$ part of 
Eq.~(\ref{gpert2}). This price is, however, rather modest, 
given that Eq.~(\ref{gpert2}) may be approximated 
introducing, for example, a dressed-loop expansion (see Fig.\ref{fig:H_aux}), without 
jeopardizing the transversality of $\Pi_{\alpha\beta}(q)$, given that 
$[1+G(q^2)]^2$ affects only the size of the scalar prefactor.

In going from Eq.~(\ref{newSDa}) to Eq.~(\ref{newSDb})
one essentially chooses to retain the original propagator $\Delta(q)$ as the  
unknown quantity, to be dynamically determined from the SDE.
There is, of course, an alternative strategy: one may define 
a new ``variable'' from the quantity appearing on the lhs (\ref{newSDa}),  
namely 
\be
\widehat{\Delta} (q)\equiv \left[1+G(q^2)\right]^{-2} {\Delta}(q),
\label{BQI}
\ee
which leads to  a new form for (\ref{newSDa}),
\be
\widehat\Delta^{-1}(q^2)P_{\alpha\beta}(q) = q^2 P_{\alpha\beta}(q) + i
\sum_{i=1}^{11}(d_i)_{\alpha\beta} \,.
\label{newSDa2}
\ee
Obviously, the special transversality properties established above holds as well for 
Eq.~(\ref{newSDa2}); for example, one may truncate it gauge-invariantly as
\be
\widehat\Delta^{-1}(q^2) P_{\alpha\beta}(q) = 
q^2 P_{\mu\nu}(q) + i[(d_1)+(d_2)]_{\alpha\beta}.
\label{trub}
\ee

Should one opt for treating  $\widehat{\Delta} (q)$ as the 
new unknown quantity, then  an additional step must be carried out:
one must 
use (\ref{BQI}) to rewrite  the entire rhs of (\ref{newSDa2}) 
in terms of $\widehat{\Delta}$ instead of $\Delta$, \ie
carry out the replacement  
\mbox{$\Delta \to \left[1+G\right]^2 \,\widehat{\Delta}$}
 {\it inside} every diagram on the rhs of Eq.~(\ref{newSDa2})
that contains $\Delta$'s.

Let us discuss further these two versions of the SDE.
Eq.~(\ref{trua}) furnishes a gauge-invariant approximation
for the conventional gluon self-energy $\Delta(q)$, 
whereas Eq.~(\ref{newSDa2})
is the gauge-invariant approximation for the effective 
PT self-energy $\widehat{\Delta}$. The crucial point is that
one may switch from one to the other by means of Eq.~(\ref{BQI}).
For practical purposes this means for example, that one
may get a gauge-invariant approximation not just for the PT quantity
(background Feynman gauge)
but also for the 
{\it conventional} self-energy computed in the Feynman gauge.
Eq.~(\ref{BQI}), which is the all-order generalization of the one-loop 
relation given in Eq.~(\ref{BQIoneloop}),  
plays an instrumental
role in this entire construction, allowing one 
to convert the SDE series into a dynamical equation for either $\widehat{\Delta} (q)$ 
or $\Delta(q)$.

\section{The formal machinery \label{fm}} 

The extension of the  PT algorithm to
the SDEs  of QCD is a  challenging exercise, mainly due  to the large
amount of  different Green's functions one needs  to manipulate in
the  process. 
Most of these Green's functions are generated when longitudinal momenta trigger  
the  STIs satisfied  by  specific subsets  of  fully dressed  vertices
appearing  in  the   ordinary  perturbative  expansion. 
Due to  the  non-linearity  of the  BRST
transformation [see Eq.~(\ref{BRSTtrans}) below] 
they involve
composite  operators; specifically,  they are  of  the type  
$\langle0\vert T[s\Phi(x)\cdots]\vert0\rangle$ with $s$  the BRST operator and $\Phi$
a generic QCD field.

It turns out that the most efficient framework for dealing with this type of quantities is the BV formalism, 
allowing the construction of these auxiliary (ghost) Green's functions in terms of
 a well-defined set of Feynman rules. In addition, this formalism furnishes
a set of useful identities (the BQIs mentioned in the previous section), 
relating Green's functions involving background fields to 
Green's functions involving quantum fields. 

In this  section, we fix  our conventions, present the QCD
Lagrangian  and   its  gauge-  fixing   procedure  (concentrating,  in
particular, on the  conventional $R_\xi$ gauges and the  BFM), and 
briefly  review the  BV formalism.   Then,  we proceed  to describe how one can extract
from the master equations the all-order STIs and BQIs needed in the coming PT construction,
postponing their actual derivation to the Appendix~\ref{Appendix:STIs} and~\ref{Appendix:BQIs}. We will also describe how 
to derive the so-called Faddeev-Popov equations (see also Appendix~\ref{Appendix:FPEs}). 
Finally, in  the process of describing  all the above topics, we  will introduce a
particularly  compact notation  for Green's  functions,  which encodes
unambiguously  all relevant  information (\ie  the  particle content,
Lorentz and color structure, and momenta flow).

\subsection{QCD Lagrangian and gauge fixing schemes}

Throughout the paper we will adopt the conventions of the book by Peskin \& Schr\"oder ~\cite{Peskin:1995ev}.
 The QCD Lagrangian density is given by
\begin{equation}
{\cal L} = {\cal L}_{\mathrm I} + {\cal L}_{\mathrm{GF}} + {\cal L}_{\mathrm{FPG}}.
\label{QCD_lag}
\end{equation}
${\cal L}_{\mathrm I}$ represents the gauge invariant $SU(3)$ Lagrangian, namely
\begin{equation}
{\cal L}_{\mathrm I} = -\frac14 F_a^{\mu\nu}F^a_{\mu\nu}+\bar{\psi}^i_\mathrm{f}
\left(i\gamma^\mu{\cal D}_\mu-m\right)_{ij}\psi^j_\mathrm{f} ,
\label{Linv}
\end{equation}
where $a=1,\dots,8$ (respectively $i,j=1,2,3$) is the color index for the adjoint (respectively fundamental) 
representation, while ``f'' represents the flavor index. 
The field strength is
\begin{equation}
F^a_{\mu\nu}=\partial_\mu A^a_\nu-\partial_\nu A^a_\mu+gf^{abc}A^b_\mu A^c_\nu,
\end{equation}
and the covariant derivative is defined according to
\begin{equation}
({\cal D}_\mu)_{ij}=\partial_\mu (\mathbb{I})_{ij}-ig A^a_\mu (t^a)_{ij},
\end{equation}
with $g$ the (strong) coupling constant. Finally, the $SU(3)$ generators $t^a$ satisfy the commutation relations
\begin{equation}
[t^a,t^b]=if^{abc}t^c,
\end{equation}
with $f^{abc}$ the totally antisymmetric $SU(3)$ structure constants. \\
${\cal L}_{\mathrm{GF}}$ and ${\cal L}_{\mathrm{FPG}}$ represent respectively the (covariant) gauge fixing Lagrangian and its associated Faddeev-Popov ghost term. The most general way of writing these terms is through the expressions
\begin{eqnarray}
{\cal L}_{\mathrm{GF}}&=&-\frac\xi2(B^a)^2+B^a{\cal F}^a,\\
{\cal L}_{\mathrm{FPG}} &=&-\bar c^a s{\cal F}^a.
\label{FPG_Lag}
\end{eqnarray}
In the formulas above ${\cal F}^a$ is the gauge fixing function, and the $B^a$ are auxiliary, 
non-dynamical fields (the so called Nakanishi-Lautrup multipliers) that can be eliminated through their (trivial) equations of motion; $c^a$ (respectively, $\bar c^a$) are the ghost (respectively, anti-ghost) fields, and, finally, $s$ is the BRST operator, with the BRST transformations of the QCD fields given by
\begin{eqnarray}
sA^a_\mu=\partial_\mu c^a+gf^{abc}A^b_\mu c^c
&\qquad& s c^a=-\frac12g f^{abc}c^bc^c, 
\nonumber \\
s\psi^i_\mathrm{f}=ig c^a(t^a)_{ij}\psi^j_\mathrm{f} &\qquad& s\bar c^a=B^a, \nonumber \\
s\bar\psi^i_\mathrm{f}=-ig c^a\bar\psi^j_\mathrm{f} (t^a)_{ji}  &\qquad& sB^a=0.
\label{BRSTtrans}
\end{eqnarray}
We thus see that the sum of the gauge fixing and Faddeev-Popov terms can be written as   a total BRST variation:
\begin{eqnarray}
{\cal L}_{\mathrm{GF}}+{\cal L}_{\mathrm{FPG}}=s\left(\bar c^a\mathcal{F}^a-\frac\xi2\bar c^a B^a\right).
\label{gffer}
\end{eqnarray}
This is of course expected, since it is well known that total BRST variations cannot appear in the physical spectrum of the theory.
For our purposes, the gauge-fixing functions of interest are the ones corresponding to the $R_\xi$ (renormalizable $\xi$ gauges) and the BFM, which we describe in what follows. 

\begin{enumerate}

\item In the usual $R_\xi$ gauges, the gauge fixing function is chosen to be ${\cal F}^a_{R_\xi}=\partial^\mu A^a_\mu$; therefore one finds
\begin{eqnarray}
{\cal L}_{\mathrm{GF}}&=&\frac1{2\xi}(\partial^\mu A^a_\mu)^2,\\
{\cal L}_{\mathrm{FPG}} &=&\partial^\mu\bar c^a \partial_\mu c^a+gf^{abc}(\partial^\mu\bar c^a)A^b_\mu c^c
\end{eqnarray}

\item In the case of the BFM, one starts by splitting the gluon field into a 
background part, $\widehat{A}^a_\mu$, and a quantum part, $A^a_\mu$. 
Notice that the BRST variation of the background field will be zero, 
but the latter will enter in the variation of the quantum one, since
\begin{equation}
sA^a_\mu=\partial_\mu c^a+gf^{abc}(A^b_\mu+\widehat{A}^b_\mu)c^c.
\end{equation}
The gauge fixing function is 
\begin{eqnarray}
{\cal F}^a_\mathrm{BFM}&=&(\widehat{\cal D}^\mu A_\mu)^a\nonumber \\
&=&\partial^\mu A_\mu^a+gf^{abc}\widehat{A}^b_\mu A_c^\mu,
\label{BFMgff}
\end{eqnarray} 
which gives in turn 
\begin{eqnarray}
{\cal L}_{\mathrm{GF}}&=&\frac1{2\xi}(\partial^\mu A^a_\mu)^2+\frac1\xi gf^{abc}(\partial^\mu A^a_\mu)\widehat{A}^b_\nu A^\nu_c+\frac1{2\xi}g^2f^{abe}f^{cde}\widehat{A}^a_\mu A^\mu_b\widehat{A}^c_\nu A^\nu_d,\\
{\cal L}_{\mathrm{FPG}}&=&\partial^\mu\bar c^a \partial_\mu c^a+gf^{abc}(\partial^\mu\bar c^a)A^b_\mu c^c+gf^{abc}(\partial^\mu\bar c^a)\widehat{A}^b_\mu c^c-gf^{abc}\bar c^a\widehat{A}^b_\mu(\partial^\mu c^c)\nonumber \\
&-&g^2f^{abe}f^{cde}\bar c^a\widehat{A}^b_\mu(A_c^\mu+\widehat{A}^c_\mu)c^d.
\label{L-BFM}
\end{eqnarray}
We thus see the appearance  of the characteristic ghost sector for the interaction with background gluons, consisting in a symmetric $\widehat{A}c\bar c$ ghost vertex and a four particle $\widehat{A}Ac\bar c$ one.
\end{enumerate}

\subsection{Green's functions: conventions}

\begin{figure}[!t]
\includegraphics[width=5cm]{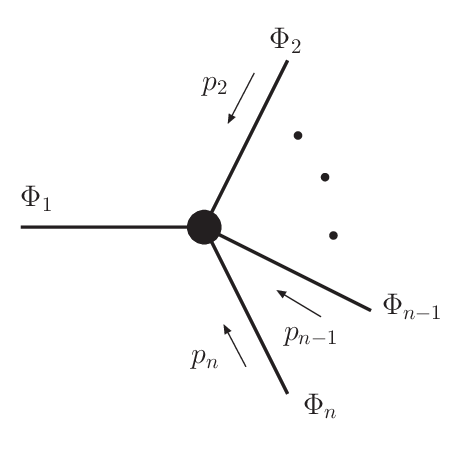}
\caption{Our conventions for the (1PI) Green's functions $\Gamma_{\Phi_1\cdots\Phi_n}(p_1,\dots,p_n)$. All momenta $p_2,\dots,p_n$ are assumed to be incoming, and are assigned to the corresponding fields starting from the rightmost one. The momentum of the leftmost field $\Phi_1$ is determined through momentum conservation ($\sum_i p_i=0$) and will be suppressed.}
\label{fig:Green_conv}
\end{figure}

The Green's functions of the theory can be constructed in terms of time-ordered products of free fields $\Phi^0_1\cdots\Phi^0_n$ and vertices of the interaction Lagrangian ${\cal L}_\mathrm{int}$ (constructed from the pieces of ${\cal L}$ which are not bilinear in the fields) through the standard Gell-Man--Low formula for the 1PI truncated Green's functions
\begin{eqnarray}
\Gamma_{\Phi_1\cdots\Phi_n}(x_1,\dots,x_n)&=&\langle T[\Phi_1(x_1)\cdots\Phi_n(x_n)]\rangle^\mathrm{1PI}\nonumber \\
&=& \langle T[\Phi^0_1(x_1)\cdots\Phi^0_n(x_n)]\exp(-i\int\!d^4x\,{\cal L}_\mathrm{int})\rangle^\mathrm{1PI}.
\label{gml}
\end{eqnarray}
The complete set of Green's functions can be handled most efficiently by introducing a generating functional, which in Fourier space reads
\begin{equation}
\Gamma[\Phi]=\sum_{n=0}^\infty\frac{(-i)^n}{n!}\int\!\prod_{i=0}^nd^4p_i\ \delta^4(\sum_{j=1}^np_j)\Phi_1(p_1)\cdots\Phi_n(p_n)\Gamma_{\Phi_1\cdots\Phi_n}(p_1,\dots,p_n),
\end{equation}
with $p_i$ the (in-going) momentum of the $\Phi_i$ field. 
Since in perturbation theory $\Gamma_{\Phi_1\cdots\Phi_n}$ is a formal power series in $\hbar$, we will denote its  $m$-loop contribution as $\Gamma^{(m)}_{\Phi_1\cdots\Phi_n}$. 
Then, in terms of the generating functional $\Gamma[\Phi]$ any of the Green's function of the theory can be obtained by means of functional derivatives:
\begin{equation}  
\Gamma_{\Phi_1\cdots\Phi_n}(p_1,\dots,p_n) = i^n\left.\frac{\delta^{n}\Gamma}{\delta\Phi_1(p_1)\delta\Phi_2(p_2)\cdots\delta\Phi_n(p_n)}\right|_{\Phi_i=0},
\label{greenfunc}
\end{equation} 
where $\Phi(p)$ denotes the Fourier transform of $\Phi(x)$ and our convention on the external momenta is  summarized in Fig.~\ref{fig:Green_conv}.
From the definition given in Eq.~(\ref{greenfunc}) it follows that the Green's functions $i^{-n}\Gamma_{\Phi_1\cdots\Phi_n}$ are simply given by the corresponding Feynman diagrams in Minkowski space. Finally, notice that upon inversion of two (adjacent) fields we have
\begin{equation}
\Gamma_{\Phi_1\cdots\Phi_i\Phi_{i+1}\cdots\Phi_n}(p_1,\dots,p_i,p_{i+1},\dots,p_n)=\pm
\Gamma_{\Phi_1\cdots\Phi_{i+1}\Phi_i\cdots\Phi_n}(p_1,\dots,p_{i+1},p_i,\dots,p_n),
\end{equation}
with the minus appearing only when both fields $\Phi_i$ and $\Phi_{i+1}$ obey Fermi statistics.

The Green's functions constructed so far are sufficient for building 
all possible amplitudes involved in the $S$-matrix computation; however, due to the non-linearity of the BRST transformations [Eq.~(\ref{BRSTtrans})], they do not cover the complete set of Green's functions appearing in the STIs of the theory (and therefore needed for its renormalization, as well as the PT construction). 

\subsection{A brief introduction to the Batalin-Vilkovisky formalism \label{BV}}

In this subsection we review briefly the BV formalism~\cite{Batalin:1977pb}, 
which allows one to obtain both the STIs as well as the BQIs of the theory at hand.
In order to simplify the notation (and since they will not play any role in what follows) 
we will suppress from now on all spinor indices (both flavor and color).

Let us  then start by introducing  for each field  $\Phi$ appearing in
the  theory  a  corresponding  anti-field, to be denoted by $\Phi^*$. 
The anti-field $\Phi^*$
has opposite  statistics   with  respect   to  $\Phi$; its ghost charge, ${\rm gh}(\Phi^*)$,
is related to the  ghost  charge ${\rm gh}(\Phi)$ of the field $\Phi$ by 
\mbox{${\rm gh}(\Phi^*)=-1- {\rm gh}(\Phi)$}.
For convenience, we summarize the ghost charges and statistics of the various QCD fields and anti-fields 
in Table~\ref{tableI}.
\begin{table}
\begin{tabular}{|r||c|c|c|c|c|c|c|c|c|c|c||c|c|}
\hline
 & $\ A^m_\mu\ $ & $\ \psi^i_\mathrm{f}\ $ & $\ \bar\psi^i_\mathrm{f}\ $ & $\ c^m\ $ & $\ \bar c^m\ $  & $\ B^m\ $ &  $\ A^{*m}_\mu\ $ & $\ \psi^{*i}_\mathrm{f}\ $ & $\ \bar\psi^{*i}_\mathrm{f}\ $ & $\ c^{*m}\ $ & $\ \bar c^{*m}\ $ & $\ \widehat{A}^m_\mu\ $ & $\ \Omega^m_\mu\ $\\
\hline\hline
Ghost ch. & 0 & 0 & 0 & 1 & -1  & 0  & -1 & -1 &  -1 & -2 & 0 & 0 & 1\\
\hline
Stat.  & B & F & F & F & F  & B & F & B & B & B & B & B & F\\
\hline
Dim. & 1 & $\frac32$ & $\frac32$ & 0 & 2 & 2 & 3 & $\frac52$ & $\frac52$ & 4 & 2 & 1 & 1 \\
\hline
\end{tabular} 
\caption{Ghost charge, statistics (B for Bose, F for Fermi), and mass dimension of the QCD
fields,  anti-fields and background sources. \label{tableI}}
\end{table}
Next, we add to the original gauge invariant Lagrangian a 
term coupling the anti-fields with the BRST variation of the corresponding fields, to get
\begin{eqnarray}
{\cal L}_{\mathrm{BV}}&=&{\cal L}_{\mathrm I}+{\cal L}_{\mathrm{BRST}}, \nonumber \\
{\cal L}_{\mathrm{BRST}}&=& \sum_{\Phi}\Phi^*s\Phi\nonumber \\
&=&A^{*a}_\mu(\partial^\mu c^a+gf^{abc}A_b^\mu c^c)-\frac12gf^{abc}c^{*a}c^bc^c+ig\bar\psi^*c^at^a\psi-igc^a\bar\psi t^a\psi^*\nonumber \\
&+&\bar c^{*a} B^a
\label{BRST_Lag}
\end{eqnarray}
Then, the action $\Gamma^{(0)}[\Phi,\Phi^*]$ constructed from ${\cal L}_{\mathrm{BV}}$, 
will satisfy the master equation
\begin{equation}
\int\!\!d^4x\sum_\Phi\frac{\delta\Gamma^{(0)}}{\delta\Phi^*}\frac{\delta\Gamma^{(0)}}{\delta\Phi}=0.
\label{master_eq}
\end{equation}
To verify this, observe that, on one hand, the terms in $\delta\Gamma^{(0)}/{\delta\Phi}$ that are independent from the anti-fields $\Phi^*$ are zero due the BRST (actually the gauge) invariance of the action
\begin{equation}
\int\!\!d^4x\sum_\Phi s\Phi\frac{\delta\Gamma^{(0)}_\mathrm{I}}{\delta\Phi}=\int\!\!d^4x (s\Gamma^{(0)}_\mathrm{I}[\Phi])=0\,;
\end{equation} 
on the other hand, terms in $\delta\Gamma^{(0)}/{\delta\Phi}$ that are linear in the anti-fields vanish due to the nihilpotency of the BRST operator
\begin{equation}
\int\!\!d^4x\sum_{\Phi,\Phi'} s\Phi'\frac{\delta (s\Phi)}{\delta\Phi'}=\int\!\!d^4x \sum_{\Phi} s^2\Phi=0.
\end{equation}

Now, since the anti-fields are external  sources, we must constrain them to
suitable values before we can use the action $\Gamma^{(0)}$ in calculations of $S$-matrix
elements~\cite{Weinberg:1996kr}. 
To  that end, we  introduce an  arbitrary
fermionic  functional, $\Psi[\Phi]$, with ghost charge -1, and set for all the anti-fields $\Phi^*$
\begin{equation}
\Phi^*=\frac{\delta\Psi[\Phi]}{\delta\Phi}.
\end{equation}
Then the action becomes
\begin{eqnarray}
\Gamma^{(0)}[\Phi,\delta\Psi/\delta\Phi]&=&\Gamma^{(0)}_\mathrm{I}[\Phi]+(s\Phi)\frac{\delta\Psi[\Phi]}{\delta\Phi}\nonumber \\
&=& \Gamma^{(0)}_\mathrm{I}[\Phi]+s\Psi[\Phi],
\end{eqnarray}
and therefore, choosing the functional $\Psi$ to satisfy the relation
\begin{equation}
s\Psi=\int\!d^4x\left({\cal L}_{\mathrm{GF}}+{\cal L}_{\mathrm{FPG}}\right),
\end{equation}  
we see that the action $\Gamma^{(0)}$ (obtained from ${\cal L}_{\mathrm{BV}}$) 
is equivalent to the gauge-fixed action obtained from the original Lagrangian ${\cal L}$ of Eq.~(\ref{QCD_lag}). The functional $\Psi$ is often referred to  as the ``gauge fixing fermion''.

The BRST symmetry is crucial for endowing a theory  with a unitary $S$-matrix 
and gauge-independent physical observables; therefore, it must be implemented to all orders. 
For doing so we establish the quantum corrected version of the master equation (\ref{master_eq}) in the form of the STI functional
\begin{eqnarray}
{\cal S}(\Gamma)[\Phi]&=&\int\!d^4x\sum_\Phi\frac{\delta\Gamma}{\delta\Phi^*}\frac{\delta\Gamma}{\delta\Phi}\nonumber \\
&=&\int\!d^4x\left\{\frac{\delta\Gamma}{\delta A^{*\mu}_m}\frac{\delta\Gamma}{\delta A^{m}_\mu}+\frac{\delta\Gamma}{\delta c^{*m}}\frac{\delta\Gamma}{\delta c^{m}}+\frac{\delta\Gamma}{\delta \psi^{*}}\frac{\delta\Gamma}{\delta\bar \psi}+\frac{\delta\Gamma}{\delta \psi}\frac{\delta\Gamma}{\delta\bar \psi^*}+B^m\frac{\delta\Gamma}{\delta\bar c^m}\right\}\nonumber \\
&=&0,
\label{STIfunc_nm}
\end{eqnarray}
where $\Gamma[\Phi,\Phi^*]$ is now the effective action. 

In order to simplify the structure of the STI generating functional of Eq.~(\ref{STIfunc_nm}), 
let us notice that the anti-ghost  $\bar c^{a}$ and the multiplier $B^a$ have {\it linear} BRST transformations; 
therefore they do not present the usual complications (due to non-linearity) of the other QCD fields. 
Together with their corresponding anti-field, they enter bi-linearly in the action, and one can write the complete action (which we now explicitly indicate it with a C subscript) as a sum of a minimal and non-minimal sector
\begin{equation}
\Gamma_\mathrm{C}^{(0)}[\Phi,\Phi^*]=\Gamma^{(0)}[A,A^*,\psi,\psi^*,\bar\psi,\bar\psi^*,c,c^{*}]+\bar c^{*a}B^a.
\end{equation}
The last term has no effect on the master equation (\ref{master_eq}), which is satisfied by
$\Gamma^{(0)}$ alone; the fields $\{A^a_\mu,A^{*a}_\mu,\psi,\psi^*,\bar\psi,\bar\psi^*,c^a,c^{*a}\}$ are then often called {\it minimal variables} while  $\bar c^{a}$ and $B^a$ are referred to as non-minimal variables or ``trivial pairs''. 
Equivalently one can introduce the minimal (or reduced) action by subtracting from the complete one the local term corresponding to the gauge-fixing Lagrangian, \ie
\be
\Gamma=\Gamma_\mathrm{C}-\int\!d^4x\,{\cal L}_\mathrm{GF}.
\ee 
In either cases, the result is that the STI functional is now written as
\be
{\cal S}(\Gamma)[\Phi]
=\int\!d^4x\left\{\frac{\delta\Gamma}{\delta A^{*\mu}_m}\frac{\delta\Gamma}{\delta A^{m}_\mu}+\frac{\delta\Gamma}{\delta c^{*m}}\frac{\delta\Gamma}{\delta c^{m}}+\frac{\delta\Gamma}{\delta \psi^{*}}\frac{\delta\Gamma}{\delta\bar \psi}+\frac{\delta\Gamma}{\delta \psi}\frac{\delta\Gamma}{\delta\bar \psi^*}\right\}=0.
\label{STIfunc}
\ee
In practice, the STIs generated from the functional of Eq.~(\ref{STIfunc}) coincide with the one obtained by the complete one after the implementation of the Faddeev-Popov equation described in the next subsection \cite{Itzykson:1980rh}. 
One should also keep in mind that the Green's functions involving unphysical fields generated by the minimal functional coincide with the ones generated by the complete functional only up to constant terms proportional to the gauge fixing parameter, \eg $\Gamma_{A_\mu A_\nu}(q)=\Gamma^\mathrm{C}_{A_\mu A_\nu}(q)-i\xi^{-1}q_\mu q_\nu$. We will discuss further the differences between employing the complete and minimal generating functionals in the Appendix~\ref{Appendix:STIs} and~\ref{Appendix:BQIs}.

Taking functional derivatives
of  ${\cal S}(\Gamma)[\Phi]$  and  setting afterwards  all fields  and
anti-fields to  zero will generate  the complete set of  the all-order
STIs of the theory; this is in exact analogy to what happens with  the effective action,
where  taking  functional derivatives  of  $\Gamma[\Phi]$ and  setting
afterwards all fields  to zero generates the Green's  functions of the
theory, see Eq.~(\ref{greenfunc}).  
However, in order to reach meaningful expressions, one needs to keep in mind that:
\begin{enumerate} 
\item ${\cal S}(\Gamma)$ has ghost charge 1; 
\item functions with non-zero ghost charge vanish, since the ghost charge is a conserved quantity.
\end{enumerate}
Thus, in order to extract
non-zero identities from Eq.~(\ref{STIfunc}) one needs to differentiate the latter with respect to a  combination of fields, containing either one ghost field, or two ghost fields and one anti-field. The only exception to this rule is when differentiating with respect to a ghost anti-field, which needs to be compensated by three ghost fields. In particular, identities 
involving one or more gauge fields are obtained by differentiating Eq.~(\ref{STIfunc})
with respect to the set of fields in which one gauge boson has been replaced by the corresponding ghost field. This is due to the fact that  the linear part of the BRST transformation of the gauge field is proportional to the ghost field: $sA^a_\mu|_\mathrm{linear}=\partial_\mu c^a$. For completeness we notice that, for obtaining STIs involving Green's functions that contain ghost fields, one ghost field must be replaced by two ghost fields, due to the non linearity of the BRST ghost field transformation [$sc^a\propto f^{abc}c^bc^c$, see Eq.~(\ref{BRSTtrans})].
The last technical point to be clarified is  the dependence of the STIs on the (external) momenta. 
One should notice that the integral over $d^4x$ present in Eq.~(\ref{STIfunc}), together with the conservation of momentum flow of the Green's functions, 
implies that no momentum integration is left over; as a result, the STIs will be expressed as a sum of products of (at most two) Green's functions. 

An advantage of working with the BV formalism is the fact that the STI functional of Eq.~(\ref{STIfunc}) is valid in any gauge, \ie
it will not be affected when switching from one gauge to another. In particular, if we want to consider the BFM gauge, the only additional step we need to take  
is to implement the equations of motion for the background fields at the quantum level. 
This latter step is achieved most efficiently by extending the BRST symmetry to the background gluon field, through the relations
\begin{equation}
s\widehat{A}_\mu^m=\Omega^m_\mu, \qquad s\Omega^m_\mu=0,
\label{extBRST}
\end{equation}
where $\Omega_\mu^m$ represents a (classical) vector field with the same quantum numbers as the gluon, ghost charge $+1$ and Fermi statistics (see also Table~\ref{tableI}). The dependence of the Green's  functions on the background fields is then controlled by  the modified STI functional
\be
{\cal S}'(\Gamma')[\Phi]={\cal S}(\Gamma')[\Phi]+\int\!d^4x\
\Omega_m^\mu\left(\frac{\delta\Gamma'}{\delta\widehat{A}^m_\mu}-\frac{\delta\Gamma'}{\delta A^m_\mu}\right)=0,
\label{STIfunc_BFM}
\ee
where $\Gamma'$ denotes the effective action that depends on the background sources $\Omega^m_\mu$ (with $\Gamma\equiv\Gamma'\vert_{\Omega=0}$), and ${\cal S}(\Gamma')[\Phi]$ is the STI functional of Eq.~(\ref{STIfunc}). 
Differentiation of the STI functional (\ref{STIfunc_BFM}) with respect to the background source and background or quantum fields will 
then provide the so called BQIs, which relate 1PI Green's functions involving background fields with the ones involving quantum fields. 
The BQIs are particularly useful in the PT context, since they  allow for a direct  comparison between PT  and BFM Green's functions.

Finally, the background gauge invariance of the BFM effective action implies that Green's functions involving background fields satisfy 
linear WIs when contracted with the momentum corresponding to a background leg [see, {\it e.g.}, Eq.s~(\ref{3gl}) --~(\ref{4gh})]. 
These WIs are generated by taking functional differentiations of the WI functional
\be
{\cal W}_{\vartheta}[\Gamma']=\int\!d^4x\,\sum_{\Phi,\Phi^*}\left(\delta_{\vartheta(x)}\Phi\right)\frac{\delta\Gamma'}{\delta\Phi}=0,
\label{WI_gen_funct}
\ee 
where $\vartheta^a(x)$ are the local infinitesimal parameters corresponding to the $SU(3)$ generators $t^a$ that now play the role of the ghost field. The transformations $\delta_{\vartheta}\Phi$ are thus given by
\bea
\delta_{\vartheta}A^a_\mu=gf^{abc}A^b_\mu \vartheta^c &\qquad& \delta_{\vartheta}\widehat{A}^a_\mu=\partial_\mu \vartheta^a+gf^{abc}\widehat{A}^b_\mu \vartheta^c,\nonumber \\
\delta_{\vartheta} c^a=-g f^{abc}c^b\vartheta^c &\qquad& \delta_{\vartheta} \bar c^a=-g f^{abc}\bar c^b\vartheta^c,
\nonumber \\
\delta_{\vartheta}\psi^i_\mathrm{f}=ig \vartheta^a(t^a)_{ij}\psi^j_\mathrm{f} &\qquad& 
\delta_{\vartheta}\bar\psi^i_\mathrm{f}=-ig \vartheta^a\bar\psi^j_\mathrm{f} (t^a)_{ji}, 
\label{theta_trans}
\eea
and the background transformations of the anti-fields  $\delta_{\vartheta}\Phi^*$ coincide with the gauge transformations 
of the corresponding quantum gauge fields according to their specific representation. 
Notice that, in order to obtain the WI satisfied by the Green's functions involving background gluons $\widehat{A}$, one has to differentiate the functional (\ref{WI_gen_funct}) with respect to the corresponding parameter $ \vartheta$. 

All the  STIs and BQIs needed  for the PT construction  carried out in
the  rest of  this  paper, together with the  method  of constructing  the
auxiliary functions  appearing in  these identities, are
reported   in   Appendix~\ref{Appendix:STIs}   and~\ref{Appendix:BQIs}, respectively.

\subsection{Faddeev-Popov equation(s)\label{FPEs}}

The final ingredient needed for carrying out the PT program for SDEs is the derivation of the so-called Faddev-Popov equation (FPE). 
The FPE depends crucially on the form of the ghost Lagrangian, which, in turn, depends on the gauge fixing function [see Eq.~(\ref{FPG_Lag})]. 
In what follows we will first present the corresponding derivation in the $R_\xi$ gauges, and then in the BFM.

To derive the FPE in the $R_\xi$ gauges, one observes that in the QCD action the only term proportional to the anti-ghost fields comes from the Faddeev-Popov lagrangian density, which can be rewritten as
\begin{equation}
{\cal L}^{R_\xi}_\mathrm{FPG}=-\bar c^m\partial^\mu(sA^m_\mu)=-\bar c^m\partial^\mu\frac{\delta\Gamma}{\delta A^{*m}_\mu}.
\label{Rxi}
\end{equation}
Differentiation of the action with respect to $\bar c^a$ then yields the FPE in the form of the identity
\begin{equation}
\frac{\delta\Gamma}{\delta \bar c^m}+\partial^\mu\frac{\delta\Gamma}{\delta A^{*m}_\mu}
=0,
\end{equation}
so that, taking the Fourier transform, we arrive at 
\begin{equation}
\frac{\delta\Gamma}{\delta \bar c^m}+iq^\mu\frac{\delta\Gamma}{\delta A^{*m}_\mu}
=0.
\label{FPeqRxi}
\end{equation}
Thus,  in the $R_\xi$ case, the FPE amounts to the simple statement that the contraction 
of a leg corresponding to a 
gluon anti-field ($A^{*m}_\mu$)  by its own momentum ($q^\mu$) converts it to an anti-ghost leg ($\bar c^m$). 
Functional differentiation of this identity with respect to QCD fields (but not background sources and fields, see below) 
furnishes useful identities, that will be used extensively in our construction. 

For obtaining FPEs for Green's functions involving BFM gluons and sources, 
one has to modify Eq.~(\ref{FPeqRxi}), in order to account for the presence of extra terms in the BFM gauge fixing function (and therefore in the BFM Faddeev-Popov ghost Lagrangian). Eq.~(\ref{Rxi}) gets then modified into
\begin{equation}
\frac{\delta\Gamma'}{\delta \bar c^m}+\left(\widehat{{\cal D}}^\mu\frac{\delta\Gamma'}{\delta A^*_\mu}\right)^m
-\left({\cal D}^\mu\Omega_\mu\right)^m-gf^{mrs}\widehat{A}^r_\mu\Omega^\mu_s=0.
\label{FPeqBFM}
\end{equation}
Notice that by undoing the splitting of the field $A$ into background and quantum parts (that is using $A+\widehat{A}\to A$) the equation above assumes the more compact form
\be
\frac{\delta\Gamma'}{\delta \bar c^m}+\left(\widehat{{\cal D}}^\mu\frac{\delta\Gamma'}{\delta A^*_\mu}\right)^m
-\left({\cal D}^\mu\Omega_\mu\right)^m=0.
\label{FPeqBFM-1}
\ee

The specific FPEs needed for the PT construction are reported in Appendix~\ref{Appendix:FPEs}.

\subsection{The (one-loop) PT algorithm in the BV language\label{PT_BV_1l}} 

Before entering into  the intricacies of the SDEs,  it is important to
make contact  between the PT algorithm  and the BV  formalism. This is
best done at  the one-loop level, since in  this case all calculations
are  rather straightforward  and  it  is relatively  easy  to compare  the
standard  diagrammatic  results  with  those coming  from  the  BV
formalism.  This comparison will ({\it i})  help us identify the pieces that
will be generated  when applying the PT algorithm,  and ({\it ii}) establish
the rules for distributing the  pieces obtained in ({\it i})
among the different Green's functions appearing in the calculation.

\begin{figure}[!t]
\includegraphics[width=13.5cm]{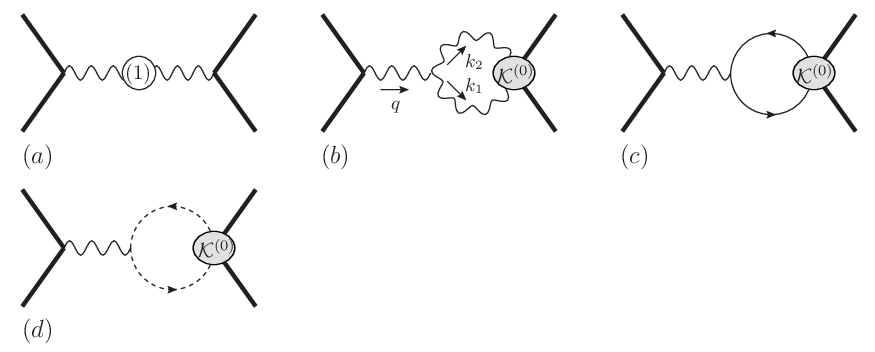}
\caption{The $S$-matrix one-loop PT setting for constructing the gluon propagator. The external particles are left unspecified since they can be both quarks as well as gluons. 
The kernels ${\cal K}^{(0)}$ appearing in diagram $(b)$  are the tree-level version of the one shown in Fig.s~\ref{fig:ggff_SDker} and~\ref{fig:gggg_SDker} (depending on the external particles chosen) and therefore contains only the 1PR terms shown there. Three diagrams having the kernels on the opposite side are not shown.}
\label{fig:1l_PT}
\end{figure}

The starting point is the embedding of the (one-loop) gluon propagator 
into an $S$-matrix element (Fig.~\ref{fig:1l_PT}), exactly as done in subsection~\ref{1lpt}.   
Then, carrying out the  PT decomposition $\Gamma=\Gamma^{\rm P}+\Gamma^{\rm F}$ 
on the tree-level three-gluon vertex of diagram $(b)$
[see Eq.~(\ref{PTDEC})], we find
\bea
(b)&=&(b)^{\rm F}+(b)^{\rm P},\\
(b)^{\rm P}&=&-\frac12gf^{am'n'}\int_{k_1}(g_{\alpha\nu'}k_{1\mu'}-g_{\alpha\mu'}k_{2\nu'})\Delta_{m'm}^{(0)\mu'\mu}(k_1)\Delta_{n'n}^{(0)\nu'\nu}(k_2){\cal K}^{(0)}_{A^m_\mu A^n_\nu\psi\bar\psi}(k_2,p_2,-p_1).\nonumber \\
\eea
Of couse, $k_1$ and $k_2$ are not independent, since $k_2=q-k_1$; thus, we have
\begin{eqnarray}
& & f^{am'n'}g_{\alpha\mu'}\int_{k_1}k_{2\nu'}\Delta_{m'm}^{(0)\mu'\mu}(k_1)\Delta_{n'n}^{(0)\nu'\nu}(k_2){\cal K}^{(0)}_{A^m_\mu A^n_\nu\psi\bar\psi}(k_2,p_2,-p_1)\nonumber \\
&=&  f^{am'n'}g_{\alpha\mu'}\int_{k_1}k_{1\nu'}\Delta_{m'm}^{(0)\mu'\mu}(k_2)\Delta_{n'n}^{(0)\nu'\nu}(k_1){\cal K}^{(0)}_{A^m_\mu A^n_\nu\psi\bar\psi}(k_1,p_2,-p_1)\nonumber \\
&=& -f^{am'n'}g_{\alpha\nu'}\int_{k_1}k_{1\mu'}\Delta_{m'm}^{(0)\mu'\mu}(k_1)\Delta_{n'n}^{(0)\nu'\nu}(k_2){\cal K}^{(0)}_{A^m_\mu A^n_\nu\psi\bar\psi}(k_2,p_2,-p_1),
\end{eqnarray}
and we see that the contributions of the two longitudinal momenta add up, thus removing the $1/2$ symmetry factor 
(this is clearly an all-order result, since the above derivation does not depend on the various Green's functions and kernels being at tree-level). 
Therefore we obtain
\be
(b)^{\rm P}=-gf^{amn}g_\alpha^\nu\int_{k_1}\frac1{k_1^2}\frac1{k_2^2}k_1^\mu{\cal K}^{(0)}_{A^m_\mu A^n_\nu\psi\bar\psi}(k_2,p_2,-p_1).
\ee
On the other hand, using the results
\be
k_1^2D^{(0)}_{mm'}(k_1)=\delta^{mm'},\qquad
\Gamma^{(0)}_{c^{m'} A^n_\nu A^{*\rho}_d}=gg^\rho_\nu f^{m'dn}, \qquad
\Gamma^{(0)}_{c^{m'}A^{*\rho}_d\psi\bar\psi}=0,
\ee
we find that the STI of Eq.~(\ref{STISDggff}) reduces to
\bea
k_1^\mu{\cal K}_{A^m_\mu A^n_\nu\psi\bar\psi}^{(0)}(k_2,p_2,-p_1)&=&-gg_\nu^\gamma f^{dmn}\Gamma^{(0)}_{A^d_\gamma\psi\bar\psi}(p_2,-p_1)+
\Gamma^{(0)}_{\psi\bar\psi}(p_1){\cal K}^{(0)}_{A^n_\nu\psi c^{m}\bar\psi^*}(p_2,k_1,-p_1)\nonumber \\
&+&{\cal K}^{(0)}_{A^n_\nu\psi^*\bar\psi c^m}(p_2,-p_1,k_1)\Gamma^{(0)}_{\psi\bar\psi}(p_2).
\eea
At this point the calculation is over and one needs to reshuffle the pieces generated.
 First of all, notice that when the external legs are on-shell the last two terms of the above STI drop out, by virtue of the (all-order) equations of motion
\bea
\left.\Gamma^{(0)}_{\psi\bar\psi}(p_2)u(p_2)\right|_{\psmp_2=m}&=&0,\\
\left.\bar u(p_1)\Gamma^{(0)}_{\psi\bar\psi}(p_1)\right|_{\psmp_1=m}&=&0.
\label{eom}
\eea
Thus, making use of Eq.~(\ref{BQI:auxOmAs}) we are finally left with the result
\bea
(b)^{\rm P}&=& g^2C_A\delta^{ad}g_\alpha^\gamma\int_{k_1}\frac1{k_1^2}\frac1{k_2^2}\Gamma^{(0)}_{A^d_\gamma\psi\bar\psi}(p_2,-p_1)\nonumber\\
&=&-\Gamma^{(1)}_{\Omega^a_\alpha A^{*\gamma}_d}(-q)\Gamma^{(0)}_{A^d_\gamma\psi\bar\psi}(p_2,-p_1).
\eea
Notice that, as explicitly shown in Appendix~\ref{Appendix:BQIs}, the auxiliary function  $\Gamma_{\Omega_\alpha A^*_\beta}$ [see Eq.~(\ref{BQI:auxOmAs}) and Fig.~\ref{fig:Composite_ope}] coincide with the function $\Lambda_{\alpha\beta}$ [see Eq.~(\ref{gpert2}) and Fig.~\ref{fig:Lambda_aux}] to all orders
\be
\Lambda_{\alpha\beta}(q)\equiv\Gamma_{\Omega_\alpha A^*_\beta}(q).
\label{Lambda_eq_GaOmAs}
\ee
Thus the scalar function $G(q^2)$ introduced earlier in Section~\ref{nsds} corresponds also to ($i$ times) the $g_{\alpha\beta}$ part of $\Gamma_{\Omega_\alpha A^*_\beta}$.

We can now define the PT (on-shell) quark-gluon vertex, by considering the corresponding Green's function embedded in the diagrams
\bea
& & (b)^{\mathrm F}+(c)=(b)+(c)-(b)^{\mathrm P}\nonumber \\
&\Rightarrow& i\widehat{\Gamma}^{(1)}_{A^a_\alpha\psi\bar\psi}(p_2,-p_1)
=i\Gamma^{(1)}_{A^a_\alpha\psi\bar\psi}(p_2,-p_1)+\Gamma^{(1)}_{\Omega^a_\alpha A^{*\gamma}_d}(-q)\Gamma^{(0)}_{A^d_\gamma\psi\bar\psi}(p_2,-p_1),
\label{PT:gffvertex}
\eea
while the PT self-energy will given by the combination
\be
(a)+2(b)^{\mathrm P}\ \Rightarrow\
\widehat{\Pi}^{(1)}_{\alpha\beta}(q)=\Pi^{(1)}_{\alpha\beta}(q)+\Pi^{{\rm P}(1)}_{\alpha\beta}(q).
\ee
The factor of 2 comes from the mirror diagram of $(b)$ having the kernel on the left side, and we have defined [with the aid of Eq.~(\ref{ga_tree_lev})]
\bea
\delta^{ab}\Pi^{{\rm P}(1)}_{\alpha\beta}(q)&=&-2\Gamma^{(1)}_{\Omega^a_\alpha A^{*\gamma}_d}(q)q^2P_{\gamma_\beta}(q)\delta^{bd} 
\nonumber\\
&=&2i\Gamma^{(1)}_{\Omega^a_\alpha A^{*\gamma}_d}(q)\Gamma^{(0)}_{A^d_\gamma A^b_\beta}(q). 
\eea

We can now proceed to the comparison of the PT Green's function with that of the BFG, by resorting to the BQIs.
Clearly, Eq.~(\ref{PT:gffvertex}) represents the one-loop version of the BQI of Eq.~(\ref{BQI:gff}), and we immediately conclude that
\be
\widehat{\Gamma}^{(1)}_{A^a_\alpha\psi\bar\psi}(p_2,-p_1)\equiv\Gamma^{(1)}_{\widehat{A}^a_\alpha\psi\bar\psi}(p_2,-p_1).
\ee
For the self-energy we have instead [recall that $-\Gamma_{A^m_\mu A^n_\nu}(k)=\Pi_{\mu\nu}(k)$]
\be
\delta^{ab}\widehat{\Pi}^{(1)}_{\alpha\beta}(q)=-\Gamma^{(1)}_{A^a_\alpha A^b_\beta}(q)+
2i\Gamma^{(1)}_{\Omega^a_\alpha A^{*\gamma}_d}(q)\Gamma^{(0)}_{A^d_\gamma A^b_\beta}(q)\,,
\ee
which represents the one-loop version of the BQI of Eq.~(\ref{BQI:gg}), \ie we have
\be
\delta^{ab}\widehat{\Pi}^{(1)}_{\alpha\beta}(q)=-\Gamma^{(1)}_{\widehat{A}^a_\alpha\widehat{A}^b_\beta}(q).
\ee

The procedure just described goes through almost unaltered when choosing the external legs of the process to be gluons. In this case
\be
(b)^{\rm P}=-gf^{amn}g_\alpha^\nu\int_{k_1}\frac1{k_1^2}\frac1{k_2^2}k_1^\mu{\cal K}^{(0)}_{A^m_\mu A^n_\nu A^r_\rho A^s_\sigma}(k_2,p_2,-p_1).
\ee
and one has from the STI of Eq.~(\ref{STISDgggg}) the result
\bea
k_1^\mu{\cal K}^{(0)}_{A^m_\mu A^n_\nu A^r_\rho A^s_\sigma}(k_2,p_2,-p_1)&=&-gg_\nu^\gamma f^{dmn}\Gamma^{(0)}_{A^d_\gamma A^r_\rho  A^s_\sigma}(p_2,-p_1)\nonumber \\
&+&{\cal K}^{(0)}_{c^{m} A^n_\nu A^s_\sigma A^{*\gamma}_d}(k_2,-p_1,p_2)\Gamma^{(0)}_{A^d_\gamma A^r_\rho}(p_2)\nonumber \\
&+&{\cal K}^{(0)}_{c^{m}A^n_\nu A^r_\rho  A^{*\gamma}_d}(k_2,p_2,-p_1)\Gamma^{(0)}_{A^d_\gamma A^s_\sigma}(p_1)\nonumber \\
&+&{\cal K}^{(0)}_{ c^{m}A^r_\rho A^s_\sigma A^{*\gamma}_d}(p_2,-p_1,k_2)\Gamma^{(0)}_{A^d_\gamma A^n_\nu}(k_2).
\eea
As before, the second and third terms drop out when the external gluons are taken to be on-shell; thus we are left with the terms
\bea
(b)^{\rm P}&=& -\Gamma^{(1)}_{\Omega^a_\alpha A^{*\gamma}_d}(-q)\Gamma^{(0)}_{A^d_\gamma A^r_\rho A^s_\sigma}(p_2,-p_1)
-gf^{amn}g_\alpha^\nu\!\int_{k_1}\!\frac1{k_1^2}\frac1{k_2^2}{\cal K}^{(0)}_{ c^{m}A^r_\rho A^s_\sigma A^{*\gamma}_d}(p_2,-p_1,k_2)\Gamma^{(0)}_{A^d_\gamma A^n_\nu}(k_2)\nonumber \\
&=&-\Gamma^{(1)}_{\Omega^a_\alpha A^{*\gamma}_d}(-q)\Gamma^{(0)}_{A^d_\gamma A^r_\rho A^s_\sigma}(p_2,-p_1)
+(b').
\label{PT_contr_ggg}
\eea
The first term is exactly the PT propagator-like piece encountered in the quark case; 
this is the essence of the process independence of the PT. Notice, however, that the second term was not present before. 
The action of this term will be discussed in detail in Section~\ref{tgv}; here it suffices to note that it is a vertex-like piece 
(as is evident from the structure of the kernel appearing in it) and, therefore, it ought to be allotted to the PT three-gluon vertex.
Thus we can define the PT (on-shell) three-gluon vertex and propagator as before, \ie
\bea
& & (b)^{\mathrm F}+(b')+(c)+(d)=(b)+(c)+(d)+(b')-(b)^{\mathrm P}\nonumber \\
&\Rightarrow& i\widehat{\Gamma}^{(1)}_{A^a_\alpha A^r_\rho A^s_\sigma}(p_2,-p_1)
=i\Gamma^{(1)}_{A^a_\alpha A^r_\rho A^s_\sigma}(p_2,-p_1)+\Gamma^{(1)}_{\Omega^a_\alpha A^{*\gamma}_d}(-q)\Gamma^{(0)}_{A^d_\gamma A^r_\rho A^s_\sigma}(p_2,-p_1),\nonumber \\
& & (a)+2(b)^{\mathrm P}\ \Rightarrow\
\widehat{\Pi}^{(1)}_{\alpha\beta}(q)=\Pi^{(1)}_{\alpha\beta}(q)+\Pi^{{\rm P}(1)}_{\alpha\beta}(q).
\eea 
The comparison with the BQI of Eq.~(\ref{BQI:ggg}) shows then that
\be
\widehat{\Gamma}^{(1)}_{A^a_\alpha A^r_\rho A^s_\sigma}(p_2,-p_1)=\Gamma^{(1)}_{\widehat{A}^a_\alpha A^r_\rho A^s_\sigma}(p_2,-p_1),
\ee
and again we find 
\be
\delta^{ab}\widehat{\Pi}^{(1)}_{\alpha\beta}(q)=-\Gamma^{(1)}_{\widehat{A}^a_\alpha\widehat{A}^b_\beta}(q).
\ee

The  (one-loop)  procedure  described above carries  over
practically unaltered to the corresponding SDEs. This is due to the fact
that: ({\it  i}) the pinching  momenta will be always  determined from
the  tree-level decomposition of  Eq.~(\ref{PTDEC}); ({\it  ii}) their
action is completely  fixed by the structure of  the STIs they trigger
[Eq.s~(\ref{STISDggff})  and~(\ref{STISDgggg})  for  the  vertices  at
hand]; ({\it  iii}) the kernels appearing  in these STIs  are the same
appearing in  the corresponding BQIs;  thus, it is always  possible to
write  the  result of  the  action of  pinching  momenta  in terms  of
auxiliary Green's functions appearing in the BQIs.

The only operational difference is that, in the case of the SDEs for
the quark-gluon vertex and the three-gluon vertex,  
{\it all three} external legs  will be {\it off-shell}. 
This is of course unavoidable, given that these (fully dressed) vertices 
are nested in the SDE of the off-shell gluon self-energy 
[see Fig.\ref{fig:PT_newSDE}, diagrams $(d_{11})$ and $(d_1)$, respectively], 
and their legs inside the diagrams are 
irrigated by the virtual off-shell momenta. As a result, 
the equations of motion employed above [{\it viz.} Eq.~(\ref{eom})]
should not be used in this case; therefore, 
the corresponding terms, proportional to inverse 
self-energies, do not drop out, and form part of the resulting BQI.

Thus the  PT rules for the  construction of SDEs may be 
summarized as follows: 

\begin{itemize}
\item[{\it i}.]
   For the SDEs  of vertices, with {\it all  three} external legs {\it
  off-shell},  the pinching  momenta,  coming from  the only  external
  three-gluon  vertex   undergoing  the  decomposition  (\ref{PTDEC}),
  generate four types of terms: one of them, corresponding to the term
  $(b')$  in   Eq.~(\ref{PT_contr_ggg}),  is  a   genuine  vertex-like
  contribution that  must be included in  the final PT  answer for the
  vertex under construction, while the remaining three-terms will form
  part  of  the  emerging  BQIs 
  (and thus would be discarded from the PT vertex). 
  These  latter  terms  have  a  very
  characteristic structure, which facilitates their identification in
  the calculation.   Specifically, one of them  is always proportional
  to the auxiliary function $\Gamma_{\Omega A^*}$, while the other two
  are proportional  to the inverse propagators of  the fields entering
  into  the  two  legs  that  did not  undergo  the  decomposition  of
  (\ref{PTDEC}).

\item[{\it ii}.]In the case of the new SDE for the gluon propagator
the  pinching  momenta  will  only  generate  pieces  proportional  to
$\Gamma_{\Omega A^*}$,  which should be  discarded from the  PT answer
for the  gluon 
two-point function 
(since they  are exactly those  that cancel against
the contribution coming from the corresponding vertices).

\end{itemize}

\section{PT Green's functions from Schwinger-Dyson equations\label{SDE}}

After  the introduction  of the  useful tools and  basic rules required  for
the application of the PT program to the (non-perturbative) case of SDEs, we are
ready to describe in detail 
the actual construction,
starting  from the corresponding SDEs written in the Feynman gauge of the $R_\xi$.
We   will first derive the new SDEs  for the two vertices,
$\widehat{\Gamma}_{A\psi\bar\psi}$  and $\widehat{\Gamma}_{AAA}$, given that 
the calculations are easier to work out, and will then address the more
complicated case of the SDE for the PT gluon propagator $\widehat{\Gamma}_{AA}$.

\subsection{Quark-Gluon vertex}

\begin{figure}[!t]
\includegraphics[width=14.5cm]{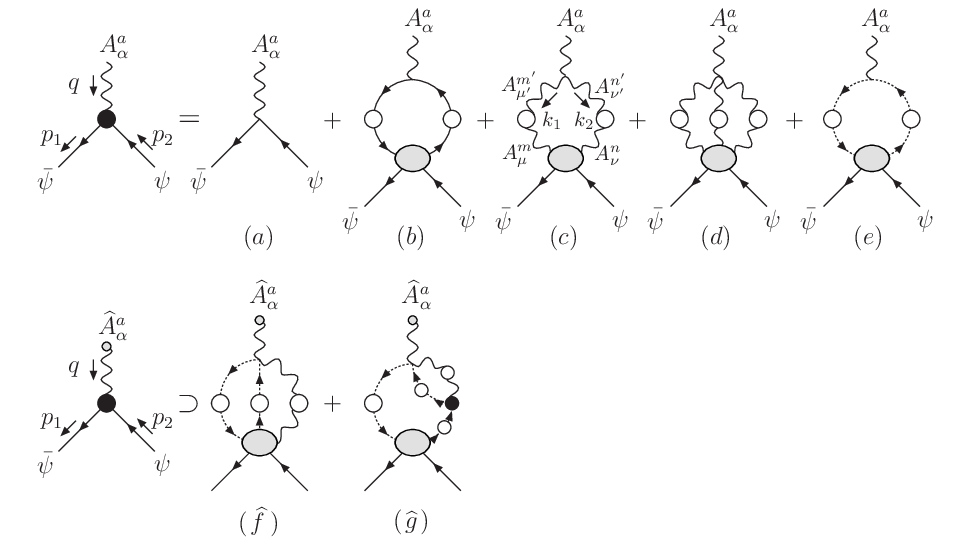}
\caption{The SDE for the quark-gluon vertex. 
The symmetry factors of the $R_\xi$ diagrams (first line) are: $s(a,b)=1$. $s(c)=1/2$, $s(d)=1/6$ and $s(e)=-1$. 
For the key diagram $(c)$ we show explicitly the kinematics chosen. 
In the second line we show the additional topologies present in the BFM version of the equation 
[$s({\widehat{f}},{\widehat{g}})=-1$], 
generated dynamically by the PT procedure.}
\label{fig:gff_SDE}
\end{figure}

The SDE of the quark-gluon vertex, 
shown in  Fig.~\ref{fig:gff_SDE},
is the simplest one as far as the PT  construction is concerned,
capturing at the same time several  of the
essential steps that appear 
during the application of the PT algorithm to the SDEs of QCD. 

\begin{figure}[!t]
\includegraphics[width=15cm]{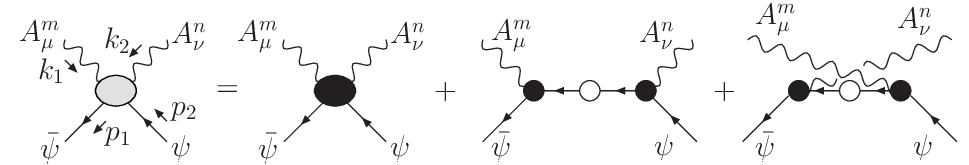}
\caption{Skeleton expansion of the kernel appearing in the SDE for the trilinear quark-gluon vertex [see diagram $(c)$ of Fig.~\ref{fig:gff_SDE}]. 
Black, white, and  gray blobs denote 1PI functions, connected functions, and SD kernels, respectively.}
\label{fig:ggff_SDker}
\end{figure}

We start by carrying out the decomposition of Eq.~(\ref{PTDEC}) on the tree-level vertex appearing in ($c$), 
the only diagram we will touch in our construction. Let us concentrate on the $\Gamma^\mathrm{P}$ part; one has
\bea
(c)^\mathrm{P}&=&-\frac12gf^{am'n'}\int_{k_1}(g_{\alpha\nu'}k_{1\mu'}-g_{\alpha\mu'}k_{2\nu'})\Delta_{m'm}^{\mu'\mu}(k_1)\Delta_{n'n}^{\nu'\nu}(k_2){\cal K}_{A^m_\mu A^n_\nu\psi\bar\psi}(k_2,p_2,-p_1)\nonumber \\
&=&gf^{amn'}g_{\alpha\nu'}\int_{k_1}\frac1{k_1^2}\Delta_{n'n}^{\nu\nu'}(k_2)k_1^{\mu}{\cal K}_{A^m_\mu A^n_\nu\psi\bar\psi}(k_2,p_2,-p_1),
\label{k1}
\eea
where the kernel ${\cal K}_{AA\psi\bar\psi}$ is shown in Fig.~\ref{fig:ggff_SDker}.
Using the STI of the kernel ${\cal K}_{A^m_\mu A^n_\nu\psi\bar\psi}$ given in Eq.~(\ref{STISDggff}), we obtain 
from (\ref{k1}) four terms, to be denoted by $(s_1)$, $(s_2)$, $(s_3)$ and $(s_4)$, \ie
\be
(c)^\mathrm{P} = (s_1)+(s_2)+(s_3)+(s_4),
\ee
with
\begin{eqnarray}
(s_1)&=& gf^{am'n'}g_{\alpha\nu'}\int_{k_1}D^{m'm}(k_1)\Delta_{n'n}^{\nu'\nu}(k_2)\Gamma_{c^{m} A^n_\nu A^{*\gamma}_d}(k_2,-k_1-k_2)
\Gamma_{A^d_\gamma\psi\bar\psi}(p_2,-p_1),\nonumber \\
(s_2)&=& gf^{am'n'}g_{\alpha\nu'}\Gamma_{\psi\bar\psi}(p_1)\int_{k_1}D^{m'm}(k_1)\Delta_{n'n}^{\nu'\nu}(k_2){\cal K}_{A^n_\nu\psi c^{m}\bar\psi^*}(p_2,k_1,-p_1), 
\nonumber \\
(s_3)&=& gf^{am'n'}g_{\alpha\nu'}\int_{k_1}D^{m'm}(k_1)\Delta_{n'n}^{\nu'\nu}(k_2){\cal K}_{A^n_\nu\psi^*\bar\psi c^{m}}(p_2,-p_1,k_1)\Gamma_{\psi\bar\psi}(p_2),
\nonumber \\
(s_4)&=& gf^{am'n'}g_{\alpha\nu'}\int_{k_1}D^{m'm}(k_1)\Delta_{n'n}^{\nu'\nu}(k_2)\Gamma_{c^{m} A_d^{*\gamma}\psi\bar\psi}(k_2,p_2,-p_1)\Gamma_{A^d_\gamma A^n_\nu}(k_2).
\label{s1234}
\end{eqnarray}

Next, using Eq.s~(\ref{BQI:auxOmAs}), (\ref{Aux-gff:1}) and (\ref{Aux-gff:2}), it is fairly straightforward to 
demonstrate that
\begin{eqnarray}
(s_1)&=& -\Gamma_{\Omega^a_\alpha A^{*\gamma}_d}(-q)\Gamma_{A^d_\gamma\psi\bar\psi}(p_2,-p_1),
\nonumber \\
(s_2)&=& -\Gamma_{\psi\bar\psi}(p_1)\Gamma_{\psi\Omega^a_\alpha\bar\psi^*}(q,-p_1),
\nonumber \\
(s_3)&=& -\Gamma_{\psi^*\bar\psi\Omega^a_\alpha}(-p_1,q)\Gamma_{\psi\bar\psi}(p_2).
\label{s123b}
\end{eqnarray}
Evidently, $(s_1)$ gives rise to the PT propagator-like term, 
while  $(s_2)$ and $(s_3)$ generate the terms
that in the usual $S$-matrix PT would vanish on-shell,  
due to the (all-order) spinor equations of motion
$\left.\Gamma_{\psi\bar\psi}(p_2)u(p_2)\right|_{\psmp_2=m}=0$,
and 
$\left.\bar u(p_1)\Gamma_{\psi\bar\psi}(p_1)\right|_{\psmp_1=m}=0$.
Of course, in our case we are not allowed to use the equations of motion, given that 
the quark legs are considered to be off-shell.

\begin{figure}
\includegraphics[width=12cm]{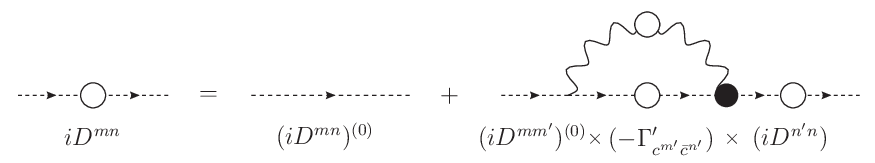}
\caption{The Schwinger-Dyson equation~(\ref{ghost_SDE}) satisfied by the ghost propagator.}
\label{fig:ghost_SDE}
\end{figure}

Let us finally look at the term $(s_4)$, and show how it 
combines with the remaining $R_\xi$ diagrams to generate the BFM quark-gluon vertex $\Gamma_{\widehat{A}\psi\bar\psi}$. 
To this end, using  Eq.~(\ref{gainvprop}) and the FPE satisfied by the 1PI function $\Gamma_{cA^*\psi\bar\psi}$, 
we write $(s_4)=(s_{4a})+(s_{4b})$,  with 
\begin{eqnarray}
(s_{4a})&=&-igf^{am'd}g_{\alpha\gamma}\int_{k_1}D^{m'm}(k_1)\Gamma_{c^{m} A_d^{*\gamma}\psi\bar\psi}(k_2,p_2,-p_1),\nonumber \\
(s_{4b})&-&gf^{am'n'}g_{\alpha\nu'}\int_{k_1}\delta^{dn'}\frac{k_2^{\nu'}}{k_2^2}D^{m'm}(k_1)\Gamma_{c^{m}\bar c^d\psi\bar\psi}(k_2,p_2,-p_1).
\label{s4terms}
\end{eqnarray}
The general structure of these two terms suggests that $(s_{4a})$ 
should give rise to the ghost quadrilinear vertex,
 while  $(s_{4b})$, when added to diagram $(e)$, should symmetrize the trilinear ghost gluon coupling. 
It turns out that this expectation is essentially correct, but its 
realization is not immediate, mainly due to  
the fact that $(s_{4b})$ contains a tree-level instead of a full
ghost propagator [$(k_2^2)^{-1}$ instead of  $D(k_2)$], while
$(s_{4a})$ can  reproduce, at most,  diagram $(\widehat{f})$ of Fig.~\ref{fig:gff_SDE}, but not  
$(\widehat{g})$. 
The solution to this apparent mismatch is rather subtle: one must employ the SDE satisfied by the {\it ghost propagator}, shown in 
Fig~\ref{fig:ghost_SDE}. This SDE   
is common to both the $R_{\xi}$-gauge and the BFM, given that there are no background ghosts.

\begin{figure}[!t]
\includegraphics[width=12cm]{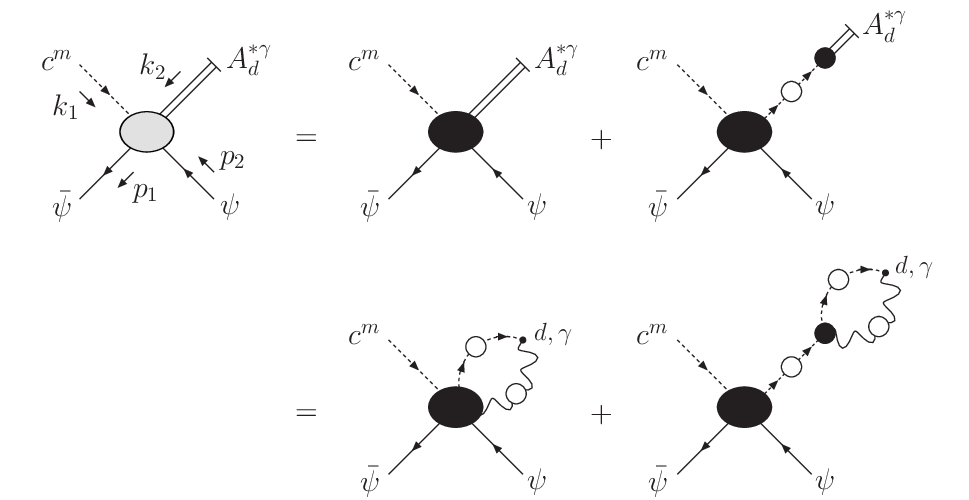}
\caption{Diagrammatic decomposition of the kernel ${\cal K}_{c^{m}A^{*\gamma}_d\psi\bar\psi}$. 
The second term is the one added (and subtracted) to the original sum $(s_{4a})+(s_{4b})$ of Eq.~(\ref{s4terms}). 
After replacing the gluon anti-field  $A^{*\gamma}_d$ by the corresponding composite operator (second line),  
taking into account the extra structure provided by the $(c)^\mathrm{P}$ term, this kernel furnishes the BFM terms $(\widehat{f})+(\widehat{g})$ [see Eq.~(\ref{s4a}).] }
\label{fig:cffAstar_SDaux}
\end{figure}

To see how this works in detail,  
add and subtract to Eq.~(\ref{s4terms}) the missing term  (see Fig.~\ref{fig:cffAstar_SDaux}), obtaining
\begin{eqnarray}
(s_{4a})&=&-igf^{am'd}g_{\alpha\gamma}\int_{k_1}D^{m'm}(k_1)\left[\Gamma_{c^{m} A_d^{*\gamma}\psi\bar\psi}(k_2,p_2,-p_1)\right.\nonumber \\
&-&\left.\Gamma'_{c^gA_d^{*\gamma}}(k_2)iD^{gg'}(k_2)\Gamma_{c^{m}\bar c^{g'}\psi\bar\psi}(k_2,p_2,-p_1)\right]\nonumber\\
&=&-igf^{am'd}g_{\alpha\gamma}\int_{k_1}D^{mm'}(k_1){\cal K}_{c^{m'}A^{*\gamma}_d\psi\bar\psi}(k_2,p_2,-p_1)\\
(s_{4b})&=&-gf^{am'n'}g^{\nu'}_\alpha\int_{k_1}\left[\delta^{dn'}\frac{k_{2\nu'}}{k_2^2}
-\Gamma'_{c^gA_{\nu'}^{*n'}}(k_2)D^{gd}(k_2)
\right]\times\nonumber \\
&\times&D^{m'm}(k_1)\Gamma_{c^{m}\bar c^d\psi\bar\psi}(k_2,p_2,-p_1),
\end{eqnarray}
where the auxiliary function $\Gamma'_{cA^*}$ has been defined in Eq.~(\ref{BQI:auxcAs}), and is given by $\Gamma_{cA^*}$ minus its tree-level part. Using Eq.~(\ref{Astrick}), we can then rewrite ($s_{4a}$) as
\begin{eqnarray}
(s_{4a})&=&ig^2f^{am'd}f^{dse}g_{\alpha\sigma}\int_{k_1}\int_{k_3}D^{m'm}(k_1)D^{ee'}(k_3)\Delta_{ss'}^{\sigma\sigma'}(k_4)\times\nonumber \\
&\times&\left[\Gamma_{c^{m}A^{s'}_{\sigma'}\bar c ^{e'}\psi\bar\psi}(k_3,k_4,p_2,-p_1)
-i\Gamma_{c^gA^{s'}_{\sigma'}\bar c^{e'}}(k_3,k_4)D^{gg'}(k_2)\Gamma_{c^m\bar c^{g'}\psi\bar\psi}(k_2,p_2,-p_1)\right]\nonumber \\
&=& (\widehat{f})+(\widehat{g}).
\label{s4a}
\end{eqnarray}

We next turn to $(s_{4b})$ and consider the ghost SD equation of Fig.~\ref{fig:ghost_SDE}. One has
\begin{equation}
iD^{dn'}(k_2)=i\frac{\delta^{dn'}}{k_2^2}+i\frac{\delta^{dg}}{k_2^2}\left[-\Gamma'_{c^g \bar c^{g'}}(k_2)\right]iD^{g'n'}(k_2),
\label{ghost_SDE}
\end{equation}
where, as before, $\Gamma'_{c^g \bar c^{g'}}$ is given by  $\Gamma_{c^g \bar c^{g'}}$ minus its tree-level part.
Multiplying the above equation by $k_2^2$, using the FPE~(\ref{FPE:ghprop}) and factoring out a $k_{2\nu'}$ we get the relation 
\begin{eqnarray}
k_{2\nu'}D^{dn'}(k_2)&=&\delta^{dn'}\frac{k_{2\nu'}}{k_2^2}-\Gamma'_{c^dA_{\nu'}^{*g}}(k_2)D^{gn'}(k_2)\nonumber\\
&=&\delta^{dn'}\frac{k_{2\nu'}}{k_2^2}-\Gamma'_{c^gA_{\nu'}^{*n'}}(k_2)D^{gd}(k_2)\,.
\label{ghSDE2}
\end{eqnarray}
Therefore, we obtain
\begin{equation}
(s_{4b})=-gf^{am'n'}\int_{k_1}k_{2\alpha}D^{m'm}(k_1)D^{n'n}(k_2)\Gamma_{c^{m}\bar c^n\psi\bar\psi}(k_2,p_2,-p_1).
\end{equation}
Adding this last contribution to diagram $(e)$ we finally arrive at 
\begin{eqnarray}
(e)+(s_{4b})&=&gf^{am'n'}\int_{k_1}(k_1-k_2)_\alpha D^{m'm}(k_1)D^{n'n}(k_2){\cal K}_{c^{m}\bar c^{n}\psi\bar\psi}(k_2,p_2,-p_1)\nonumber \\
&=&(\widehat{e}).
\end{eqnarray}

Next, observe that the graphs $(a),\ (b),$ and $(d)$ of Fig.~\ref{fig:gff_SDE} can be converted to hatted ones automatically
(see the  corresponding tree-level Feynman in Appendix~\ref{Frules}), and that $(c)^\mathrm{F}=(\widehat{c})$
since in the BFG $\Gamma^\mathrm{F} = \Gamma^{(0)}_{\widehat{A}AA}$.
Thus,
\begin{eqnarray}
i\Gamma_{A^a_\alpha \psi\bar\psi}(p_2,-p_1)&=&
-\Gamma_{\Omega^a_\alpha A^{*\gamma}_d}(-q)\Gamma_{A^d_\gamma\psi\bar\psi}(p_2,-p_1)-\Gamma_{\psi^*\bar\psi\Omega^a_\alpha}(-p_1,q)\Gamma_{\psi\bar\psi}(p_2)\nonumber \\
&-&\Gamma_{\psi\bar\psi}(p_1)\Gamma_{\psi\Omega^a_\alpha\bar\psi^*}(q,-p_1)
+[(\widehat{a})+(\widehat{b})+
 (\widehat{c})  +
(\widehat{d})+(\widehat{e})+
(\widehat{f})+(\widehat{g})]^a_\alpha.\nonumber \\
\label{PT-gff-res}
\end{eqnarray}
The sum of diagrams in the brackets is nothing but the kernel expansion of the SDE governing the vertex  $\Gamma_{\widehat{A}\psi\bar\psi}$, \ie
\be
i\Gamma_{\widehat{A}^a_\alpha \psi\bar\psi}(p_2,-p_1)=[(\widehat{a})+(\widehat{b})+
 (\widehat{c})  +
(\widehat{d})+(\widehat{e})+
(\widehat{f})+(\widehat{g})]^a_\alpha.
\label{final_gff}
\ee
After this identification,  it is clear that Eq.~(\ref{PT-gff-res}) coincides with the full BQI of Eq.~(\ref{BQI:gff}), namely
\begin{eqnarray}
i\Gamma_{\widehat{A}^a_\alpha\psi\bar\psi}(p_2,-p_1)&=&[ig^\gamma_\alpha\delta^{ad}+\Gamma_{\Omega^a_\alpha A^{*\gamma}_d}(-q)]\Gamma_{A^d_\gamma\psi\bar\psi}(p_2,-p_1)\nonumber \\
&+&\Gamma_{\psi^*\bar\psi\Omega^a_\alpha}(-p_1,q)\Gamma_{\psi\bar\psi}(p_2)+\Gamma_{\psi\bar\psi}(p_1)\Gamma_{\psi\Omega^a_\alpha\bar\psi^*}(q,-p_1).
\label{BQI_gff_repr}
\end{eqnarray}

In summary, the application of the PT to the conventional SDE for the quark-gluon vertex ({\it i}) has converted the initial  kernel expansion [graphs ({\it a}) to ({\it e}) in Fig.~\ref{fig:gff_SDE}] into the graphs corresponding to the kernel expansion of the vertex $\Gamma_{\widehat{A}\psi\bar\psi}$; ({\it ii}) all other pinching terms extracted from the original diagram $(c)$ are precisely the combinations of auxiliary Green's functions appearing in the BQI that relates the initial vertex $\Gamma_{A\psi\bar\psi}$ with the final vertex $\Gamma_{\widehat{A}\psi\bar\psi}$.

Notice at this point that the skeleton expansion of the multi-particle kernels appearing in the SDE for $\Gamma_{\widehat{A}\psi\bar\psi}$ is still written in terms of the conventional fully dressed vertices and propagators (involving only quantum fields). 
Thus, Eq.~(\ref{final_gff}) is not manifestly dynamical, \ie it does not involve the same unknown quantities on the right and left hand side; this situation is exactly analogous to the gluon propagator case discussed in subsection~\ref{nsds}. 
Specifically, in order to convert (\ref{final_gff}) into a genuine SDE, one has two possibilities, both involving the use of the above BQI: ({\it i}) substitute the lhs of Eq.~(\ref{BQI_gff_repr}) into the rhs of Eq.~(\ref{final_gff}) and solve for the conventional $\Gamma_{A\psi\bar\psi}$ vertex, or ({\it ii}) invert Eq.~(\ref{BQI_gff_repr}) and use it to convert every $\Gamma_{A\psi\bar\psi}$ vertex appearing in the rhs of Eq.~(\ref{final_gff}) into a $\Gamma_{\widehat{A}\psi\bar\psi}$ vertex. 
It would seem that the latter option is operationally more cumbersome, especially taking into account that a similar procedure has to be followed for all the Green's functions that appear in the coupled system of SDEs that one considers.

\subsection{Three-gluon vertex\label{tgv}}

\begin{figure}[!t]
\includegraphics[width=14.5cm]{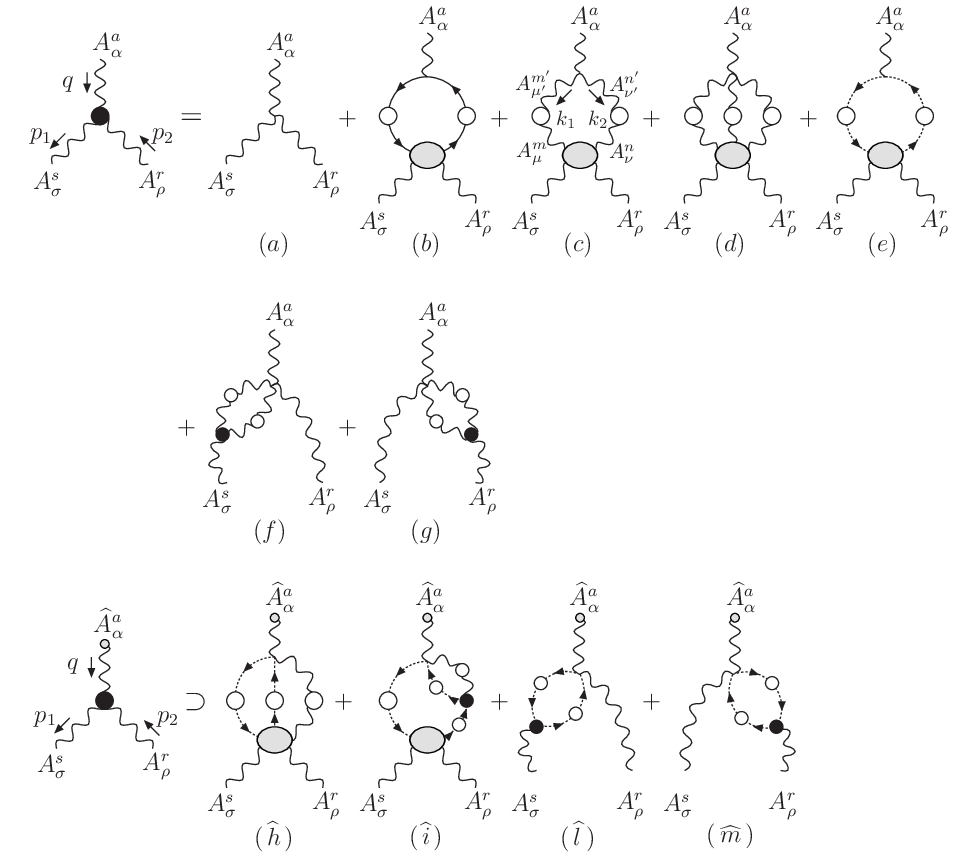}
\caption{The SDE of the  three-gluon
vertex. The symmetry factors of  the $R_\xi$ (first and second line of
the   figure)  diagrams   are   $s(a,b)=1$.  $s(c)=1/2$,   $s(d)=1/6$,
$s(e)=-1$,  $s(f,g)=1/2$. In  the third  line we  show  the additional
topologies  present  in the  BFM  version  of  the equation
[$s(\widehat{h},\widehat{i},\widehat{l},\widehat{m})=-1$], 
generated during the PT procedure.}
\label{fig:ggg_SDE}
\end{figure}

The construction  of the  PT three-gluon vertex  proceeds in a very similar way, with
some additional subtleties that we will spell out in detail in what follows.
We emphasize that the purpose of this exercise is to generate dynamically the vertex $\Gamma_{\widehat{A}AA}$ 
and {\it not} the fully Bose-symmetric vertex $\Gamma_{\widehat{A}\widehat{A}\widehat{A}}$ studied in~\cite{Cornwall:1989gv,Binger:2006sj}. 
The reason is that it is the former vertex that appears in the SDEs for the gluon propagator 
[see, {\it e.g.}, diagram $(d_1)$ in Fig.~\ref{fig:PT_newSDE}], making it the relevant object to consider at this level.

We start by considering the conventional~\cite{Marciano:1977su} SDE for  the three gluon
vertex  (Fig.~\ref{fig:ggg_SDE}), and 
carry out the standard $\Gamma^\mathrm{P}+\Gamma^\mathrm{F}$  
decomposition to the  tree-level vertex of diagram $(c)$, 
which is the only one we will modify in our construction. 

\begin{figure}[!t]
\includegraphics[width=15cm]{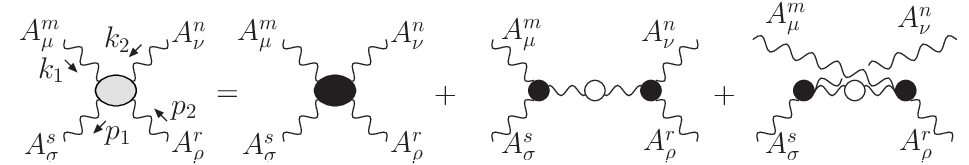}
\caption{Skeleton expansion of the kernel appearing in the SDE for the three-gluon vertex [diagram $(c)$ of Fig.~\ref{fig:ggg_SDE}]}
\label{fig:gggg_SDker}
\end{figure}

We then find
\begin{eqnarray}
(c)^\mathrm{P}&=&-\frac12gf^{am'n'}\int_{k_1}(g_{\alpha\nu'}k_{1\mu'}-g_{\alpha\mu'}k_{2\nu'})\Delta_{m'm}^{\mu'\mu}(k_1)\Delta_{n'n}^{\nu'\nu}(k_2){\cal K}_{A^m_\mu A^n_\nu A^r_\rho A^s_\sigma}(k_2,p_2,-p_1)\nonumber \\
&=& gf^{amn'}g_{\alpha\nu'}\int_{k_1}\frac1{k_1^2}\Delta_{n'n}^{\nu'\nu}(k_2)k_1^{\mu}{\cal K}_{A^m_\mu A^n_\nu A^r_\rho A^s_\sigma}(k_2,p_2,-p_1),
\end{eqnarray}
and the kernel $K_{AAAA}$ is shown in Fig.~\ref{fig:gggg_SDker}.

The next step is to apply the STI of Eq.~(\ref{STISDgggg}), and scrutinize the various terms,  
denoted again by $(s_1)$, $(s_2)$, $(s_3)$, and $(s_4)$.

For the first three terms, we get the following results
\begin{eqnarray}
(s_1)&=& gf^{am'n'}g_{\alpha\nu'}\int_{k_1}D^{m'm}(k_1)\Delta_{n'n}^{\nu'\nu}(k_2)
\Gamma_{c^m A^n_\nu A^{*\gamma}_d}(k_2,-k_1-k_2)\Gamma_{A^d_\gamma A^r_\rho A^s_\sigma}(p_2,-p_1)\nonumber \\
&=&-\Gamma_{\Omega^a_\alpha A^{*\gamma}_d}(-q)\Gamma_{A^d_\gamma A^r_\rho A^s_\sigma}(p_2,-p_1),\nonumber\\
(s_2) & = &gf^{am'n'}g_{\alpha\nu'}\int_{k_1}D^{m'm}(k_1)\Delta_{n'n}^{\nu'\nu}(k_2){\cal K}_{c^m A^n_\nu A^s_\sigma A^{*\gamma}_d}(k_2,-p_1,p_2)\Gamma_{A^d_\gamma A^r_\rho}(p_2)\nonumber \\
&=&-\Gamma_{\Omega^a_\alpha A^s_\sigma A^{*\gamma}_d}(-p_1,p_2)\Gamma_{A^d_\gamma A^r_\rho}(p_2),\nonumber\\
(s_3)& = & gf^{am'n'}g_{\alpha\nu'}\int_{k_1}D^{m'm}(k_1)\Delta_{n'n}^{\nu\nu'}(k_2){\cal K}_{c^m A^n_\nu A^r_\rho A^{*\gamma}_d}(k_2,p_2,-p_1)\Gamma_{A^d_\gamma A^s_\sigma}(p_1)\nonumber\\
&=&-\Gamma_{\Omega^a_\alpha A^r_\rho A^{*\gamma}_d}(p_2,-p_1)\Gamma_{A^d_\gamma A^s_\sigma}(p_1).
\label{gggs3}
\end{eqnarray}

As  in  the case of the quark-gluon vertex, $(s_1)$  represents  the  propagator-like
contribution  that in the $S$-matrix PT would  be allotted  to  the new two point  function
$\widehat{\Gamma}_{AA}$.  Notice that  
this term  is  equal (modulo the external vertex) 
to the one found in the quark-gluon vertex
case; this is the manifestation of the   
well-known  property  of  the process-independence  
of the  PT algorithm (as already noticed in our previous one-loop analysis): the propagator-like 
contributions do not depend on the details of the 
external (embedding) particles.  
As for $(s_2)$  and
$(s_3)$, they correspond again to terms that would vanish on-shell, but now are retained in the final answer due o the off-shell condition of the external legs. 

\begin{figure}[t]
\includegraphics[width=11cm]{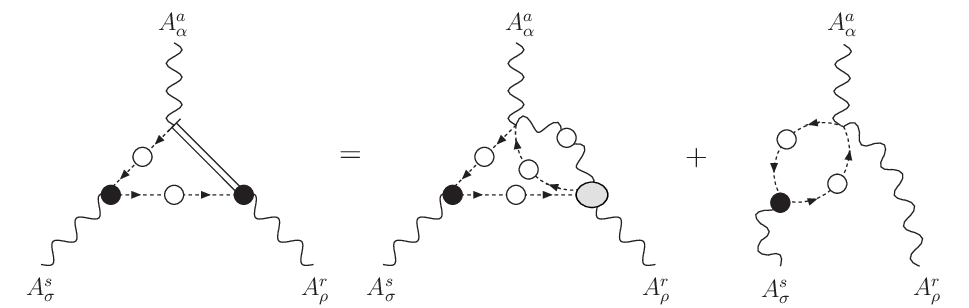}
\caption{The 1PR terms appearing in ${\cal K}_{c^mA^{*\gamma}_dA^r_\rho A^s_\sigma}$ 
contain a tree-level contribution  
generating the missing BFM topologies. 
Here we show  the case for $(\widehat{l})$;
 a symmetric term  generates $(\widehat{m})$. 
The first term on the rhs is part of the skeleton 
expansion of diagram $(\widehat{h})$ of Fig.~\ref{fig:ggg_SDE}. 
Notice that the lhs is simply a pictorial representation of 
the rhs, taking advantage of 
the notation introduced in Fig.\ref{fig:cffAstar_SDaux}; 
the anti-fields are static sources and do not propagate.}
\label{fig:1PR_treelevel}
\end{figure}

Finally, one has to consider the term $(s_4)$, given by
\begin{eqnarray}
(s_4)&=& gf^{am'n'}g_{\alpha\nu'}\int_{k_1}D^{m'm}(k_1)\Delta
_{n'n}^{\nu'\nu}(k_2){\cal K}_{c^{m} A_d^{*\gamma} A^r_\rho A^s_\sigma}(p_2,-p_1,k_2)\Gamma_{A^d_\gamma A^n_\nu}(k_2),\hspace{1cm}
\end{eqnarray}
which again can be written as the sum of the following two terms
\begin{eqnarray}
(s_{4a})&=&-igf^{am'd}g_{\alpha\gamma}\int_{k_1}D^{m'm}(k_1){\cal K}_{c^{m} A_d^{*\gamma}A^r_\rho A^s_\sigma}(k_2,p_2,-p_1), \\
(s_{4b})&=&-gf^{am'n'}g_{\alpha\nu'}\int_{k_1}\delta^{dn'}\frac{k_2^{\nu'}}{k_2^2}D^{m'm}(k_1){\cal K}_{c^{m}\bar c^dA^r_\rho A^s_\sigma}(k_2,p_2,-p_1).
\end{eqnarray}
The kernel ${\cal K}_{c\bar c AA}$ is defined by 
replacing in Eq.~(\ref{1PRker:gggg3}) every anti-field leg $A^*$ by 
the corresponding anti-ghost field $\bar c$.
As before,
$(s_{4b})$ has a tree-level ghost propagator, while $(s_{4a})$ misses a 
diagram that we need to add and subtract to solve 
the two problems simultaneously. Even so, we are still missing the diagrams $(\widehat{l})$ and $(\widehat{m})$ of Fig.~\ref{fig:ggg_SDE}; they will be  
generated by the tree-level contribution appearing in the SDE of the auxiliary function $\Gamma_{cAA^*}$. 
In order to isolate this contribution as early as possible, let us write
\begin{eqnarray}
{\cal K}_{c^{m} A_d^{*\gamma}A^r_\rho A^s_\sigma }(k_2,p_2,-p_1)
&=&
{\cal K}'_{c^{m} A_d^{*\gamma}A^r_\rho A^s_\sigma }(k_2,p_2,-p_1)\nonumber \\
&+&i\Gamma_{c^{m}A^s_\sigma\bar c^{e'}}(-p_1,\ell)iD^{ee'}(\ell)i\Gamma^{(0)}_{c^{e'}A^r_\rho A^{*\gamma}_d}(p_2,k_2)\nonumber \\
&+&i\Gamma^{(0)}_{c^eA^s_\sigma A^{*\gamma}_d}(-p_1,k_2)iD^{ee'}(\ell')i\Gamma_{c^{m}A^r_\rho \bar c^{e'}}(p_2,-\ell')\nonumber \\
&=&{\cal K}'_{c^{m} A_d^{*\gamma}A^r_\rho A^s_\sigma }(k_2,p_2,-p_1)-igf^{dre'}g_\rho^\gamma\Gamma_{c^{m}A^s_\sigma\bar c^{e'}}(-p_1,\ell)D^{ee'}(\ell)\nonumber \\
&-&igf^{dse}g^\gamma_\sigma D^{ee'}(\ell')\Gamma_{c^{m}A^r_\rho \bar c^{e'}}(p_2,-\ell'),
\end{eqnarray}
where the prime denotes that the $\Gamma_{cAA^*}$ that appears in the 1PR terms starts at one-loop. We then find (see also Fig.~\ref{fig:1PR_treelevel})
\begin{eqnarray}
(s_{4a})&=&-igf^{am'd}g_{\alpha\gamma}\int_{k_1}D^{m'm}(k_1){\cal K}'_{c^{m}A_d^{*\gamma}A^r_\rho A^s_\sigma }(k_2,p_2,-p_1)\nonumber \\
&-&g^2f^{am'd}f^{dre'}g_{\alpha\rho}\int_{k_1}D^{m'm}(k_1)D^{e'e}(\ell)\Gamma_{c^{m}A^s_\sigma\bar c^{e'}}(-p_1,\ell)\nonumber\\
&-&g^2f^{am'd}f^{dse'}g_{\alpha\sigma}\int_{k_1}D^{m'm}(k_1)D^{e'e}(\ell')\Gamma_{c^{m}A^r_\rho\bar c^{e'}}(p_2,-\ell')\nonumber \\
&=&(s'_{4a})+(\widehat{l})+(\widehat{m}).
\end{eqnarray}

\begin{figure}[!t]
\includegraphics[width=12cm]{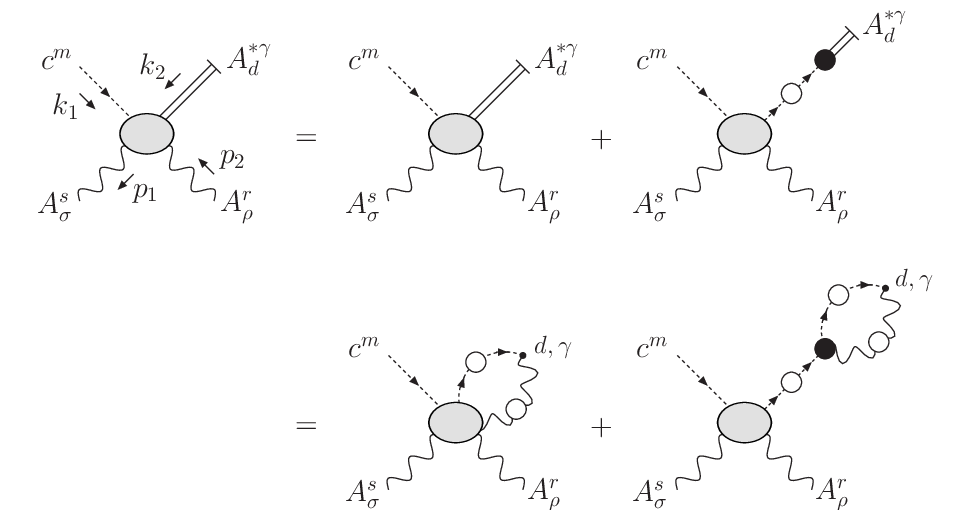}
\caption{Diagrammatic decomposition of the SD kernel ${\cal K}^{\mathrm{full}}_{c^{m}A^{*\gamma}_d A^r_\rho A^s_\sigma}$ defined in Eq.~(\ref{Kfull}). 
The first term represents the kernel ${\cal K}'_{c^{m} A_d^{*\gamma} A^r_\rho A^s_\sigma}$; 
therefore the $\Gamma_{cAA^*}$ appearing in the corresponding 1PR terms start at one-loop. 
The second term is the one added (and subtracted) to the original sum $(s'_{4a})+(s_{4b})$.
After replacing the gluon anti-field  $A^{*\gamma}_d$ with the corresponding composite operator (second line),  
this kernel generates the BFM terms $(\widehat{h})+(\widehat{i})$.}
\label{fig:cggAstar_SDaux}
\end{figure}

For generating the remaining terms one proceeds as in the quark case, writing (see Fig.~\ref{fig:cggAstar_SDaux})
\begin{eqnarray}
(s'_{4a})&=&-igf^{am'd}g_{\alpha\gamma}\int_{k_1}D^{m'm}(k_1)\left[{\cal K}'_{c^{m}A_d^{*\gamma}A^r_\rho A^s_\sigma }(k_2,p_2,-p_1)\right.\nonumber \\
&-&\left.\Gamma'_{c^gA_d^{*\gamma}}(k_2)iD^{gg'}(k_2){\cal K}_{c^{m}\bar c^{g'}A^r_\rho A^s_\sigma}(k_2,p_2,-p_1)\right]\nonumber\\
&=&-igf^{am'd}g_{\alpha\gamma}\int_{k_1}D^{m'm}(k_1){\cal K}^{\mathrm{full}}_{c^{m}A^{*\gamma}_dA^r_\rho A^s_\sigma }(k_2,p_2,-p_1)
\label{Kfull}\\
(s_{4b})&=&-gf^{am'n'}g^{\nu'}_\alpha\int_{k_1}\left[\delta^{dn'}\frac{k_{2\nu'}}{k_2^2}
-\Gamma'_{c^eA_{\nu'}^{*n'}}(k_2)D^{ed}(k_2)
\right]\times\nonumber \\
&\times&D^{m'm}(k_1){\cal K}_{c^{m}\bar c^dA^r_\rho A^s_\sigma}(k_2,p_2,-p_1).
\end{eqnarray}
Then, using Eq.~(\ref{Astrick}) (which can be safely done now, since tree-level contribution
 has been already taken into account) and Eq.~(\ref{ghSDE2}), one finds
\begin{eqnarray}
(s'_{4a})&=&(\widehat{h})+(\widehat{i})\\
(s_{4b})&=&-gf^{am'n'}\int_{k_1}k_{2\alpha}D^{m'm}(k_1)D^{n'n}(k_2){\cal K}_{c^{m}\bar c^nA^r_\rho A^s_\sigma}(k_2,p_2,-p_1),
\end{eqnarray}
so that
\begin{equation}
(s_{4b})+(e)=(\widehat{e}).
\end{equation}

Using the tree-level Feynman rules (see Appendix~\ref{Frules}),
it is straightforward  to establish that the graphs $(b)$, $(d)$, $(f)$, and $(g)$
can be converted to hatted ones automatically, and that 
$(c)^\mathrm{F}=(\widehat{c})$.
Thus, collecting all the pieces we have, and using the standard PT decomposition (\ref{PTDEC}) on the tree-level contribution $(a)$, we get 
\begin{eqnarray}
i\Gamma_{A^a_\alpha A^r_\rho A^s_\sigma}(p_2,-p_1)&=&-\Gamma_{\Omega^a_\alpha A^{*\gamma}_d}(-q)\Gamma_{A^d_\gamma A^r_\rho A^s_\sigma}(p_2,-p_1)-\Gamma_{\Omega^a_\alpha A^s_\sigma A^{*\gamma}_d}(-p_1,p_2)\Gamma_{A^d_\gamma A^{r}_\rho}(p_2)\nonumber \\
&-&\Gamma_{\Omega^a_\alpha A^r_\rho A^{*\gamma}_d}(p_2,-p_1)\Gamma_{A^d_\gamma A^{s}_\sigma}(p_1)+[(\widehat{a})+(\widehat{b})
+ (\widehat{c})
+(\widehat{d})+(\widehat{e})\nonumber\\
&+&(\widehat{f})+(\widehat{g})+
(\widehat{h})+(\widehat{i})+
(\widehat{l})+(\widehat{m})]^{ars}_{\alpha\rho\sigma}-igf^{ars}\Gamma^\mathrm{P}(p_2,-p_1).\hspace{1cm}
\label{PT-ggg-res}
\end{eqnarray}
As in the previous case, the sum of diagrams in the brackets is nothing but the kernel expansion of the SDE governing the vertex $\Gamma_{\widehat{A}AA}$, {\it i.e.}, 
\bea
i\Gamma_{\widehat{A}^a_\alpha A^r_\rho A^s_\sigma}(p_2,-p_1)&=&[(\widehat{a})+(\widehat{b})
+ (\widehat{c})
+(\widehat{d})+(\widehat{e})\nonumber \\
&+&(\widehat{f})+(\widehat{g})+(\widehat{h})+(\widehat{i})+
(\widehat{l})+(\widehat{m})]^{ars}_{\alpha\rho\sigma}.
\eea
This in turn implies that Eq.~(\ref{PT-ggg-res}) represents the BQI of Eq.~(\ref{BQI:ggg}) up to the last (tree-level) term in the rhs.
Of course this tree-level discrepancy is to be expected since the  PT
algorithm cannot possibly work at  tree-level if the external legs are amputated, 
as is the case in the SDEs we are considering. To be sure, if we start  from the  tree-level $\Gamma^{(0)}_{AAA}$  only, \ie without hooking (two of) the external legs to (conserved) external currents, we can
still carry out  the decomposition of  Eq.~(\ref{PTDEC}), but the $\Gamma^\mathrm{P}$ term will have nothing to act upon.

Notice finally that the discussion following Eq.~(\ref{BQI_gff_repr}) the SDE for the gluon-quark vertex applies with minimal modification to the three-gluon vertex case discussed here.

\subsection{The gluon propagator}

In  this  section  we  turn to the SDE of the  gluon
self-energy. From the 
technical point of view the construction is somewhat more involved  
compared to that presented for the vertices, simply   
because the PT decomposition of Eq.~(\ref{PTDEC}) must be carried out 
on both sides of the self-energy diagram. Put in a different way,  
now we must convert to background gluons not one 
but two external gluons. To the best of our knowledge,
the most efficient procedure   to   follow    consists   of the three  
basic steps described below~\cite{Binosi:2007pi}.

\subsubsection{First step}

The starting point is diagram $(a_1)$ of Fig.\ref{fig:gg_SDE-1}.
Following the PT procedure, we decompose the tree-level 
three-gluon vertex according to (\ref{PTDEC})
and concentrate on the pinch part. We then get
\begin{eqnarray}
(a_1)^\mathrm{P}&=&-\frac i2gf^{am'n'}\int_{k_1}(g_{\alpha\nu'}k_{1\mu'}-g_{\alpha\mu'}k_{2\nu'})\Delta_{m'm}^{\mu'\mu}(k_1)\Delta_{n'n}^{\nu'\nu}(k_2)
\Gamma_{A^m_\mu A^n_\nu A^b_\beta}(k_2,-q)\nonumber \\
&=& igf^{amn'}g_{\alpha\nu'}\int_{k_1}\frac1{k_1^2}\Delta_{n'n}^{\nu'\nu}(k_2)k_1^{\mu}\Gamma_{A^m_\mu A^n_\nu A^b_\beta}(k_2,-q).
\end{eqnarray}
At this point the application of the STI of Eq.~(\ref{STI:ggg}) together with Eq.~(\ref{gainvprop}) and the FPE~(\ref{FP:gcc}), results in the following terms
\begin{eqnarray}
(a_1)^\mathrm{P}&=&igf^{am'n'}g_{\alpha\nu'}\int_{k_1}D^{m'm}(k_1)\Delta_{n'n}^{\nu'\nu}(k_2)\Gamma_{c^m A^n_\nu A^{*\gamma}_d}(k_2,-q)\Gamma_{A^d_\gamma A^b_\beta}(q)\nonumber \\
&+&gf^{am'd}g_{\alpha\gamma}\int_{k_1}D^{m'm}(k_1)\Gamma_{c^m A^b_\beta A^{*\gamma}_d}(-q,k_2)\nonumber \\
&-&igf^{am'n'}g_{\alpha\nu'}\int_{k_1}\delta^{dn'}\frac{k_2^{\nu'}}{k_2^2}D^{m'm}(k_1)\Gamma_{c^m A^b_\beta\bar c^d}(-q,k_2)\nonumber \\
&=& (s_1)+(s_2)+(s_3).
\end{eqnarray}
Clearly, using the SDE of the auxiliary function $\Gamma_{\Omega A^*}$, shown in Eq.~(\ref{BQI:auxOmAs}), one has immediately that
\begin{equation}
(s_1)=-i\Gamma_{\Omega^a_\alpha A^{*\gamma}_d}(q)\Gamma_{A^d_\gamma A^b_\beta}(q).
\end{equation}
This would be half of the pinching contribution coming from the vertex in the $S$-matrix PT.

\begin{figure}[!t]
\includegraphics[width=16.4cm]{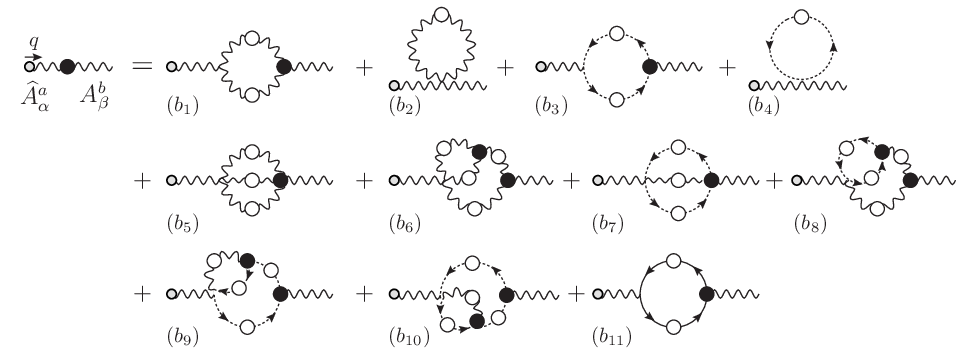}
\caption{Schwinger-Dyson equation satisfied by the gluon self-energy $-\Gamma_{\widehat{A} A}$. The symmetry factors of the diagrams are $s(b_1,b_2,b_6)=1/2$, $s(b_5)=1/6$, and all the remaining diagrams have $s=-1$.}
\label{fig:ghatg_SDE}
\end{figure}

As far as the $(s_2)$ and $(s_3)$ terms are concerned, 
let us start by adding and subtracting to them the expression needed to convert 
the tree-level ghost propagator of $(s_{3})$ into a full one; making use of the ghost SDE~(\ref{ghSDE2}) we then get
\begin{eqnarray}
(s_{2})&=&-gf^{am'd}g_{\alpha\gamma}\int_{k_1}iD^{m'm}(k_1)\left[i\Gamma_{c^m A^b_\beta A^{*\gamma}_d}(-q,k_2)+\Gamma'_{c^{g'}A^{*\gamma}_d}(k_2)D^{g'g}(k_2)\Gamma_{c^m A^b_\beta\bar c^g}(-q,k_2)\right]\nonumber \\
(s_{3})&=&-igf^{am'n'}\int_{k_1}k_{2\alpha}D^{m'm}(k_1)D^{n'n}(k_2)\Gamma_{c^mA^b_\beta\bar c^n}(-q,k_2).
\label{s4}
\end{eqnarray}
The second term symmetrizes the  trilinear ghost-gluon coupling, and one has
\begin{equation}
(s_{3})+(a_3)=(b_3),
\end{equation}
where $(b_3)$ is shown in Fig.~\ref{fig:ghatg_SDE}. The term $(s_{2})$ will finally generate all the remaining terms.
To see how this happens, we denote by $(s_{2a})$ and $(s_{2b})$ 
the two terms appearing in the square brackets of $(s_{2a})$, and concentrate on the first one. 
Making use of the SDE~(\ref{BQI:auxcAAs}) satisfied by the auxiliary function $\Gamma_{cAA^*}$ 
and the decomposition~(\ref{SDE_kercAAbarc}) of the kernel 
appearing in the latter, we get
\begin{eqnarray}
(s_{2a}) &=&g^2f^{am'd}f^{mdb}g_{\alpha\beta}\int_{k_1}D^{m'm}(k_1)\nonumber \\
&+&g^2f^{am'd}f^{dn's'}g_{\alpha\sigma'}\int_{k_1}\int_{k_3}D^{m'm}(k_1)\Delta_{s's}^{\sigma'\sigma}(k_3)D^{n'n}(k_4){\cal K}_{c^mA^b_\beta A^s_\sigma\bar c^n}(-q,k_3,k_4)\nonumber \\
&=&(b_4)+(b_7)+(b_8)+(b_{10}).
\end{eqnarray}
Using instead the SDE satisfied by $\Gamma_{cA^*}$, shown in Eq.~(\ref{BQI:auxcAs}), we obtain
\begin{eqnarray}
(s_{2b}) &=&ig^2f^{am'd}f^{dse}g_\alpha^\sigma\int_{k_1}\int_{k_3}D^{m'm}(k_1)\Delta_{\sigma\sigma'}^{ss'}(k_3)D^{ee'}(k_4)\Gamma_{c^{g'} A_{s'}^{\sigma'}\bar c^{e'}}(k_3,k_4)D^{g'g}(k_2)\times\nonumber\\
&&\hspace{4.1cm}\times\Gamma_{c^m A^b_\beta\bar c^g}(-q,k_2)\nonumber\\
&=&(b_9).
\end{eqnarray}
Finally, since diagrams $(a_2)$, $(a_4)$ $(a_5)$ and $(a_6)$ carry over to the corresponding 
BFM ones $(b_2)$, $(b_5)$, $(b_6)$ and $(b_{11})$ and $(a_1)^\mathrm{F}=(b_1)$, we have the final identity
\begin{equation}
(s_{2})+(s_{3})+\left[(a_1)^\mathrm{F}+\sum_{i=2}^6(a_i)\right]
=\sum_{i=1}^{11}(b_i),
\end{equation}
and therefore 
\begin{equation}
-\Gamma_{A^a_\alpha A^b_\beta}(q)=-i\Gamma_{\Omega^a_\alpha A^{*\gamma}_d}(q)\Gamma_{A^d_\gamma A^b_\beta}(q)-\Gamma_{\widehat{A}^a_\alpha A^b_\beta}(q),
\end{equation}
which is the BQI of Eq.~(\ref{twoBQI1}).

\subsubsection{Second step\label{step_2}}

\begin{figure}[!t]
\includegraphics[width=16cm]{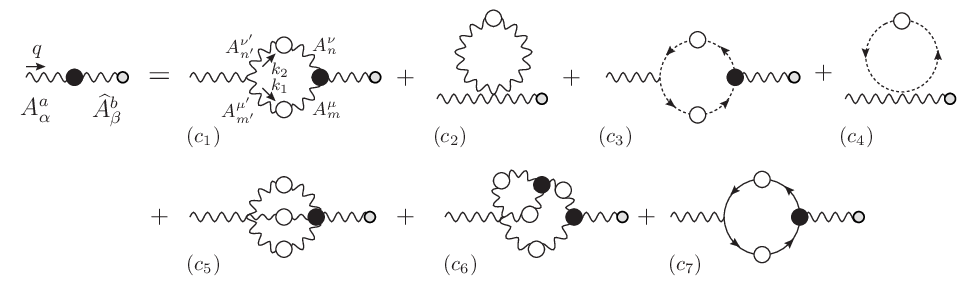}
\caption{Schwinger-Dyson equation satisfied by the gluon self-energy $-\Gamma_{A\widehat{A}}$. The symmetry factors of the diagrams are $s(c_1,c_2,c_6)=1/2$, $s(c_3,c_4,c_7)=-1$, $s(c_5)=1/6$.}
\label{fig:gghat_SDE}
\end{figure}

The second step in the propagator construction is to employ the obvious relation
\begin{equation}
\Gamma_{\widehat{A}^a_\alpha A^b_\beta}(q)=\Gamma_{A^a_\alpha \widehat{A}^b_\beta}(q),
\end{equation}
that is to  interchange the  background  and quantum
legs (the SDE for the self-energy  $-\Gamma_{A\widehat{A}}$ is shown in Fig.~\ref{fig:gghat_SDE}).
This apparently trivial operation introduces a considerable simplification. First of all, 
it allows for the identification of the  pinching  momenta from the usual PT decomposition of the (tree-level) 
$\Gamma$ appearing in diagram $(c_1)$ of Fig.~\ref{fig:gghat_SDE} 
[something not directly possible from diagram $(b_1)$]; thus, from the operational 
point of view, we remain on familiar ground.
In addition,  it  avoids the need to employ the 
(formidably complicated) BQI for the four-gluon vertex;
 indeed, the equality between diagrams $(c_5)$, $(c_6)$, $(c_7)$  
of Fig.~\ref{fig:gghat_SDE} and $(d_5)$, $(d_6)$, $(d_{11})$ 
of Fig.~\ref{fig:PT_newSDE}, 
respectively, is now immediate [as it was before, 
between the diagrams $(a_4)$, $(a_5)$, $(a_6)$ and $(b_5)$, $(b_6)$, $(b_{11})$, respectively].

\subsubsection{Third step}

We now turn to diagram $(c_1)$ and concentrate on its pinching part, given by
\begin{equation}
(c_1)^\mathrm{P}= igf^{amn'}g_{\alpha\nu'}\int_{k_1}\frac1{k_1^2}\Delta_{n'n}^{\nu'\nu}(k_2)k_1^{\mu}\Gamma_{A^m_\mu A^n_\nu \widehat{A}^b_\beta}(k_2,-q).
\end{equation}
Notice the appearance of  
the full BFM vertex $\Gamma_{AA\widehat A}$
instead of the standard $\Gamma_{AAA}$ (in the $R_{\xi}$). 
The STI satisfied by the former vertex has been derived in Eq.~(\ref{STI:mixed1}). 
Now, the first three terms, $(s_1)$, $(s_2)$ and $(s_3)$, appearing in this STI, 
will give rise to PT contributions exactly equal to those encountered 
in first step described above, the only difference being that 
the $A^b_\beta$ field appearing there is now a background field $\widehat{A}^b_\beta$. 
Thus, following exactly the reasoning described before, 
we find [see again Fig.~\ref{fig:PT_newSDE} for the diagrams corresponding to each $(d_i)$]
\bea
(s_1)&\to&-i\Gamma_{\Omega_\alpha^aA^{*e}_\epsilon}(q)\Gamma_{ A^e_\epsilon\widehat{A}^b_\beta}(q),\\
(s_2)+(s_3)+(c_3)&=&(d_3)+(d_4)+(d_7)+(d_8)+(d_9)+(d_{10}).
\eea
For the term $(s_4)$ we have instead
\be
(s_4)\to g^2f^{am'e}f^{ebm}g_{\alpha\mu'}g_{\beta\mu}\int_{k_1}\Delta_{m'm}^{\mu'\mu}(k_1).
\ee
Clearly this has a tadpole-like structure; in particular, it is immediate to prove 
that when added to $(c_2)$ it will convert it into $(d_2)$
\begin{equation}
(s_4)+(c_2)=(d_2).
\end{equation}
Thus, since as always $(c_1)^\mathrm{F}=(d_1)$ we get
\be
(s_{2})+(s_{3})+(s_4)+\left[(c_1)^\mathrm{F}+\sum_{i=2}^7(c_i)\right]
=\sum_{i=1}^{11}(d_i),
\ee
and therefore
\begin{equation}
-\Gamma_{A^a_\alpha \widehat{A}^b_\beta}(q)=-i\Gamma_{\Omega_\alpha^aA^{*e}_\epsilon}(q)\Gamma_{A^e_\epsilon \widehat{A}^b_\beta}(q)-\Gamma_{\widehat{A}^a_\alpha \widehat{A}^b_\beta}(q),
\end{equation}
which is the BQI of Eq.~(\ref{twoBQI2}). This concludes our proof.
\begin{figure}[!t]
\includegraphics[width=8.5cm]{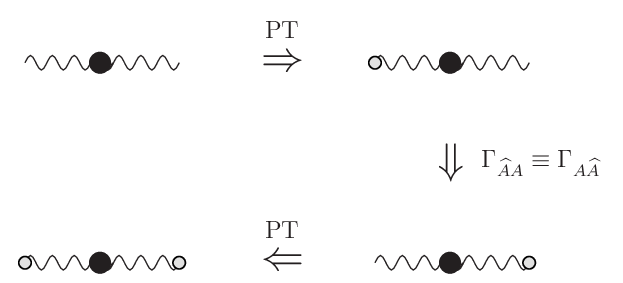}
\caption{Summary of the PT procedure employed in the text in order to construct the new PT SDE of the gluon propagator.}
\label{fig:4steps}
\end{figure}

In Fig.~\ref{fig:4steps} we summarize the steps that allowed the successful construction of the 
SDE for the PT propagator; putting together the three steps above, we have been able to generate the complete BQI of Eq.~(\ref{BQI:gg}), namely
\be
i\Gamma_{\widehat{A}_\alpha^a\widehat{A}_\beta^b}(q)=i\Gamma_{A^a_\alpha A^b_\beta}(q)+2\Gamma_{\Omega_\alpha^a A^{*\gamma}_d}(q)\Gamma_{A^d_\gamma A^b_\beta}(q)-i\Gamma_{\Omega_\alpha^a A^{*\gamma}_d}(q)\Gamma_{A^d_\gamma A^{e}_{\epsilon}}(q)\Gamma_{\Omega_\beta^b A^{*\epsilon}_{e}}(q).
\label{BQI:gg-1}
\ee
According to the PT rules put forward in Section~\ref{PT_BV_1l}, on the one hand the PT gluon two-point function $i\widehat{\Gamma}_{AA}$ would coincide with the rhs of Eq.~(\ref{BQI:gg-1}) after dropping the terms proportional to the auxiliary function $\Gamma_{\Omega A^*}$, since these would cancel anyway after adding the 
contribution coming from the corresponding vertices.
On the other hand, recalling that $\Gamma_{\Omega_\alpha A^*_\beta}$ coincide with $\Lambda_{\alpha\beta}$, and observing that the only relevant part in the identity above of such functions is the one proportional to the metric tensor (due to the transversality of the gluon two-point function $\Gamma_{AA}$), it is immediate to show, using 
Eq.~(\ref{inv_prog}) and the relation $-\Gamma_{A_\alpha A_\beta}=\Pi_{\alpha\beta}$, that Eq.~(\ref{BQI:gg-1}) can be cast in the form of the SDE shown in Eq.~(\ref{newSDb}).

\subsection{How to truncate the new SDEs}

After constructing the new SDE series, let us focus 
on its truncation properties, both from the theoretical 
as well as the practical point of view. 

As has been stated repeatedly, the main theoretical advantage 
of the new SD series is that it allows for 
a  systematic gauge-invariant truncation, in the sense described in subsection~\ref{nsds}.
There  we focused on how to truncate the SDE for the gluon self-energy, 
shown in Fig.\ref{fig:PT_newSDE}, by exploiting the fact that 
the various fully dressed graphs organize themselves into gauge-invariant subsets  
[those appearing in Eq.(\ref{4tr})].
The practical importance of this property 
is the following: one can reduce the number of coupled SDE that one must include 
in order to maintain the gauge (or BRST) symmetry  
of the theory intact, as reflected, for example in the validity of Eq.(\ref{fundtrans}). Thus, 
in the case of pure Yang-Mills, within this new formulation,   
the minimum number of 
equations that one must consider is only two: The SDE for the gluon self-energy, given by the 
first gauge-invariant subset {\it only} ({\it i.e.},  
$\left[ (d_1) + (d_2) \right]_{\alpha\beta}$ in Fig.\ref{fig:PT_newSDE}) and 
the SDE for the full three-gluon vertex, shown in Fig.\ref{fig:ggg_SDE} 
(which is instrumental in assuring the gauge invariance of the subset chosen).  
This is to be contrasted to what happens within the conventional formulation: there the SDEs for {\it all} 
vertices must be considered, or else  Eq.~(\ref{fundtrans}) is violated (which is what usually happens).

Notice an important point, however: the present analysis does {\it not} furnish a simple diagrammatic 
truncation, analogous to that of the gluon self-energy,
for the SDE of the 
three gluon  vertex $\Gamma_{\widehat{A}_\alpha A_\mu A_\nu}(k_1,k_2)$, shown in Fig.\ref{fig:ggg_SDE}. 
Thus, if one were to truncate the SDE for the three-gluon vertex by keeping any subset of the graphs appearing 
in Fig.\ref{fig:ggg_SDE}, 
one would violate the validity of the {\it all-order} WI of Eq.~(\ref{3gl}); 
this, in turn, would lead immediately to the violation of Eq.~(\ref{fundtrans}), thus 
making the entire truncation scheme collapse.

The strategy one should adopt is instead the following. Given that the proposed 
truncation scheme hinges crucially on the validity of  Eq.~(\ref{3gl}), one should 
start out with an approximation that manifestly preserves it. 
The way to enforce this, familiar to the SDE practitioners already from the time of QED, is to resort 
to the ``gauge-technique''~\cite{Salam:1963sa}, namely ``solve'' the WI of Eq.~(\ref{3gl}). 
Specifically, one must  express  the three-gluon vertex 
as a functional of the corresponding self-energies, in such a way that (by construction) 
its WI is automatically satisfied. For example, an Ansatz with this property would be 
\be
\Gamma_{\widehat{A}_\alpha A_\mu A_\nu}(k_1,k_2)=\Gamma^{(0)}_{\widehat{A}_\alpha A_\mu A_\nu}(k_1,k_2) - 
i\frac{(k_2-k_1)_\alpha}{k_2^2-k_1^2}\left[\Pi_{\mu\nu}(k_2)-\Pi_{\mu\nu}(k_1)\right]; 
\label{GTA}
\ee
contracting the rhs with $q_{\alpha}=(k_1+k_2)_{\alpha}$ yields automatically the WI of Eq.~(\ref{3gl}). 
Thus, the minimum amount of ingredients for initiating a {\it self-consistent} non-perturbative 
treatment is the SD for the gluon self-energy, consisting of $\left[ (d_1) + (d_2) \right]_{\alpha\beta}$, 
supplemented by an Ansatz for the three-gluon vertex like the one given in (\ref{GTA}). 
Note that the  ``gauge-technique'' leaves the transverse (\ie automatically conserved) part 
of the vertex undetermined. This is where the SDE for the vertex enters; 
it is used precisely to determine the transverse parts. 
Specifically, following standard techniques~\cite{Ball:1980ax,Binger:2006sj}, 
one must expand the vertex into a suitable tensorial basis, consisting of fourteen independent 
tensors, and then isolate the transverse subset. This procedure will lead to a large number 
of coupled integral equations, one for each of the form-factors multiplying 
the corresponding tensorial structures, which may or may not be tractable.
However, at this point, one may simplify the resulting equations (\eg linearize, etc)  
without jeopardizing the transversality of $\Pi_{\mu\nu}$, which only depends on the 
``longitudinal'' part of the vertex, \ie the one determined by (\ref{GTA}).
Thus, the transverse parts will be approximately determined, but gauge invariance, 
as captured by $q^{\mu}\Pi_{\mu\nu}=0$, will remain exact. 

Note by the way that the methodology described above constitutes, even to date,  
the standard procedure  even in the context of QED, where the structure of the SDE is much simpler, given 
that the SDE for the photon contains one single graph [diagram $(a_6)$ in Fig.\ref{fig:gg_SDE-1}], 
and  the photon-electron vertex satisfies automatically a naive all-order WI.
Thus, while  the PT approach described here 
replicates QED-like properties at the level of the SDEs of QCD, in our opinion a striking fact in itself, 
does not make QCD easier to solve than QED.  

The reader should appreciate one additional point: any attempt to apply the approach described above 
in the context of the conventional SDE is bound to lead 
to the violation of the transversality of $\Pi_{\mu\nu}$, because ({\it i}) 
the vertices satisfy complicated STI's instead of the WIs of Eq.~(\ref{3gl})--(\ref{4gh}), 
a fact that makes the application of the  ``gauge-technique'' impractical, and ({\it ii})  
even if one came up with the analogue of Eq.~(\ref{GTA}) for all vertices, one should still 
keep all self-energy diagrams in Fig.\ref{fig:gg_SDE-1} to guarantee that $q^{\mu}\Pi_{\mu\nu}=0$. 
From this point of view, the improvement of the present approach over the standard formulation becomes evident.

Finally, one should be aware of the fact that there is no a-priori 
guarantee that the gauge-invariant subset kept ({\it i.e.},  $\left[ (d_1) + (d_2) \right]_{\alpha\beta}$) 
capture necessarily most of the dynamics, or, in other words, that 
they represent the numerically dominant contributions  
(however, for a variety of cases it seems to be true~\cite{Aguilar:2008xm}). 
But, the point is that one can {\it systematically} improve the picture by including more terms, 
without worrying that the initial approximation is plagued 
with artifacts, originating from the  violation of the gauge invariance or of the BRST symmetry.

\section{Conclusions\label{Concl}}

In this  article we have presented  a detailed derivation of  a new SD
series  for  non-Abelian gauge  theories,  based  on  the PT  and  its
correspondence  with  the  BFM.   
The procedure we followed for constructing the PT SDE 
is identical to that followed in the perturbative construction,
{\it without} any additional new assumption.

Our starting point is 
the conventional SDEs for the vertices and the gluon self-energy
written  in the Feynman gauge.
The first step in the 
derivation is to simply
carry  out  the PT  decomposition  of  the elementary  (tree-level)
three-gluon   vertex,   $\Gamma_{\alpha\mu\nu}$  given   in
Eq.~(\ref{PTDEC}),  to the external  three-gluon vertices  appearing in
the corresponding  diagrams of  the standard SD  series.  The  part of
$\Gamma_{\alpha\mu\nu}$     denoted     by    $\Gamma^{{\rm
P}}_{\alpha\mu\nu}$ contains longitudinal momenta which get contracted
with  the kernels  or  the fully-dressed  Green's functions  appearing
inside       the      diagram       containing       the      original
$\Gamma_{\alpha\mu\nu}$,   triggering   the   corresponding
STIs.  These  STIs,  in  turn,  contain pieces  that, according to 
the well-established PT criteria, either form part of the 
answer, in this case the diagrammatic expansion of 
the SDE for the corresponding Green's function,
 or they are discarded from it.  
In Section~\ref{SDE} we  have worked out in detail three
cases: ({\it i})  the quark-gluon  vertex, ({\it ii}) the three-gluon  vertex, 
and  ({\it iii}) the technically more involved case of the gluon self-energy.
It turns out that the diagrams comprising the PT answer
are identical to those corresponding to the BFG. 
Thus, the  resulting  new   SDEs,  generated  after  the characteristic  
PT rearrangements have taken place,  
and the PT criterion for identifying the answer has been employed, 
correspond to the BFM SDEs, written in  the BFG.   This 
is an important result, because it 
proves the PT-BFG correspondence 
at the level of the SDE of the theory; obviously, all
results on this point 
presented in the literature so far  
are included in the result presented here,
given that any order in perturbation theory is already contained
in the SDEs we consider.

 An additional important 
result, obtained from the same procedure, is the 
diagrammatic derivation of the BQIs, which, to date, have 
only been formally derived in the context of the BV formalism.  
The way that the terms comprising the BQIs 
appear in the present analysis is automatic:
they are simply the leftovers of the PT construction, \ie the 
pieces that have been discarded from the PT answer.

As explained in detail in~\cite{Aguilar:2006gr, Binosi:2007pi}, and mentioned also in Section~\ref{General},
the new SDE series for the gluon self-energy 
contains fully dressed vertices that satisfy simple,
QED-like (\ie tree-level-like) WIs, instead of STIs.
This fact allows for the truncation of the SDE series while maintaining the 
transversality of the answer at any step.  

Returning to the construction of the gluon SDE, a crucial ingredient for the proof has been the interchange of background and quantum legs done in subsection~\ref{step_2}, \ie  \mbox{$\Gamma_{\widehat{A}A}\to\Gamma_{A\widehat{A}}$} (see also Fig.~\ref{fig:4steps}). As we have mentioned there this allowed us to ({\it i}) identify the  pinching  momenta from the usual PT decomposition, and ({\it ii}) avoid the use of some otherwise indispensable multi-leg BQIs. But more importantly it unveils a recursive pattern that can be used to generalize the construction to $n$-point Green's functions with $n$ arbitrary. Work in this direction is already in progress.
 
As we  have emphasized 
in the Introduction, the BFG is singled out dynamically when
carrying  out the  PT rearrangement  of a  physical quantity,  such as an
$S$-matrix element or a Wilson  loop. In particular, after  the full
cancellation  of  all  (effectively  propagator-like)  gauge-dependent
pieces has  taken place,  and after the  vertices have been  forced to
obey  Abelian  WIs,  the  resulting self-energy  contribution  (to  be
identified with the PT self-energy) {\it coincides} with the BFM gluon
self-energy, calculated in the BFG. In that sense the BFG is 
very special, because it captures the net gauge-independent and   
universal (\ie process-independent) contribution 
contained in any physical quantity.
In practice, however, one would like to be able to truncate  
gauge-invariantly (\ie maintaining transversality) sets of 
SDEs written in different gauges.
This becomes particularly relevant, for example,  
when one attempts to compare SDE predictions with 
lattice simulations, carried out usually in the Landau gauge.
One of the most powerful features of this formalism,
not explored in this article,
is that it can be generalized to any other gauge choice.
In particular, Eq.~(\ref{newSDb}) maintains the same form,
regardless of the gauge chosen. 
The way to accomplish this is to use the 
``generalized'' PT, developed in~\cite{Pilaftsis:1996fh}. The generalized PT 
modifies the  starting point of the PT algorithm, namely Eq.~(\ref{PTDEC}), distributing  
differently the longitudinal momenta between $\Gamma_{\alpha \mu \nu}^{{\rm F}}$
and $\Gamma_{\alpha \mu \nu}^{{\rm P}}$.
Specifically, the non-pinching part, \ie the analogue of 
$\Gamma_{\alpha \mu \nu}^{{\rm F}}$, 
must satisfy, instead of (\ref{WI2B}), 
a WI whose rhs is the difference of two inverse tree-level propagators 
in the gauge one wishes to consider.
The way this works is the following. 
One starts out with the conventional SDE in the chosen gauge, 
carrying out the generalized  PT vertex decomposition.
Then, the action of the corresponding $\Gamma_{\alpha \mu \nu}^{{\rm P}}$
projects one to the corresponding BFM gauge; this 
includes the covariant gauges, such as the  Landau gauge,  
or even non-covariant gauges, such as axial or light-cone gauges (for the way how to use non-covariant gauges
within the BFM framework, see~\cite{Pilaftsis:1996fh}). 
This new SD series contains full vertices 
that, even though they are in a different gauge,
satisfy the QED-like WIs given in Eq.s~(\ref{3gl}) --~(\ref{4tr}).
Therefore, the truncation properties of this SDE 
are the same as those discussed in Section~\ref{General}
for the case of the Feynman gauge.
The analogy is completed by realizing that 
the BQIs in the corresponding gauge allow 
one to switch back and forth 
from the conventional to the BFM Green's function.
Thus, one may obtain, for example, 
transverse approximations for
the gluon propagator 
in the conventional Landau gauge by studying the SDE written 
in the BFM Landau gauge, computing the 
$[1+G(q^2)]^2$ in the same gauge, \ie employ
 Eq.~(\ref{newSDb}) using for the diagrams on its rhs 
the BFM Feynman rules in the Landau gauge (see Appendix).
This Landau gauge SDE has already been used in ~\cite{Aguilar:2008xm}, 
in order to derive results for the gluon and ghost propagators that are in qualitative agreement 
with recent lattice data; as explained there,
particular care is needed when taking the limit $\xi_Q \to 0$.  

It would be important, both from the theoretical point of view
as well as for the practical applications, 
to study in detail the renormalization properties 
of the new SDEs (for a general discussion see the Appendix \ref{Appendix:REn}). 
For addressing this problem
the intrinsically non-perturbative renormalization 
method known as ``displacement operator formalism''~\cite{Binosi:2005yk} 
may prove to be particularly suitable. 

It is also very appealing to believe that the SDE derived here 
may be actually obtained from a variational principle, \ie 
as a result of the extremization of an appropriate effective action,
as happens in the case of the CJT formalism~\cite{Cornwall:1974vz}. 
Calculations in this directions are already in progress.

Finally, it would be interesting to establish connections between the dynamics obtained from the SDEs 
derived here and results based on the  
non-perturbative BFM formalism developed in~\cite{Simonov:1993kt}.  

\acknowledgments{D.B. acknowledges the hospitality of the Physics Department of the University of 
Valencia, where part of this work was carried out. 
The research of J.P. is supported by the Fundaci\'on General of the UV and by the 
MEC grant FPA 2005-01678.\\
Feynman diagrams have been drawn using \verb+JaxoDraw+~\cite{Binosi:2003yf}.}

\newpage

\appendix

\section{Which way to pinch and why\label{Appendix:howtopinch}}

\renewcommand{\theequation}{A.\arabic{equation}}

\begin{figure}[!t]
\bce
\includegraphics[width=15cm]{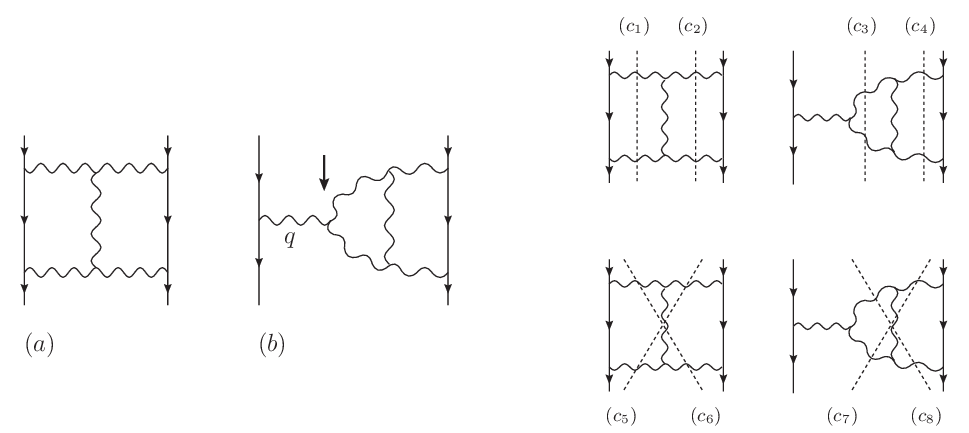}
\ece
\caption{Left panel: Some examples of external ( indicated with an arrow) and 
internal  vertices (all the remaining). 
Diagram $(a)$ has only internal three-gluon vertices, while diagram $(b)$ 
has two internal vertices and one external, indicated by the arrow.
Right panel: The two- and three-particle Cutkosky cuts (cutting through gluons only).}
\label{TWL0}
\end{figure}

Historically,  the  basic  conceptual  difficulty associated  with  the
generalization of  the PT  beyond one loop  has been to  determine the
origin of the  pinching momenta.  Let us assume  that, without loss of
generality, one  chooses from  the beginning the  conventional Feynman
gauge.  Then,  the only  sources of possible pinching  momenta are
the three-gluon  vertices. The question  is whether all  such vertices
must  be somehow forced  to pinch,  or, in  other words,  whether the
standard PT decomposition of  Eq.~(\ref{PTDEC}) should be carried out
to  all available  three-gluon  vertices.  The  problem  with such  an
operation, however,  is the following:  for the case of  a three-gluon
vertex nested  inside a Feynman diagram,  how does one  choose which is
the  ``special'' momentum?   Or,  in  other words,  which  way is  one
supposed to brake the Bose-symmetry of the vertex?  Turns out that the
solution  to these  questions is  very simple~\cite{Papavassiliou:1999az}:  
one should apply Eq.~(\ref{PTDEC}) {\it only} 
to the  vertices that  have the  physical momentum
incoming (or  outgoing) in one of  their legs (not  mixed with virtual
momenta); the special  leg is precisely the one  carrying $q$, 
{\it i.e.},  the  {\it physical}  momentum transfer appearing in the problem.
We will call such  a vertex  ``external''.  All other  vertices are not  to be
touched, {\it i.e.},   they should not be decomposed in any way; such vertices
have virtual momenta entering into every one  of their three legs, 
and are called ``internal'' (see Fig.\ref{TWL0}).

The reason why all other  three-gluon vertices inside the loops should
remain unchanged (no splitting) can  be best understood by resorting to the
{\it absorptive construction} 
of the PT~\cite{Papavassiliou:1995fq,Papavassiliou:1996fn,Papavassiliou:1996zn}.  
The basic philosophy behind the absorptive construction 
is to emulate as much as possible 
the text-book reconstruction of the real part of the
vacuum polarization of QED (containing say a muon-loop)   
from the tree-level cross-section for
\mbox{$e^{+}e^{-} \rightarrow \mu^{+} \mu^{-}$}, {\it i.e.},  the optical theorem, and a 
(once-subtracted) dispersion relation. 
As in QED, 
in the case of the PT 
the basic observation also happens already at one loop:  
the PT subamplitudes (self-energies, vertices, boxes) 
satisfy the optical theorem {\it individually}, in a way similar to 
what happens with scalar theories  and QED.  

Specifically, let us write $S=1+iT$, and 
consider the forward  scattering   process  
$q(p_1)  {\bar q}(p_2)\!\to\! q(p_1){\bar q}(p_2)$, 
with $s=q^2=(p_1+p_2)^2$. 
Restricting ourselves to only 
gluonic intermediate states, the PT amplitudes,  
at lowest order satisfy   
\begin{equation}
\Im m \langle q\bar{q}|T^{[4]}|q\bar{q}\rangle_{\ell} = \frac12\times
\frac12\int_{\mathrm{PS}_{2g}}  
[\langle q\bar{q}|T^{[2]}|gg\rangle \langle gg|T^{[2]}|q\bar{q}\rangle^{*}]_{\ell}, 
\label{OTgg}
\end{equation}
with $\int_{\mathrm{PS}_{2g}}$ denoting the two-body phase space for massless gluons.
In the equation above the superscript $[n]$ denotes the order of the corresponding 
amplitude in the coupling constant $g$ (when counting powers of 
$g$ remember the couplings coming from the vertices with the external particles);
the subscript $\ell=1,2,3$ denotes, respectively, the 
propagator-, vertex-, and box-like parts of either side 
(to recover the full optical theorem, one simply sums both sides over $\ell$).
Finally, the extra factor of $\frac12$ is statistical, since 
the final state gluons are considered as identical particles in the total rate. 

The meaning of propagator-, vertex-, and box-like is clear as far as the 
lhs of  Eq.~(\ref{OTgg}) is concerned: one must determine the imaginary (absorptive)
parts of the 
three one-loop PT 
subamplitudes obtained after casting the amplitude 
$q(p_1)  {\bar q}(p_2)\!\to\! q(p_1){\bar q}(p_2)$ into the PT form, following the 
standard PT rules. To get these absorptive parts one 
may carry the corresponding Cutkosky cuts to the various integrals, 
(including ``unphysical'' contributions coming from ghost loops) 
or, equivalently, 
study where the various logarithms develop imaginary parts.

Let us see now what  propagator-, vertex-, and box-like means on the  
rhs, consisting of the squared amplitude for the tree-level process 
$q(p_1)  {\bar q}(p_2)\!\to\! g(k_1) g(k_2)$ (with $k_1$ and $k_2$ integrated over 
all available phase space).  
A  PT-rearranged squared amplitude means the following.
Consider a normal squared amplitude, {\it i.e.},  the product of two 
regular amplitudes [remember that 
``product'' means that they are also connected (multiplied) 
by the corresponding polarization 
tensors]. Then each amplitude must be first cast into its PT form, by again simply following the standard PT rules.
However, this is not the end of the story as far as 
the  PT-rearrangement of the square is concerned. 
One must go through the additional exercise  
of letting the longitudinal momenta coming from the 
polarization vectors trigger a particular cancellation 
between the $s$-channel and the $t$-channel graphs 
(known in the literature as the ``$s$-$t$ cancellation''). 
That will finally identify the genuine propagator-, vertex-, and box-like 
pieces of the entire product.

Now the important step is the following: Suppose that one starts out with the 
rhs of   Eq.~(\ref{OTgg}), {\it i.e.},  one works at the level of the physical 
squared amplitude. The PT rearrangement of the rhs may furnish the 
PT rearranged amplitudes on the lhs, through an (appropriately subtracted) dispersion 
relation. Thus, the absorptive PT construction means to 
({\it i}) PT-rearrange the rhs, 
({\it ii}) impose the optical theorem (individually for each $\ell$), and ({\it iii}) use analyticity 
to get the real parts of the PT amplitudes. 

Let us now see how the PT absorptive construction gets generalized to higher orders, and, in 
particular, how it can furnish a unique way for defining the PT construction 
without any a-priori reference to the BFM and its special vertices.

\begin{figure}[!t]
\bce
\includegraphics[width=16cm]{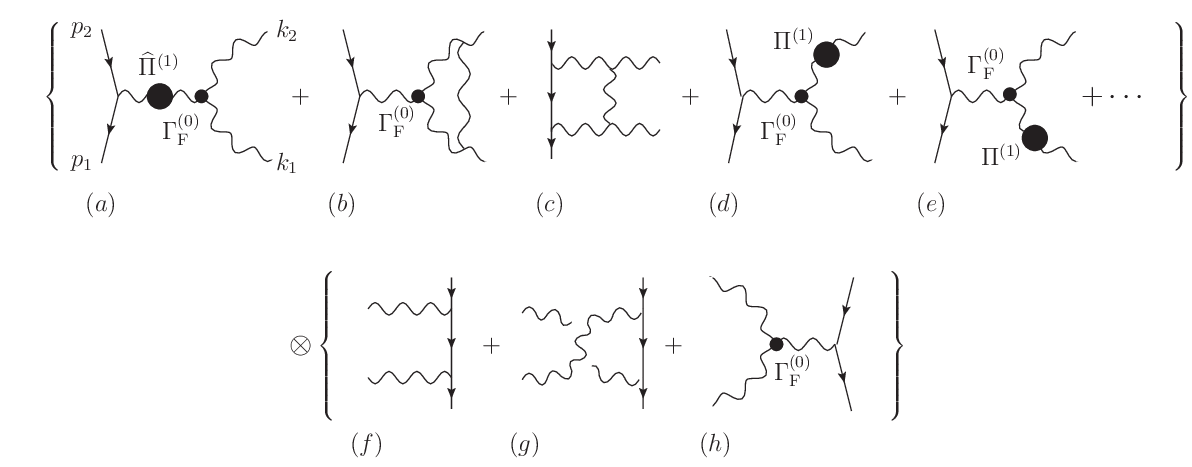}
\ece
\caption{The product of the PT-rearranged amplitudes 
of the process $q\bar{q}\to q\bar{q}$ at one-loop (up) and tree-level (down). 
The longitudinal momenta from the polarization tensors will produce 
additional cancellations between $s$-channel and $t$-channel graphs~\cite{Papavassiliou:1999az}, 
furnishing finally the first term on the rhs of Eq.~(\ref{abq}).}
\label{PTabs2}
\end{figure}

At the next order in $g^2$ Eq.~(\ref{OTgg}) becomes 
\bea
\Im m 
\langle q \bar{q}| T^{[6]} | q\bar{q} \rangle_{\ell} &=&
\frac{1}{2} \left(\frac{1}{2!}\right)\int_{\mathrm{PS}_{2g}}\!
2 \,\Re e [ {\langle gg | T^{[4]} |q\bar{q} \rangle}^{*}
\langle gg | T^{[2]} |q\bar{q} \rangle ]_{\ell}
\nonumber\\
&+& \frac{1}{2} \left(\frac{1}{3!}\right)\int_{\mathrm{PS}_{3g}}\!
[{\langle ggg | T^{[2]} |q\bar{q} \rangle}^{*}
\langle ggg | T^{[2]} |q\bar{q} \rangle]_{\ell}\,,
\label{abq}
\eea
where now $\int_{\mathrm{PS}_{3g}}$ denotes the three-body 
phase-space for massless gluons.

\begin{figure}[!t]
\bce
\includegraphics[width=10cm]{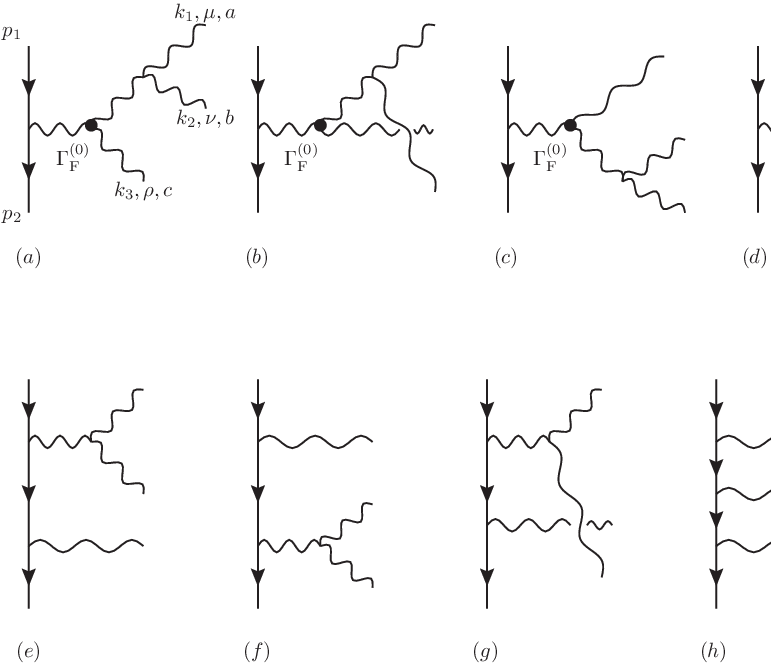}
\ece
\caption{The PT-rearranged tree-level amplitude for the process $q\bar{q} \to ggg$; 
squaring and further pinching triggered by the longitudinal momenta inside the polarization tensors~\cite{Papavassiliou:1999az}
will furnish the second term on the rhs of Eq.~(\ref{abq}).}
\label{PTabs3}
\end{figure}

According to Eq.~(\ref{abq}) then
the   imaginary parts of the
two-loop PT Green's  functions (under construction) are related by  the optical theorem  to
precisely  identifiable   and very  special  parts  of 
the squared amplitudes for the  processes
$q\bar{q} \to  gg$ and $q\bar{q} \to ggg$.  In  particular, the two-particle Cutkosky cuts of
the two-loop PT self-energy are related to the propagator-like part of the
{~\it PT-rearranged}
one-loop squared amplitude for $q\bar{q} \to gg$, while, at the same time,
the three-particle Cutkosky  cuts of the same  quantity are related to
the  propagator-like part of the 
{~\it PT-rearranged} tree-level squared amplitude for $q\bar{q} \to ggg$.
The same holds for vertex- and box-like contributions.
The processes appearing on the rhs of  Eq.~(\ref{abq}) 
are shown in Fig.s~\ref{PTabs2} and~\ref{PTabs3}.
The advantage of this formulation is the following:
all the PT-rearranged 
(squared) amplitudes appearing on the rhs are at least one loop lower 
than the amplitude on the lhs. Therefore, one can actually reconstruct  
the lhs, by working directly on the rhs, 
because one knows how to pinch at lower orders. 

To see how all this analysis makes finally contact with the main question at hand, 
namely which way to pinch in higher orders, let us focus on  Fig.~\ref{PTabs2}.
There it is clear that the product 
involves the PT-rearranged {\it one-loop} on-shell amplitude for the process 
$q(p_1)  {\bar q}(p_2)\!\to\! g(k_1) g(k_2)$, whose construction 
is absolutely fixed and well defined, and has been described in great detail 
in the literature~\cite{Watson:1994tn,Binosi:2002ez} 
In fact, it was the first explicit example~\cite{Watson:1994tn} 
demonstrating the universality (process-independence) 
of the PT gluon self-energy: the resulting gluon self-energy  
does not depend on the embedding process 
(quarks to quarks or quarks to gluons or gluons to gluons, etc).
The PT-rearranged {\it one-loop} $q(p_1)  {\bar q}(p_2)\!\to\! g(k_1) g(k_2)$
is obtained following exactly the same PT procedure as for the 
process with only quarks as external particles.  
In particular, the three-gluon vertices in graphs $(b)$ and $(c)$ in Fig.~\ref{PTabs2} must be exactly as shown, {\it i.e.},  
the one injected with $q$ has undergone the  PT decomposition 
(and has become $\Gamma^{{\rm F}}$), while 
the ones injected  with $k_1$ and $k_2$ remain unchanged. 
Let us now return to the two representative two-loop diagrams of Fig.\ref{TWL0}.
After their PT rearrangement, the two-particle Cutkosky cut on $(a)$ denoted by $(c_2)$ in Fig.~\ref{TWL0} 
must reproduce 
$(c)\otimes [(f)+(g)]$ in  Fig.~\ref{PTabs2}, and  cut $(c_4)$ on $(b)$ must reproduce  $(b)\otimes [(f)+(g)]$. 
Obviously, if we were to modify the internal three-gluon vertices of $(a)$ or $(b)$ in Fig.~\ref{TWL0}
in any way, this identification would not work: one must modify {\it only} the vertex injected with $q$ 
(turning it to $\Gamma^{{\rm F}}$).
This argument may be generalized to include all remaining two-particle and   
three-particle cuts, making  the above conclusion completely airtight. 
Under the light of these observations it should be clear why, for example, the relevant 
full three-gluon vertex 
 entering into the SD equation for the gluon self-energy ({\it viz.} Fig.\ref{fig:PT_newSDE}) is indeed  
$\Gamma_{\widehat{A}AA}$ (constructed in subsection \ref{tgv} )
and {\it not} the vertex $\Gamma_{\widehat{A}\widehat{A}\widehat{A}}$.
The latter could not be consistently constructed inside loops, because any preferred direction 
(i.e. the direction determining the ``would-be'' background field) is immediately at odds 
with the unitarity-cut arguments developed above.

The arguments presented here do not
postulate at any point the existence of any relation between the PT and the 
BFM. On the other hand, on hindsight, all conclusions drawn 
(for example, the $\Gamma_{\widehat{A}AA}$ versus $\Gamma_{\widehat{A}\widehat{A}\widehat{A}}$ issue) 
are in complete agreement with the known PT/BFM correspondence.
Specifically, switching now to the BFM language, 
the fact that internal vertices should not be touched is precisely what the unique set of BFM Feynman rules dictates:
since one cannot have background fields propagating inside loops, all internal vertices 
have three {\it quantum} gluons merging. 
This is exactly what one finds, {\it e.g.}, when computing the two-loop gluon self-energy~\cite{Abbott:1980hw}: 
as a subset of the calculation one will have to consider the one-loop 
vertex $\Gamma^{(1)}_{\widehat{A}AA}$, but will never encounter the one-loop vertex 
$\Gamma^{(1)}_{\widehat{A}\widehat{A}\widehat{A}}$ (constructed in~\cite{Cornwall:1989gv} and studied in~\cite{Binger:2006sj}).

\section{A brief discussion of renormalization \label{Appendix:REn}}
\renewcommand{\theequation}{B.\arabic{equation}}

This  appendix is meant  to outline the  general  framework 
for dealing with the issue of renormalization in the context of the PT.  
The emphasis is put on the various
conceptual  and   methodological  issues  involved,   rather  than  an
explicit proof  of renormalizability. In particular, we  consider  this
discussion necessary  for convincing the  reader that renormalization
poses  no problem  whatsoever for the self-consistent implementation of the PT.  

The analysis presented thus far assumes implicitly that the 
theory is renormalizable (as is QCD in $d=4$) 
or superrenormalizable (as is QCD in $d=3$), 
and that all momentum space integrals have been regularized by resorting to 
a regularization scheme that preserves the 
gauge symmetry (obviously there is little 
point in applying the PT to a theory that is ill-defined to begin with).
Specifically, throughout 
the paper we have adopted the most widely used such scheme,
namely dimensional regularization. 
Given that the original theory is renormalizable (by assumption) 
it should be clear 
 that there is no step throughout the PT procedure that could jeopardize 
renormalizability. Indeed, all that the PT really does is to 
trigger STIs. The latter are a direct 
consequence of the  original BRST 
symmetry of the  theory;  therefore, 
within a suitable regularization scheme (such as dimensional regularization)
they will be  preserved by renormalization ({\it i.e}, they will not get deformed).
It is important to emphasize that the latter property  
holds true for the BQIs as well; 
they too are a consequence of the 
 BRST symmetry and (under the same assumptions) do not get deformed either.
Notice that  this  is  completely different  from  the case  of the 
Nielsen  identities \cite{Nielsen:1975fs}, describing the   gauge  
fixing  parameter dependence of the  bare 
Green's functions (we do not use them here).  In this latter case, one needs to
extend the BRST symmetry to  include the variation of the gauge fixing
parameter. This, in turn,  will spoil  the original BRST invariance  of the
theory,  implying   that  the   latter  identities  get   deformed  by
renormalization already at the one-loop level~\cite{Binosi:2005yk}.

For concreteness let us assume that we start the PT procedure  
in the renormalizable (linear) Feynman gauge (RFG),
as we have done throughout this article.
Let us denote by  
$Z_A$ the gluon wave-function renormalization, 
by $Z_3$ the vertex renormalization constant for the 
three-gluon vertex $\Gamma_{\alpha\mu\nu}$, 
by ${\bar Z}_2$ the usual ghost wave-function renormalization,
and by ${\bar Z}_1$ the ghost-gluon vertex renormalization 
constant. 
Notice also that, the BRST symmetry demands that
$Z_3/{Z_A}  = \bar{Z}_1/\bar{Z}_2$.
Then, the fundamental STIs employed when carrying out the PT  
survives renormalization, simply because all 
counterterms necessary to render it finite are already furnished 
by the usual counterterms of the RFG Lagrangian.
This is, of course, a direct result of the basic assumption 
the the theory in the RFG is renormalizable:  
once all counterterms have been supplied in the RFG, the STI
which is studied in the same gauge, will continue being valid.

The actual implementation of the renormalization procedure 
proceeds along the lines described in~\cite{Papavassiliou:1999az} for the two-loop 
case. One should start out with the counterterms
that  are  necessary  to  renormalize individually  the  conventional
Green's functions in the RFG.
Then, one should show  that, by simply rearranging these counterterms
following the PT rules, one can renormalize the PT Green's
functions.  
Notice also that, due to the validity of the Abelian WIs, 
the renormalization constants 
before and after the PT rearrangements 
are related to the 
gauge coupling renormalization as follows:  
\bea
Z^2_g &=& Z_1^2 Z_{2}^{-2} Z_A^{-1} 
=   \widehat{Z}_1^{2} \widehat{Z}_2^{-2}\widehat{Z}_A^{-1}
= \widehat{Z}_A^{-1}.
\label{R3}
\eea

After rearranging the original RFG counterterms  in such a way as to render the PT Green's functions finite, one  
should be able to verify 
that the resulting counterterms  are  in fact identical to those
obtained  when  carrying out   the BFM 
renormalization  program as explained by Abbott~\cite{Abbott:1980hw}, 
i.e. by  renormalizing  only  the  background
gluons,  the   coupling constant $g$, and the
quantum gauge-fixing parameter $\xi_Q$.
Thus, the relevant renormalization constants are given by
\be 
g_0 = Z_g g, \qquad
\widetilde{A}_0 = Z_{\widetilde{A}}^{1/2} \widetilde{A}, \qquad
\xi_Q^{0} = Z_{\xi_Q} \xi_Q, \qquad Z_{\xi_Q}=Z_A.
\ee 
The renormalization of $\xi_Q$ is necessary 
due to the fact that 
the longitudinal part of the quantum gluon propagator is not 
renormalized. As pointed out 
by Abbott, in the context of the BFM
this step may be avoided if the calculation is carried 
out with an arbitrary $\xi_Q$ rather than the BFG $\xi_Q=1$. 
Of course, as we have seen,
the PT brings us effectively to $\xi_Q=1$; 
thus, when interpreting the resulting counterterm 
from the BFM point of view, one should 
keep in mind that   
gauge-fixing parameter renormalization is necessary. 
The renormalization of $\xi_Q$ 
not only affects the propagator-lines, but also the 
longitudinal parts of the external vertices; it renormalizes
precisely the $\Gamma^{\rm P}$ part, as can be seen from 
the corresponding BFM Feynman rule 
for the three-gluon vertex (see Appendix~\ref{Frules}). 

All the above ingredients must be combined appropriately in order to 
demonstrate the renormalizability of the new SDE; we shall not pursue this 
point any further. 

\section{Faddeev-Popov Equations\label{Appendix:FPEs}}

\renewcommand{\theequation}{C.\arabic{equation}}

As a first example of the use of the FPE introduced in Section~\ref{FPEs}, let us 
differentiate the functional equation (\ref{FPeqRxi}) with respect to the ghost field $c^b$; after setting the fields/anti-fields to zero we get (relabeling the color and Lorentz indices)
\begin{equation}
\Gamma_{c^m\bar c^n}(q)+iq^\nu\Gamma_{c^mA^{*n}_\nu}(q)=0,
\label{FPE:ghprop}
\end{equation}
which can be used to relate the auxiliary function $\Gamma_{c^mA^{*n}_\nu}(q)$ 
with the full ghost propagator $D^{ab}(q)$. Due to Lorentz invariance, we can in fact write $\Gamma_{c^mA^{*n}_\nu}(q)=q_\nu\Gamma_{c^mA^{*n}}(q)$, and therefore
\begin{equation}
\Gamma_{c^m\bar c^n}(q)=-iq^\nu\Gamma_{c^mA^{*n}_\nu}(q)=-iq^2\Gamma_{c^mA^{*n}}(q).
\end{equation}
On the other hand, due to our definition of the Green's functions [see Eq.~(\ref{greenfunc})], one has that 
\begin{equation}
iD^{mr}(q)\Gamma_{c^r\bar c^n}(q)=\delta^{mn}, 
\label{twop_inverse}
\end{equation}
and therefore we get the announced relation:
\begin{eqnarray}
\Gamma_{c^mA^{*n}_\nu}(q)&=&q_\nu\Gamma_{c^mA^{*n}}(q)\nonumber \\
&=&q_\nu[q^2 D^{mn}(q)]^{-1}.
\label{gaprop}
\end{eqnarray}

As a second example, let us differentiate  Eq.~(\ref{FPeqRxi}) twice, 
once with respect to $A^n_\nu$ and once with respect to $c^r$,
and then set the fields/anti-fields to zero; in this way we get the identity
\begin{equation}
\Gamma_{c^rA^n_\nu\bar c^m}(k,q)+iq^\mu\Gamma_{c^rA^n_\nu A^{*m}_\mu}(k,q)=0,
\label{FP:gcc}
\end{equation}
which is particularly useful for the PT construction. 
Notice that this identity is the equivalent of the one introduced 
in Section~\ref{convform} relating the conventional $H$ function with the trilinear gluon-ghost vertex.  
All these identities can be easily checked at tree-level; for example, using the Feynman rules presented in Appendix~\ref{Frules}, we have
\begin{equation}
iq^\mu\Gamma^{(0)}_{c^rA^n_\nu A^{*m}_\mu}(k,q)=igf^{mnr}q_\nu=-\Gamma^{(0)}_{c^rA^n_\nu\bar c^m}(k,q).
\end{equation}

Differentiation of the functional (\ref{FPeqBFM}) with respect to a BFM source $\Omega$ 
and a quantum gluon field $A$ or a ghost field $c$ and a background gluon $\widehat{A}$, 
provides instead the identities ($k_1+k+q=0$)
\begin{eqnarray}
\Gamma_{\Omega^r_\rho A^n_\nu\bar c^m}(k,q)+iq^\mu\Gamma_{\Omega^r_\rho A^n_\nu A^{*m}_\mu}(k,q)&=&
gf^{mnr}g_{\nu\rho}, \label{FPE_Om1} \\
\Gamma_{c^r \widehat{A}^n_\nu\bar c^m}(k,q)+iq^\mu\Gamma_{c^r\widehat{A}^n_\nu A^{*m}_\mu}(k,q)&=&
-igf^{mne}\Gamma_{c^rA^{*e}_\nu}(-k_1), \label{FPE_Om2}
\end{eqnarray}
that can be easily checked at tree-level.

\section{Slavnov-Taylor Identities\label{Appendix:STIs}}

STIs are obtained by functional differentiation of the STI functional of Eq.~(\ref{STIfunc}) with respect to suitable combinations of fields chosen following the rules discussed in Section~\ref{BV}.

\renewcommand{\theequation}{D.\arabic{equation}}

\subsection{STIs for quark proper vertices}

We begin by deriving the  STI satisfied by the trilinear quark-gluon
vertex (see, {\it e.g.},~\cite{Marciano:1977su,Pascual:1984zb}). 
From our general discussion of Section~\ref{BV}, for obtaining
this identity we need to consider the functional differentiation
\begin{equation}
\left.\frac{\delta^3{\cal S}(\Gamma)}{\delta c^a(q)\delta\psi(p_2)\delta\bar\psi(-p_1)}\right|_{\Phi,\Phi^*=0}=0 \qquad q+p_2=p_1,
\end{equation}
from which we obtain the equation 
\be
\Gamma_{c^aA^{*\gamma}_d}(-q)\Gamma_{A^d_\gamma\psi\bar\psi}(p_2,-p_1) +
\Gamma_{\psi^*\bar\psi c^m}(-p_1,q)\Gamma_{\psi\bar\psi}(p_2) 
+ \Gamma_{\psi\bar\psi}(p_1)\Gamma_{\psi c^m\bar\psi^*}(q,-p_1)=0,
\ee
where the two-point function $\Gamma_{\psi\bar\psi}(p)$ is defined through the identity
\begin{equation}
iS(p)\Gamma_{\psi\bar\psi}(p)=\mathbb{I}.
\label{2pff}
\end{equation}
Using then the relation of Eq.~(\ref{gaprop}), we get the STI in its final form, namely
\begin{equation}
q^\alpha\Gamma_{A^a_\alpha\psi\bar\psi}(p_2,-p_1)=[q^2D^{aa'}(q)]\left\{\Gamma_{\psi^*\bar\psi c^{a'}}(-p_1,q)\Gamma_{\psi\bar\psi}(p_2)+\Gamma_{\psi\bar\psi}(p_1)\Gamma_{\psi c^{a'}\bar\psi^*}(q,-p_1)\right\}.
\label{STI:gff}
\end{equation}
The three-point  auxiliary functions  appearing in the  equation above
can  be  constructed using  the  Feynman  rules  reported in  Appendix
\ref{Frules}.  At  tree-level  the  above identity  can  be  trivially
checked, with  the rhs  being proportional to  two inverse
quark propagators
\begin{eqnarray}
q^\alpha\Gamma^{(0)}_{A^a_\alpha\psi\bar\psi}(p_2,-p_1)&=&
[q^2D^{aa'}(q)]^{(0)}\left\{\Gamma^{(0)}_{\psi^*\bar\psi c^{a'}}(-p_1,q)\Gamma^{(0)}_{\psi\bar\psi}(p_2)+\Gamma^{(0)}_{\psi\bar\psi}(p_1)\Gamma^{(0)}_{\psi c^{a'}\bar\psi^*}(q,-p_1)\right\}\nonumber\\
&=&gt^a\left[(\psm_1-m)-(\psm_2-m)\right].
\end{eqnarray}
Contrary to the case of QED, where this generalizes directly to all orders, 
in the QCD case it is valid only to lowest order, due to the non-linearity of Eq.~(\ref{STI:gff}).

The  STI satisfied  by the  quadrilinear quark-gluon vertex (induced beyond tree-level) can be
derived by considering the following functional differentiation
\begin{equation}
\left.\frac{\delta^4{\cal S}(\Gamma)}{\delta c^m(k_1)\delta A^n_\nu(k_2)\delta\psi(p_2)\delta\bar\psi(-p_1)}\right|_{\Phi,\Phi^*=0}=0 \qquad k_1+k_2+p_2=p_1.
\end{equation}
Carrying out the functional differentiation and using again Eq.~(\ref{gaprop}), we get
\begin{eqnarray}
& & k_1^\mu\Gamma_{A_\mu^mA_\nu^n\psi\bar\psi}(k_2,p_2,-p_1)=[k_1^2D^{mm'}(k_1)]\bigg\{
\Gamma_{\psi^*\bar\psi c^{m'}}(-p_1,k_1)\Gamma_{A_\nu^n\psi\bar\psi}(p_2,-p_2-k_2) \nonumber\\
&&\hspace{.3cm}+\Gamma_{A_\nu^n\psi\bar\psi}(p_1-k_2,-p_1)\Gamma_{\psi c^{m'}\bar\psi^*}(k_1,-p_2-k_1)+\Gamma_{c^{m'} A^n_\nu A^{*\gamma}_d}(k_2,-k_1-k_2)\Gamma_{A_\gamma^d\psi\bar\psi}(p_2,-p_1) \nonumber \\
&&\hspace{.3cm}+\Gamma_{A_\nu^n\psi^*\bar\psi c^{m'}}(p_2,-p_1,k_1)\Gamma_{\psi\bar\psi}(p_2) + \Gamma_{\psi\bar\psi}(p_1) 
\Gamma_{A_\nu^n\psi c^{m'}\bar\psi^*}(p_2,k_1,p_1)\nonumber \\
&&\hspace{.3cm}+\Gamma_{c^{m'} A_d^{*\gamma}\psi\bar\psi}(k_2,p_2,-p_1)\Gamma_{A^d_\gamma A^n_\nu}(k_2)\bigg\}.
\label{STI:ggff}
\end{eqnarray}
Clearly this identity starts at the one-loop level (recall that $\Gamma$s represent 1PI functions). 

\subsection{STIs for gluon proper vertices}

Let us start by deriving the well-known STI for the trilinear 
gluon vertex~\cite{Marciano:1977su,Pascual:1984zb,Ball:1980ax}. 
By considering the functional differentiation
\begin{equation}
\left.\frac{\delta^3{\cal S}(\Gamma)}{\delta c^a(q)\delta A^m_\mu(k_1)\delta A^n_\nu(k_2)}\right|_{\Phi,\Phi^*=0}=0 \qquad q+k_1+k_2=0,
\label{3g_diff}
\end{equation}
and using Eq.~(\ref{gaprop}) one obtains
\begin{equation}
q^\alpha\Gamma_{A^a_\alpha  A^m_\mu A^n_\nu}(k_1,k_2)=[q^2D^{aa'}(q)]\left\{\Gamma_{c^{a'} A^n_\nu A^{*\gamma}_d}(k_2,k_1)\Gamma_{A^d_\gamma A^m_\mu}(k_1)+\Gamma_{c^{a'} A^m_\mu A^{*\gamma}_d}(k_1,k_2)\Gamma_{A^d_\gamma A^n_\nu}(k_2)\right\}.
\label{STI:ggg}
\end{equation}
Notice that since we are working with the minimal action (see subsection~\ref{BV}), one has
\begin{equation}
\Gamma^{(0)}_{A^a_\alpha A^b_\beta}(q)=iq^2\delta^{ab}P_{\alpha\beta}(q),
\label{ga_tree_lev}
\end{equation}
and therefore
\bea
\Gamma_{A^a_\alpha A^b_\beta}(q) &=& (\Delta^{-1})^{ab}_{\alpha\beta}(q)-i\delta^{ab}q_\alpha q_\beta
\nonumber \\ 
&=&
i\delta^{ab} P_{\alpha\beta}(q) \Delta^{-1}(q^2).
\label{gainvprop}
\eea
As an overall consistency check of our definitions and conventions, 
notice that  Eq.s~(\ref{inv_prop_pi}) and (\ref{gainvprop}) 
imply that $-\Gamma_{A^a_\alpha A^b_\beta}(q)=\delta^{ab}\Pi_{\alpha\beta}(q)$, in full agreement with Eq.~(\ref{greenfunc}).
Using the above relation, we can now check the identity at tree-level; we get
\begin{eqnarray}
q^\alpha\Gamma^{(0)}_{A^a_\alpha  A^m_\mu A^n_\nu}(k_1,k_2)&=&[q^2D^{aa'}(q)]^{(0)}\left\{\Gamma^{(0)}_{c^{a'} A^n_\nu A^{*\gamma}_d}(k_2,k_1)\Gamma^{(0)}_{A^d_\gamma A^m_\mu}(k_1)\right.\nonumber \\
&+&\left.\Gamma^{(0)}_{c^{a'} A^m_\mu A^{*\gamma}_d}(k_1,k_2)\Gamma^{(0)}_{A^d_\gamma A^n_\nu}(k_2)\right\}\nonumber \\
&=&igf^{amn}\left[(g_{\mu\nu}k_1^2-k_{1\mu}k_{1\nu})-(g_{\mu\nu}k_2^2-k_{2\mu}k_{2\nu})\right].
\end{eqnarray}
Notice also that Eq.~(\ref{gainvprop}) allows us to compare the STI of Eq.~(\ref{STI:ggg}) with that 
of Eq.~(\ref{sti3gv}), which is written in the conventional formalism; in this way we 
get the identity (factoring out the color structure) 
\be
H_{\mu\gamma}(k_1,k_2)=\Gamma_{cA_\mu A^*_\gamma}(k_1,k_2).
\ee

We pause here to show what would have happened had we worked with the complete generating functional. In this case, due to the extra term appearing in the master equation (\ref{STIfunc_nm}) satisfied by the complete action, the differentiation carried out in Eq.~(\ref{3g_diff}) would generate two more terms with respect to the ones already appearing in Eq.~(\ref{STI:ggg}),  namely
\be
\delta^{dn}k_{2\nu}\Gamma_{c^a A^m_\mu\bar c^d}(k_1,k_2)+\delta^{dm}k_{2\mu}\Gamma_{c^a A^n_\nu\bar c^d}(k_2,k_1).
\ee
To get to the terms above we have used the equation of motion of the Nakanishi-Lautrup multiplier $B$ eliminating the latter in favor of the corresponding gauge-fixing function ${\cal F}$. Then, making use of the FPE (\ref{FP:gcc}), we get  
\be
-i\delta^{dn}k_{2\nu}k_{2\gamma}\Gamma_{c^a A^m_\mu A^{*\gamma}_d}(k_1,k_2)-i\delta^{dm}k_{1\mu}k_{1\gamma}\Gamma_{c^a A^n_\nu A^{*\gamma}_d}(k_2,k_1),
\ee
so that we finally would get the STI
\bea
q^\alpha\Gamma_{A^a_\alpha  A^m_\mu A^n_\nu}(k_1,k_2)&=&[q^2D^{aa'}(q)]\left\{\Gamma_{c^{a'} A^n_\nu A^{*\gamma}_d}(k_2,k_1)\left[\Gamma^\mathrm{C}_{A^d_\gamma A^m_\mu}(k_1)-i\delta^{dm}k_{1\mu}k_{1\gamma}\right]\right.\nonumber \\
&+&\left.\Gamma_{c^{a'} A^m_\mu A^{*\gamma}_d}(k_1,k_2)\left[\Gamma^\mathrm{C}_{A^d_\gamma A^n_\nu}(k_2)-i\delta^{dn}k_{2\gamma}k_{2\nu}\right]\right\},
\eea
where we have indicated explicitly that the two-point functions are to be evaluated from the completed functional (for the three point functions appearing in the STI above there is no difference). We then see that the difference amounts to a tree-level piece appearing in the two-point function, as has been anticipated in our general discussion of subsection~\ref{BV} (recall that we are using the Feynman gauge $\xi=1$). In particular notice that we correctly find the relation $\Gamma^\mathrm{C}_{A^a_\alpha A^b_\beta}(q)= (\Delta^{-1})^{ab}_{\alpha\beta}(q)$.

Another STI  that will  be needed  in the PT  construction is  the one
involving the  quadrilinear gluon vertex; carrying  out the functional
differentiation
\begin{equation}
\left.\frac{\delta^4{\cal S}(\Gamma)}{\delta c^m(k_1)\delta A^n_\nu(k_2)\delta A^r_\rho(p_2)\delta A^s_\sigma(-p_1)}\right|_{\Phi,\Phi^*=0}=0 \qquad k_1+k_2+p_2=p_1,
\end{equation}
and using Eq.~(\ref{gaprop}), we arrive at the result
\begin{eqnarray}
& & k_1^\mu\Gamma_{A_\mu^mA_\nu^n A^r_\rho A^s_\sigma}(k_2,p_2,-p_1)=[k_1^2D^{mm'}(k_1)]\bigg\{\Gamma_{c^{m'} A^s_\sigma A^{*\gamma}_d}(-p_1,k_2+p_2)\Gamma_{A^d_\gamma A^n_\nu A^r_\rho}(k_2,p_2)\nonumber \\
&&\hspace{.3cm}+\Gamma_{c^{m'} A^r_\rho A^{*\gamma}_d}(p_2,k_2-p_1)\Gamma_{A^d_\gamma A^n_\nu A^s_\sigma}(k_2,-p_1)+\Gamma_{c^{m'} A^n_\nu A^{*\gamma}_d}(k_2,p_2-p_1)\Gamma_{A^d_\gamma A^r_\rho  A^s_\sigma}(p_2,-p_1)\nonumber\\
&&\hspace{.3cm}+\Gamma_{c^{m'} A^r_\rho A^s_\sigma A^{*\gamma}_d}(p_2,-p_1,k_2)\Gamma_{A^d_\gamma A^n_\nu}(k_2)+\Gamma_{c^{m'} A^n_\nu A^s_\sigma A^{*\gamma}_d}(k_2,-p_1,p_2)\Gamma_{A^d_\gamma A^r_\rho}(p_2)\nonumber \\
&&\hspace{.3cm}+ \Gamma_{c^{m'} A^n_\nu A^r_\rho A^{*\gamma}_d}(k_2,p_2,-p_1)\Gamma_{A^d_\gamma A^s_\sigma}(p_1)\bigg\}.
\label{STI:gggg}
\end{eqnarray}
At tree-level, notice that only the first two lines of this identity are different from zero; 
then, using the Jacobi identity, we obtain 
\begin{eqnarray}
k_1^\mu\Gamma^{(0)}_{A_\mu^mA_\nu^n A^r_\rho A^s_\sigma}(k_2,p_2,-p_1)&=&[k_1^2D^{mm'}(k_1)]^{(0)}\bigg\{\Gamma^{(0)}_{c^{m'} A^s_\sigma A^{*\gamma}_d}(-p_1,k_2+p_2)\Gamma^{(0)}_{A^d_\gamma A^n_\nu A^r_\rho}(k_2,p_2)\nonumber \\
&+&\Gamma^{(0)}_{c^{m'} A^r_\rho A^{*\gamma}_d}(p_2,k_2-p_1)\Gamma^{(0)}_{A^d_\gamma A^n_\nu A^s_\sigma}(k_2,-p_1)\nonumber\\
&+&\Gamma^{(0)}_{c^{m'} A^n_\nu A^{*\gamma}_d}(k_2,p_2-p_1)\Gamma^{(0)}_{A^d_\gamma A^r_\rho  A^s_\sigma}(p_2,-p_1)\bigg\}\nonumber\\
&=&-ig^2\bigg\{f^{mse}f^{ern}\left(g_{\nu\sigma}k_{1\rho}-g_{\rho\sigma}k_{1\nu}\right)
+f^{mre}f^{esn}\left(g_{\nu\rho}k_{1\sigma}-g_{\rho\sigma}k_{1\nu}\right)\nonumber \\
&+&f^{mne}f^{esr}\left(g_{\nu\rho}k_{1\sigma}-g_{\nu\sigma}k_{1\rho}\right)\bigg\}.
\end{eqnarray}

\subsection{STIs for mixed quantum/background Green's functions}

Let us consider a Green's function  involving background as  well as
quantum fields. Clearly, when contracting such a function with  the momentum
corresponding to a background leg it will satisfy a linear WI 
[see, {\it e.g.},   Eq.s~(\ref{3gl})  --   (\ref{4gh})], whereas  
 when contracting it with  the momentum corresponding to a 
quantum leg it will satisfy  a non-linear STI.
Let us then study the particularly  interesting case of
the  STI   satisfied  by  the   vertex  $\Gamma_{\widehat{A}AA}$  when
contracted  with  the momentum  of  one  of  the quantum  fields. Taking the functional differentiation
\be
\left.\frac{\delta^3{\cal S'}(\Gamma')}{\delta c^m(k_1)\delta\widehat{A}^a_\alpha(q)\delta A^n_\nu(k_2)}\right|_{\Phi,\Phi^*,\Omega=0}=0 \qquad q+k_1+k_2=0,
\ee
we get
\bea
k_1^\mu\Gamma_{\widehat{A}^a_\alpha A^m_\mu A^n_\nu}(k_1,k_2)&=&[k_1^2D^{mm'}(k_1)]\left\{\Gamma_{c^{m'}A^n_\nu A^{*\epsilon}_e}(k_2,q)\Gamma_{\widehat{A}^a_\alpha A^e_\epsilon}(q)
+\Gamma_{c^{m'}\widehat{A}^a_\alpha A^{*\epsilon}_e}(q,k_2)\Gamma_{A^e_\epsilon A^n_\nu}(k_2)\right\}.\nonumber \\
\label{STI:mixed_wtl}
\eea
Notice that the same result can be achieved by contracting directly the BQI of Eq.~(\ref{BQI:ggg_tlc}) with the momentum of one of the quantum fields and then using the STI of Eq.~(\ref{STI:ggg}) together with the BQIs of Eq.s(\ref{twoBQI1}) and~(\ref{BQI:cgbarc}) to bring the result in the above form.

It is particularly important to correctly identify in the above identity the missing tree-level contributions (due to the use of the reduced functional, see also the discussion in Section~\ref{BQI:3p}). In order to do that, one can either work with the complete functional and use the FPE (\ref{FPE_Om2}), or add them by hand using Eq.~(\ref{BQI:ggg_tlc}), obtaining in either cases the STI 
\bea
k_1^\mu\Gamma_{\widehat{A}^a_\alpha A^m_\mu A^n_\nu}(k_1,k_2)&=&[k_1^2D^{mm'}(k_1)]\left\{\Gamma_{c^{m'}A^n_\nu A^{*\epsilon}_e}(k_2,q)\Gamma_{\widehat{A}^a_\alpha A^e_\epsilon}(q)
+\Gamma_{c^{m'}\widehat{A}^a_\alpha A^{*\epsilon}_e}(q,k_2)\Gamma_{A^e_\epsilon A^n_\nu}(k_2)\right\}\nonumber \\
&-&igf^{amn}(k_1^2g_{\alpha\nu}-k_{1\alpha}k_{2\nu}).
\label{STI:mixed}
\eea

This STI can be further manipulate by using Eq.~(\ref{gainvprop}) and the FPE~(\ref{FPE_Om2}) 
for rewriting the term proportional to $\Gamma_{AA}(k_2)$ as
\bea
\Gamma_{c^{m'}\widehat{A}^a_\alpha A^{*\epsilon}_e}(q,k_2)\Gamma_{A^e_\epsilon A^n_\nu}(k_2)&=&\Gamma_{c^{m'}\widehat{A}^a_\alpha A^{*\epsilon}_e}(q,k_2)(\Delta^{-1})^{en}_{\epsilon\nu}(k_2)+k_{2\nu}\Gamma_{c^{m'}\widehat{A}^a_\alpha\bar c^n}(q,k_2)\nonumber \\
&+&igf^{aen} k_{2\nu}\Gamma_{c^{m'}A^{*e}_\alpha}(-k_1).
\label{fmassage}
\eea
On the other hand, employing Eq.~(\ref{gaprop}) we find
\be
[k_1^2D^{mm'}(k_1)](igf^{nae}k_{2\nu})\Gamma_{c^{m'}A^{*e}_\alpha}(-k_1)=-igf^{amn}k_{1\alpha}k_{2\nu};
\ee
so, inserting Eq.~(\ref{fmassage}) back into Eq.~(\ref{STI:mixed}) we see 
that the term above partially cancels the tree level contribution, thus leaving us with the STI
\bea
k_1^\mu\Gamma_{\widehat{A}^a_\alpha A^m_\mu A^s_\nu}(k_1,k_2)&=&[k_1^2D^{mm'}(k_1)]\left\{\Gamma_{c^{m'}A^n_\nu A^{*\epsilon}_e}(k_2,q)\Gamma_{\widehat{A}^a_\alpha A^e_\epsilon}(q)
+\Gamma_{c^{m'}\widehat{A}^a_\alpha A^{*\epsilon}_e}(q,k_2)(\Delta^{-1})^{en}_{\epsilon\nu}(k_2)\right.\nonumber \\
&+&\left.k_{2\nu}\Gamma_{c^{m'}\widehat{A}^a_\alpha\bar c^n}(q,k_2)
\right\}-igf^{amn}k_1^2g_{\alpha\nu}.
\label{STI:mixed1}
\eea

\subsection{STIs for the quark SD kernel}

In addition to the STIs for 1PI (proper) vertices,  the PT construction for
SDEs requires  the  additional knowledge of the result of  the  action of  longitudinal
momenta  on   connected  kernels.  The  first one of   these  kernels  is
encountered  in  the  construction  of  the  PT  gluon-quark-quark
vertex, and can be written as follows (see Fig.~\ref{fig:ggff_SDker})
\begin{eqnarray}
{\cal K}_{A^m_\mu A^n_\nu \psi\bar\psi}(k_2,p_2,-p_1)&=&
\Gamma_{A^m_\mu A^n_\nu \psi\bar\psi}(k_2,p_2,-p_1)\nonumber\\
&+&i\Gamma_{A^m_\mu \psi\bar\psi}(\ell,-p_1)iS(\ell)i\Gamma_{A^n_\nu \psi\bar\psi}(p_2,-\ell)\nonumber \\
&+&i\Gamma_{A^n_\nu \psi\bar\psi}(\ell',-p_1)iS(\ell')i\Gamma_{A^m_\mu \psi\bar\psi}(p_2,-\ell').
\end{eqnarray}
where $\ell=k_2+p_2=p_1-k_1$ and $\ell'=k_1+p_2=p_1-k_2$. Then, using the STI of Eq.~(\ref{STI:gff}) and the relation
(\ref{2pff}), one gets the results
\begin{eqnarray}
k_1^\mu i\Gamma_{A^m_\mu \psi\bar\psi}(\ell,-p_1)iS(\ell)i\Gamma_{A^n_\nu \psi\bar\psi}(p_2,-\ell)&=&-[k_1^2D^{mm'}(k_1)] \left\{\Gamma_{\psi^*\bar\psi c^{m'}}(-p_1,k_1)\right.\nonumber \\
&+&\left.
i\Gamma_{\psi\bar\psi}(p_1)\Gamma_{\psi c^{m'}\bar\psi^*}(k_1,-p_1)S(\ell)\right\}\Gamma_{A^n_\nu\psi\bar\psi}(p_2,-\ell),\label{1PR1}\nonumber \\
\\
k_1^\mu i\Gamma_{A^n_\nu \psi\bar\psi}(\ell',-p_1)iS(\ell')i\Gamma_{A^m_\mu \psi\bar\psi}(p_2,-\ell')&=&-\Gamma_{A^n_\nu \psi\bar\psi}(\ell',-p_1)
[k_1^2D^{mm'}(k_1)]\left\{\Gamma_{\psi c^{m'}\bar\psi^*}(k_1,-\ell')\right.\nonumber\\
&+&\left.iS(\ell')\Gamma_{\psi^*\bar\psi c^{m'}}(-\ell',k_1)\Gamma_{\psi\bar\psi}(p_2)
\right\}. \label{1PR2}
\end{eqnarray}
We then see that the first term in Eq.s~(\ref{1PR1}) and (\ref{1PR2}) will cancel 
the first two terms of the STI of the 1PI vertex of Eq.~(\ref{STI:ggff}), and we finally arrive at the STI
\begin{eqnarray}
k_1^\mu{\cal K}_{A^m_\mu A^n_\nu \psi\bar\psi}(k_2,p_2,-p_1)&=&[k_1^2D^{mm'}(k_1)]\left\{\Gamma_{c^{m'} A^n_\nu A^{*\gamma}_d}(k_2,-k_1-k_2)\Gamma_{A_\gamma^d\psi\bar\psi}(p_2,-p_1)\right.\nonumber \\
&+&\Gamma_{\psi\bar\psi}(p_1){\cal K}_{A^n_\nu\psi c^{m'}\bar\psi^*}(p_2,k_1,-p_1)+{\cal K}_{A^n_\nu\psi^*\bar\psi c^{m'}}(p_2,-p_1,k_1)\Gamma_{\psi\bar\psi}(p_2)
\nonumber \\
&+&\left.\Gamma_{c^{m'} A_d^{*\gamma}\psi\bar\psi}(k_2,p_2,-p_1)\Gamma_{A^d_\gamma A^n_\nu}(k_2)\right\},
\label{STISDggff}
\end{eqnarray}
where we have defined the auxiliary kernels
\begin{eqnarray}
{\cal K}_{A^n_\nu\psi c^{m'}\bar\psi^*}(p_2,k_1,-p_1)&=&\Gamma_{A^n_\nu\psi c^{m'}\bar\psi^*}(p_2,k_1,-p_1)\nonumber \\
&+&i\Gamma_{\psi c^{m'}\bar\psi^*}(k_1,-p_1)iS(\ell)i\Gamma_{A^n_\nu\psi \bar\psi}(p_2,-\ell),
\label{1PRker:ggff1}\\
{\cal K}_{A^n_\nu\psi^*\bar\psi c^{m'}}(p_2,-p_1,k_1)&=&\Gamma_{A^n_\nu\psi^*\bar\psi c^{m'}}(p_2,-p_1,k_1)\nonumber \\
&+& i\Gamma_{A^n_\nu\psi\bar\psi}(\ell,-p_1)iS(\ell)i\Gamma_{\psi^*\bar\psi c^{m'}}(-\ell,k_1).
 \label{1PRker:ggff2}
\end{eqnarray}

\subsection{STIs for the gluon SD kernel}

In the construction of the SDEs for the gluon self-energy and three-gluon vertex, one 
needs the knowledge of the STI satisfied by the kernel (see Fig.~\ref{fig:gggg_SDker})
\begin{eqnarray}
{\cal K}_{A^m_\mu A^n_\nu A^r_\rho A^s_\sigma}(k_2,p_2,-p_1)&=&
\Gamma_{A^m_\mu A^n_\nu A^r_\rho A^s_\sigma}(k_2,p_2,-p_1)\nonumber\\
&+&i\Gamma_{A^s_\sigma A^m_\mu A^e_\epsilon}(k_1,\ell)i\Delta_{ee'}^{\epsilon\epsilon'}(\ell)i\Gamma_{A^{e'}_{\epsilon'}A^n_\nu A^r_\rho}(k_2,p_2)\nonumber \\
&+&i\Gamma_{A^s_\sigma A^n_\nu A^e_\epsilon}(k_2,\ell')i\Delta_{ee'}^{\epsilon\epsilon'}(\ell')i\Gamma_{A^{e'}_{\epsilon'}A^m_\mu A^r_\rho}(k_1,p_2).
\end{eqnarray}

Using the above relation, together with STI of Eq.~(\ref{STI:ggg}), we find the following result
\begin{eqnarray}
&&k_1^\mu i\Gamma_{A^s_\sigma A^m_\mu A^e_\epsilon}(k_1,\ell)i\Delta_{ee'}^{\epsilon\epsilon'}(\ell)i\Gamma_{A^{e'}_{\epsilon'}A^n_\nu A^r_\rho}(k_2,p_2)=-[k_1^2D^{mm'}(k_1)]\times\nonumber \\
&&\hspace{1cm}\times\Bigg\{\Gamma_{c^{m'} A^s_\sigma A^{*e'}_\epsilon}(-p_1,\ell)P^{\epsilon\epsilon'}(\ell)+i\Gamma_{c^{m'} A^e_\epsilon A^{*\gamma}_d}(\ell,-p_1)\Gamma_{A^d_\gamma A^s_\sigma}(p_1)\Delta_{ee'}^{\epsilon\epsilon'}(\ell)
\Bigg\}\Gamma_{A^{e'}_{\epsilon'} A^n_\nu A^r_\rho}(k_2,p_2).\nonumber \\
\end{eqnarray}
In this case this is, however, not the end of the story, since the 
first term in the equation above still contains (virtual) longitudinal momenta, 
which will trigger the STI of Eq.~(\ref{STI:ggg}) together with the FPE~(\ref{FP:gcc}). 
After taking this into account, we obtain
\begin{eqnarray}
&&k_1^\mu i\Gamma_{A^s_\sigma A^m_\mu A^e_\epsilon}(k_1,\ell)i\Delta_{ee'}^{\epsilon\epsilon'}(\ell)i\Gamma_{A^{e'}_{\epsilon'}A^n_\nu A^r_\rho}(k_2,p_2)=-[k_1^2D^{mm'}(k_1)]\times\nonumber \\
&&\hspace{.5cm}\times\bigg\{
\left[\Gamma_{c^{m'} A^s_\sigma A^{*\epsilon'}_{e'}}(-p_1,\ell)+
i\Gamma_{c^{m'} A^e_{\epsilon}A^{*\gamma}_d}(\ell,-p_1)\Gamma_{A^d_\gamma A^s_\sigma}(p_1)\Delta_{ee'}^{\epsilon\epsilon'}(\ell)
\right]\Gamma_{A^{e'}_{\epsilon'} A^n_\nu A^r_\rho}(k_2,p_2)\nonumber \\
&&\hspace{.5cm}+i\Gamma_{c^{m'} A^s_\sigma \bar c^e}(-p_1,\ell)D^{ee'}(\ell)\left[\Gamma_{c^{e'}A^r_\rho A^{*\gamma}_d}(p_2,k_2)\Gamma_{A^d_\gamma A^n_\nu}(k_2)+\Gamma_{c^{e'}A^n_\nu A^{*\gamma}_d}(k_2,p_2)\Gamma_{A^d_\gamma A^r_\rho}(p_2)\right]\bigg\}.\nonumber \\
\end{eqnarray}
Similarly we find
\begin{eqnarray}
&&k_1^\mu i\Gamma_{A^s_\sigma A^n_\nu A^e_\epsilon}(k_2,\ell')i\Delta_{ee'}^{\epsilon\epsilon'}(\ell')i\Gamma_{A^{e'}_{\epsilon'}A^m_\mu A^r_\rho}(k_1,p_2)=-[k_1^2D^{mm'}(k_1)]\times\nonumber \\
&&\hspace{.5cm}\times\bigg\{\Gamma_{A^s_\sigma  A^n_\nu A^{e}_{\epsilon}}(k_2,\ell')
\left[\Gamma_{c^{m'} A^r_\rho A^{*\epsilon}_e}(p_2,-\ell')+i\Delta_{ee'}^{\epsilon\epsilon'}(\ell')\Gamma_{c^{m'}A^{e'}_{\epsilon'}A^{*\gamma}_d}(-\ell',p_2)\Gamma_{A^d_\gamma A^r_\rho}(p_2)\right]\nonumber \\
&&\hspace{.5cm}+iD^{ee'}(\ell')\left[
\Gamma_{c^{e}A^n_\nu A^{*\gamma}_d}(k_2,-p_1)
\Gamma_{A^d_\gamma A^s_\sigma}(p_1)+\Gamma_{c^{e}A^s_\sigma A^{*\gamma}_d}(-p_1,k_2)\Gamma_{A^d_\gamma A^n_\nu}(k_2)\right]\times\nonumber\\
&&\hspace{.5cm}\times\Gamma_{c^{m'} A^r_\rho \bar c^{e'}}(p_2,-\ell')\bigg\}.
\end{eqnarray}

As before, after combining these results with the four-gluon 1PI vertex STI of  Eq.~(\ref{STI:gggg}) 
we arrive  at the needed STI for the four-gluon SD kernel, namely
\begin{eqnarray}
k_1^\mu{\cal K}_{A^m_\mu A^n_\nu A^r_\rho A^s_\sigma}(k_2,p_2,-p_1)&=&[k_1^2D^{mm'}(k_1)]\bigg\{\Gamma_{c^{m'} A^n_\nu A^{*\gamma}_d}(k_2,-k_1-k_2)\Gamma_{A^d_\gamma A^r_\rho  A^s_\sigma}(p_2,-p_1)\nonumber \\
&+&{\cal K}_{c^{m'} A^n_\nu A^s_\sigma A^{*\gamma}_d}(k_2,-p_1,p_2)\Gamma_{A^d_\gamma A^r_\rho}(p_2)\nonumber \\
&+&{\cal K}_{c^{m'}A^n_\nu A^r_\rho  A^{*\gamma}_d}(k_2,p_2,-p_1)\Gamma_{A^d_\gamma A^s_\sigma}(p_1)\nonumber \\
&+&{\cal K}_{ c^{m'}A^r_\rho A^s_\sigma A^{*\gamma}_d}(p_2,-p_1,k_2)\Gamma_{A^d_\gamma A^n_\nu}(k_2)\bigg\},
\label{STISDgggg}
\end{eqnarray}
where the following auxiliary kernels have been defined
\begin{eqnarray}
{\cal K}_{c^{m'} A^n_\nu A^s_\sigma A^{*\gamma}_d}(k_2,-p_1,p_2)&=&
\Gamma_{c^{m'} A^n_\nu A^s_\sigma A^{*\gamma}_d}(k_2,-p_1,p_2)\nonumber\\
&+&i\Gamma_{A^s_\sigma A^n_\nu A^{e}_\epsilon}(k_2,\ell')i\Delta^{\epsilon\epsilon'}_{ee'}(\ell')i\Gamma_{c^{m'} A^{e'}_{\epsilon'} A^{*\gamma}_d}(-\ell',p_2)\nonumber \\
&+&i\Gamma_{c^{m'} A^s_\sigma\bar c^e}(-p_1,\ell)iD^{ee'}(\ell)i\Gamma_{c^{e'}A^n_\nu A^{*\gamma}_d}(k_2,p_2),
\label{1PRker:gggg1} \\
{\cal K}_{c^{m'}A^n_\nu A^r_\rho  A^{*\gamma}_d}(k_2,p_2,-p_1)&=&
\Gamma_{c^{m'}A^n_\nu A^r_\rho  A^{*\gamma}_d}(k_2,p_2,-p_1)\nonumber \\
&+&i\Gamma_{c^{m'} A^{e}_{\epsilon}A^{*\gamma}_d}(\ell,-p_1)i\Delta^{\epsilon\epsilon'}_{ee'}(\ell)i\Gamma_{A^{e'}_{\epsilon'} A^r_\rho A^n_\nu}(k_2,p_2)\nonumber \\
&+&i\Gamma_{c^e A^n_\nu A^{*\gamma}_d}(k_2,-p_1)iD^{ee'}(\ell')i\Gamma_{c^{m'}A^r_\rho\bar c^{e'}}(p_2,-\ell'),
\label{1PRker:gggg2} \\
{\cal K}_{ c^{m'}A^r_\rho A^s_\sigma A^{*\gamma}_d}(p_2,-p_1,k_2)&=&\Gamma_{ c^{m'}A^r_\rho A^s_\sigma A^{*\gamma}_d}(p_2,-p_1,k_2)\nonumber \\
&+&i\Gamma_{c^{m'}A^s_\sigma\bar c^{e'}}(-p_1,\ell)iD^{ee'}(\ell)i\Gamma_{c^{e'}A^r_\rho A^{*\gamma}_d}(p_2,k_2)\nonumber \\
&+&i\Gamma_{c^eA^s_\sigma A^{*\gamma}_d}(-p_1,k_2)iD^{ee'}(\ell')i\Gamma_{c^{m'}A^r_\rho \bar c^{e'}}(p_2,-\ell').\label{1PRker:gggg3} 
\end{eqnarray}

\section{Background-Quantum Identities\label{Appendix:BQIs}}

BQIs are obtained by functional differentiation of the STI functional of Eq.~(\ref{STIfunc_BFM}) with respect to combinations of background fields, quantum fields and background sources.

\renewcommand{\theequation}{E.\arabic{equation}}

\subsection{BQIs for two-point functions}

The  first BQI  we can  construct is  the one  relating  the 
conventional with the BFM gluon self-energies. 
To this
end,   consider   the   following  functional   differentiation   
\bea
\left.\frac{\delta^2{\cal               S'}\left(\Gamma'\right)}{\delta
\Omega_\alpha^a(p)\delta A_\beta^b(q)}\right|_{\Phi,\Phi^*,\Omega=0}=0
&\qquad&       q+p=0,       \nonumber\\      \left.\frac{\delta^2{\cal
S'}\left(\Gamma'\right)}{\delta
\Omega_\alpha^a(p)\delta\widehat{A}_\beta^b(q)}\right|_{\Phi,\Phi^*,\Omega=0}=0
&\qquad&   q+p=0,   \eea   which   will  give   the   relations   \bea
i\Gamma_{\widehat{A}_\alpha^a    A_\beta^b}(q)&=&\left[ig_\alpha^\gamma
\delta^{ad}+                                    \Gamma_{\Omega_\alpha^a
A^{*\gamma}_d}(q)\right]\Gamma_{A^d_\gamma A^b_\beta}(q),
\label{twoBQI1}\\
i\Gamma_{\widehat{A}_\alpha^a\widehat{A}_\beta^b}(q)&=&\left[i g_\alpha^\gamma
\delta^{ad}+ 
\Gamma_{\Omega_\alpha^a A^{*\gamma}_d}(q)\right]\Gamma_{A^d_\gamma
\widehat A^b_\beta}(q).
\label{twoBQI2}
\eea

\begin{figure}[!t]
\includegraphics[width=15cm]{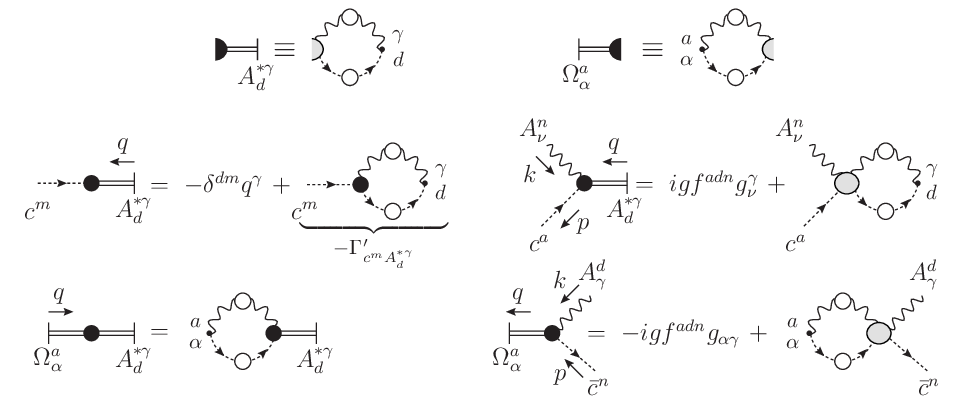}
\caption{Expansions of the gluon anti-field and BFM source in terms of
the corresponding  composite operators. Notice that  if the anti-field
or the BFM sources are attached to  a 1PI vertex, as shown in the first
line, such an expansion will in  general convert the 1PI vertex into a
(connected) SD kernel. The equivalence shown is therefore not valid at
tree-level ({\it e.g.},  in the case of three-point  functions such an
equivalence would imply that the  kernels shown on the rhs
of the  corresponding expansions would be  disconnected); when present,
the tree-level needs  to be added by hand, as  explicitly shown in the
two expansions of the second line  and the last one of the third line.
This type of expansion allows one to express the terms appearing in
the BQIs  in a  form that reveals kernels  appearing in  the STIs
[see,   {\it   e.g.},   Eq.s   (\ref{Aux-gff:1}),   (\ref{Aux-gff:2}),
(\ref{Aux-ggg:1}) and (\ref{Aux-ggg:2})]}
\label{fig:Composite_ope}
\end{figure}

We can now combine Eq.s~(\ref{twoBQI1}) and~(\ref{twoBQI2}) such that
the two-point function mixing background and quantum fields drops out, to get the BQI
\bea
i\Gamma_{\widehat{A}_\alpha^a\widehat{A}_\beta^b}(q)&=&i\Gamma_{A^a_\alpha A^b_\beta}(q)+\Gamma_{\Omega_\alpha^a A^{*\gamma}_d}(q)\Gamma_{A^d_\gamma A^b_\beta}(q)+\Gamma_{\Omega_\beta^b A^{*\gamma}_d}(q)\Gamma_{A^a_\alpha A^d_\gamma}(q)\nonumber \\
&-&i\Gamma_{\Omega_\alpha^a A^{*\gamma}_d}(q)\Gamma_{A^d_\gamma A^{e}_{\epsilon}}(q)\Gamma_{\Omega_\beta^b A^{*\epsilon}_{e}}(q)\nonumber \\
&=&i\Gamma_{A^a_\alpha A^b_\beta}(q)+2\Gamma_{\Omega_\alpha^a A^{*\gamma}_d}(q)\Gamma_{A^d_\gamma A^b_\beta}(q)-i\Gamma_{\Omega_\alpha^a A^{*\gamma}_d}(q)\Gamma_{A^d_\gamma A^{e}_{\epsilon}}(q)\Gamma_{\Omega_\beta^b A^{*\epsilon}_{e}}(q),
\label{BQI:gg}
\eea
where the last identity is due to the transversality of the $\Gamma_{AA}$ two-point function.

\begin{figure}[!t]
\includegraphics[width=15cm]{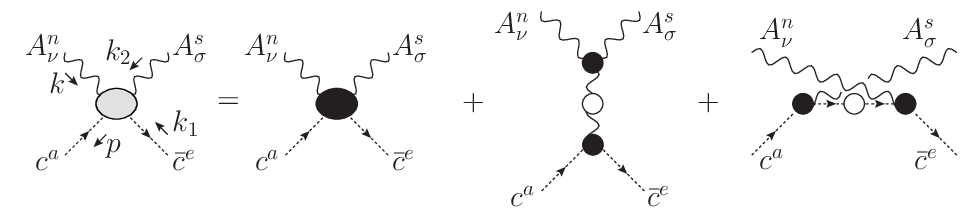}
\caption{Skeleton expansion of the kernel appearing in the SDE for the auxiliary function $\Gamma_{cAA^*}$.}
\label{fig:cggc_SDker}
\end{figure}

In order for our PT procedure to be self-contained,
it is important to express  
the  1PI auxiliary Green's function involved 
in the various STIs  and the BQIs in terms of kernels  
that also appear in  the relevant STIs. 
The key observation that  makes this possible
is  that  one may always  replace  an
anti-field  or  BFM  source  with  its  corresponding  BRST  composite
operator. Thus, for example, one has (see Fig.~\ref{fig:Composite_ope})
\begin{eqnarray}
A^{*\gamma}_d(q)&\to&i\Gamma^{(0)}_{c^{e'} A^{n'}_{\nu'} A^{*\gamma}_d}\int_{k_1}i\Delta_{n'n}^{\nu'\nu}(k_2)iD^{e'e}(k_1), \label{Astrick}\\
\Omega^a_\alpha(q)&\to&i\Gamma^{(0)}_{\Omega^{a}_{\alpha} A^{n'}_{\nu'} \bar c^{e'}}
\int_{k_1}i\Delta_{n'n}^{\nu'\nu}(k_2)iD^{e'e}(k_1), 
\label{Omtrick}
\end{eqnarray}
where $k_1$ and $k_2$ are related through $k_2=q-k_1$.
In this way we get the following SDEs (see again Fig.~\ref{fig:Composite_ope})
\begin{eqnarray}
-\Gamma_{c^mA^{*\gamma}_d}(q)&=&-\delta^{dm}q_\gamma-\Gamma'_{c^mA^{*\gamma}_d}(q)\nonumber \\
&=&-\delta^{dm}q_\gamma+gf^{dn'e'}g^\gamma_{\nu'}\int_{k_1}D^{e'e}(k_1)\Delta_{\nu'\nu}^{nn'}(k_2)\Gamma_{c^mA^{n}_{\nu}\bar c^{e}}(k_2,k_1),
\label{BQI:auxcAs} \\
i\Gamma_{c^{a} A^n_\nu A^{*\gamma}_d}(k,q)&=&igf^{adn}g^\gamma_\nu-igf^{e'ds'}g^\gamma_{\sigma'}\int_{k_1}D^{ee'}(k_1)\Delta_{ss'}^{\sigma\sigma'}(k_2){\cal K}_{c^a A^n_\nu A^s_\sigma\bar c^e}(k,k_2,k_1),
\label{BQI:auxcAAs} \\
- \Gamma_{\Omega_\alpha^a A^{*\gamma}_d}(q)&=&gf^{ae'n'}g_{\alpha\nu'}\int_{k_1}D^{e'e}(k_1)\Delta_{n'n}^{\nu'\nu}(k_2)\Gamma_{c^e A^n_\nu A^{*\gamma}_d}(k_2,-q),
\label{BQI:auxOmAs}\\
i\Gamma_{\Omega_\alpha^a A^d_\gamma\bar c^n}(k,p)&=&-igf^{adn}g_{\alpha\gamma}-igf^{ae'n'}g_{\alpha\nu'}\int_{k_1}D^{e'e}(k_1)\Delta_{n'n}^{\nu'\nu}(k_2){\cal K}_{c^e A^n_\nu A^d_\gamma\bar c^n}(k_2,k,p).\nonumber \\
\label{BQI:auxOmAbarc}
\end{eqnarray}
The kernel ${\cal K}_{cAA\bar c}$ appearing in the SDEs~(\ref{BQI:auxcAAs}) and~(\ref{BQI:auxOmAbarc})  is shown in Fig.~\ref{fig:cggc_SDker} and reads
\begin{eqnarray}
{\cal K}_{c^a A^n_\nu A^s_\sigma\bar c^e}(k,k_2,k_1)&=&\Gamma_{c^a A^n_\nu A^s_\sigma\bar c^e}(k,k_2,k_1)\nonumber\\
&+&i\Gamma_{A^n_\nu A^s_\sigma A^r_\rho}(k_2,-k-k_2)i\Delta^{\rho\rho'}_{rr'}(k+k_2)i\Gamma_{c^mA^{r'}_{\rho'}\bar c^e}(k+k_2,k_1)\nonumber \\
&+&i\Gamma_{c^a A^s_\sigma\bar c^r}(k_2,-k_1-k_2)iD^{rr'}(k_1+k_2)i\Gamma_{c^{r'} A^n_\nu\bar c^e}(k,k_1).
\label{SDE_kercAAbarc}
\end{eqnarray}
Finally notice that the auxiliary function $\Gamma_{\Omega_\alpha A^*_\beta}$ corresponds precisely to the auxiliary function $\Lambda_{\alpha\beta}$ introduced in Eq.~(\ref{gpert2}), and therefore its part proportional to  $g_{\alpha\beta}$ corresponds to the scalar function $G(q^2)$.

\subsection{BQIs for three-point functions \label{BQI:3p}}

The relation between the trilinear quantum gluon-quark vertex 
and the trilinear background gluon-quark vertex, 
can be obtained by considering the following functional differentiation
\begin{equation}
\left.\frac{\delta^3{\cal S'}(\Gamma')}{\delta \Omega^a_\alpha(q)\delta\psi(p_2)\delta\bar\psi(-p_1)}\right|_{\Phi,\Phi^*,\Omega=0}=0 \qquad q+p_2=p_1.
\label{hgff_diff}
\end{equation}
We then get
\begin{eqnarray}
i\Gamma_{\widehat{A}^a_\alpha\psi\bar\psi}(p_2,-p_1)&=&[ig^\gamma_\alpha\delta^{ad}+\Gamma_{\Omega^a_\alpha A^{*\gamma}_d}(-q)]\Gamma_{A^d_\gamma\psi\bar\psi}(p_2,-p_1)\nonumber \\
&+&\Gamma_{\psi^*\bar\psi\Omega^a_\alpha}(-p_1,q)\Gamma_{\psi\bar\psi}(p_2)+\Gamma_{\psi\bar\psi}(p_1)\Gamma_{\psi\Omega^a_\alpha\bar\psi^*}(q,-p_1).
\label{BQI:gff}
\end{eqnarray}

In order to explore further the all-order structure of these two  auxiliary Green's functions, replace the BFM source 
with the corresponding composite operator using Eq.~(\ref{Omtrick}),  thus obtaining 
\begin{eqnarray}
i\Gamma_{\psi\Omega^a_\alpha\bar\psi^*}(q,-p_1)&=&i\Gamma_{\Omega_\alpha^a A_{\nu'}^{n'}\bar c^{m'}}^{(0)}\int_{k_1}iD^{m'm}(k_1)i\Delta_{n'n}^{\nu'\nu}(k_2){\cal K}_{A^n_\nu\psi c^m\bar\psi^*}(p_2,k_1,-p_1), \label{Aux-gff:1}\\
i\Gamma_{\psi^*\bar\psi\Omega^a_\alpha}(-p_1,q)&=&i\Gamma_{\Omega_\alpha^a A_{\nu'}^{n'}\bar c^{m'}}^{(0)}\int_{k_1}iD^{m'm}(k_1)i\Delta_{n'n}^{\nu'\nu}(k_2){\cal K}_{A^n_\nu\psi^*\bar\psi c^m}(p_2,-p_1,k_1).  \label{Aux-gff:2}
\end{eqnarray}
where  the kernels ${\cal  K}_{A^n_\nu\psi c^m\bar\psi^*}$  and ${\cal
K}_{A^n_\nu\psi^*\bar\psi     c^m}$    have     been     defined    in
Eq.~(\ref{1PRker:ggff1}) and (\ref{1PRker:ggff2}).  As it is clear from
the two equations above,  while the (auxiliary) functions appearing in
the STIs and BQIs ought to be  1PI (lhs of the equations), that is not
true  for  the kernels  appearing  after  using  the substitutions  of
Eq.~(\ref{Astrick}) or (\ref{Omtrick}), which, in fact, consist of both
1PI and 1PR diagrams (rhs of the equations).

For the BQI involving the three-gluon vertex a similar result can be obtained; choosing
\begin{equation}
\left.\frac{\delta^3{\cal S'}(\Gamma')}{\delta \Omega^a_\alpha(q)\delta A^r_\rho(p_2)\delta A^s_\sigma(-p_1)}\right|_{\Phi,\Phi^*,\Omega=0}=0 \qquad q+p_2=p_1,
\label{hggg_diff}
\end{equation}
we will get
\begin{eqnarray}
i\Gamma_{\widehat{A}^a_\alpha A^r_\rho A^s_\sigma}(p_2,-p_1)&=&[ig^\gamma_\alpha\delta^{ad}+\Gamma_{\Omega^a_\alpha A^{*\gamma}_d}(-q)]\Gamma_{A^d_\gamma A^r_\rho A^s_\sigma}(p_2,-p_1)\nonumber \\
&+&\Gamma_{\Omega^a_\alpha A^s_\sigma A^{*\gamma}_d}(-p_1,p_2)\Gamma_{A^d_\gamma A^{r}_\rho}(p_2)+\Gamma_{\Omega^a_\alpha A^r_\rho A^{*\gamma}_d}(p_2,-p_1)\Gamma_{A^d_\gamma A^{s}_\sigma}(p_1).\nonumber \\
\label{BQI:ggg}
\end{eqnarray}
Again we can write
\begin{eqnarray}
i\Gamma_{\Omega^a_\alpha A^s_\sigma A^{*\gamma}_d}(-p_1,p_2)&=&i\Gamma^{(0)}_{\Omega^a_\alpha A^{n'}_{\nu'}\bar c^{m'}}\int_{k_1}iD^{m'm}(k_1)i\Delta_{n'n}^{\nu'\nu}(k_2){\cal K}_{c^m A^n_\nu A^s_\sigma A^{*\gamma}_d}(k_2,-p_1,p_2),\hspace{1cm}
\label{Aux-ggg:1}\\
i\Gamma_{\Omega^a_\alpha A^r_\rho A^{*\gamma}_d}(p_2,-p_1)&=&i\Gamma^{(0)}_{\Omega^a_\alpha A^{n'}_{\nu'}\bar c^{m'}}\int_{k_1}iD^{m'm}(k_1)i\Delta_{n'n}^{\nu'\nu}(k_2){\cal K}_{c^m A^n_\nu A^r_\rho A^{*\gamma}_d}(k_2,p_2,-p_1),
\label{Aux-ggg:2}
\end{eqnarray}
where the kernels  appearing in the above equations  have been defined
in Eq.s~(\ref{1PRker:gggg1})  and~(\ref{1PRker:gggg2}). 
Notice the  emergence  of the  pattern exploited in the application of the PT to the SDEs of QCD: namely that  the  auxiliary
functions  appearing in  the  BQI satisfied  by  a particular  Green's
function  can be written  in terms  of kernels  appearing in  the STIs
triggered when the  PT procedure is applied to  that same Green's
function.
The BQI of Eq.~(\ref{BQI:ggg}) gives at tree-level the result
\be
\Gamma_{\widehat{A}^a_\alpha A^r_\rho A^s_\sigma}(p_2,-p_1)=\Gamma_{A^a_\alpha A^r_\rho A^s_\sigma}(p_2,-p_1).
\ee
This is once again due to the use of the reduced functional: in fact in such case the two (tree-level) vertices need to coincide, since the  difference between them is proportional to the inverse of the gauge fixing parameter (see Appendix~\ref{Frules}) and therefore entirely due to the gauge fixing Lagrangian. To restore the correct tree-level terms one would have to use the complete functional; in that case the differentiation of Eq.~(\ref{hggg_diff}) shows the two additional terms 
\be
-\delta^{ds}p_{1\sigma}\Gamma_{\Omega^a_\alpha A^r_\rho \bar c^d}(p_2,-p_1)+\delta^{dr}p_{2\rho}\Gamma_{\Omega^a_\alpha A^s_\sigma  \bar c^d}(-p_1,p_2),
\ee
which, with the help of Eq.~(\ref{FPE_Om1}) become
\be
-i\delta^{ds}p_{1\sigma}p_{1\gamma}\Gamma_{\Omega^a_\alpha A^r_\rho A^{*\gamma}_d}(p_2,-p_1)-i\delta^{dr}p_{2\rho}p_{2\gamma}\Gamma_{\Omega^a_\alpha A^s_\sigma A^{*\gamma}_d}(-p_1,p_2)+gf^{ars}(q_{\alpha\rho}p_{1\sigma}+g_{\alpha\sigma}p_{2\rho}).
\ee
Therefore we get the final identity
\bea
i\Gamma^\mathrm{C}_{\widehat{A}^a_\alpha A^r_\rho A^s_\sigma}(p_2,-p_1)&=&[ig^\gamma_\alpha\delta^{ad}+\Gamma_{\Omega^a_\alpha A^{*\gamma}_d}(-q)]\Gamma_{A^d_\gamma A^r_\rho A^s_\sigma}(p_2,-p_1)+gf^{ars}(q_{\alpha\rho}p_{1\sigma}+g_{\alpha\sigma}p_{2\rho})\nonumber \\
&+&\Gamma_{\Omega^a_\alpha A^s_\sigma A^{*\gamma}_d}(-p_1,p_2)\left[\Gamma^\mathrm{C}_{A^d_\gamma A^{r}_\rho}(p_2)-i\delta^{dr}p_{2\rho}p_{2\gamma}\right]\nonumber \\
&+&\Gamma_{\Omega^a_\alpha A^r_\rho A^{*\gamma}_d}(p_2,-p_1)\left[\Gamma^\mathrm{C}_{A^d_\gamma A^{s}_\sigma}(p_1)
-i\delta^{ds}p_{1\sigma}p_{1\gamma}\right],
\eea
which gives the expected tree-level result. Once again we see that the difference between working with the reduced and complete functional lies in some constant (tree-level) terms that one recovers after applying the FPE for writing the STI/BQI at hand in the same form using $\Gamma$ or $\Gamma_\mathrm{C}$. Thus, opting for the fast way of deriving the STI/BQI  with the reduced functional and adding the correct tree-level term, we write the BQI in its final form
\begin{eqnarray}
i\Gamma_{\widehat{A}^a_\alpha A^r_\rho A^s_\sigma}(p_2,-p_1)&=&[ig^\gamma_\alpha\delta^{ad}+\Gamma_{\Omega^a_\alpha A^{*\gamma}_d}(-q)]\Gamma_{A^d_\gamma A^r_\rho A^s_\sigma}(p_2,-p_1)\nonumber \\
&+&\Gamma_{\Omega^a_\alpha A^s_\sigma A^{*\gamma}_d}(-p_1,p_2)\Gamma_{A^d_\gamma A^{r}_\rho}(p_2)+\Gamma_{\Omega^a_\alpha A^r_\rho A^{*\gamma}_d}(p_2,-p_1)\Gamma_{A^d_\gamma A^{s}_\sigma}(p_1)\nonumber \\
&+&gf^{ars}\left(p_{2\rho}g_{\alpha\sigma}+p_{1\sigma}g_{\alpha\rho}\right).
\label{BQI:ggg_tlc}
\end{eqnarray}

\subsection{BQI for the ghost-gluon trilinear vertex}

In this section we are going to derive the BQIs relating the $R_\xi$ ghost sector with the BFM ones.
We start from the trilinear ghost-gluon coupling, for which we choose the following functional differentiation 
\begin{equation}
\left.\frac{\delta^3{\cal S'}(\Gamma')}{\delta \Omega^a_\alpha(-q)\delta c^m(k_1)\delta \bar c^n(k_2)}\right|_{\Phi,\Phi^*,\Omega=0}=0 \qquad k_1+k_2=q,
\end{equation}
thus getting the result
\begin{eqnarray}
i\Gamma_{c^m\widehat{A}^a_\alpha\bar c^n}(-q,k_2)&=&[i\delta^{da}g^\gamma_\alpha+\Gamma_{\Omega^a_\alpha A^{*\gamma}_d}(q)]
\Gamma_{c^m A^d_\gamma\bar c^n}(-q,k_2)\nonumber\\
&-&\Gamma_{c^m A^{*\gamma}_d}(-k_1)\Gamma_{\Omega^a_\alpha A^d_\gamma\bar c^n }(k_1,k_2)-
\Gamma_{\Omega^a_\alpha c^m c^{*d}}(k_1,k_2)\Gamma_{c^d\bar c^n}(k_2).
\label{BQI:cgbarc}
\end{eqnarray}
At tree-level we then correctly recover the result
\begin{eqnarray}
i\Gamma^{(0)}_{c^m\widehat{A}^a_\alpha\bar c^n}(-q,k_2)&=&i\Gamma^{(0)}_{c^m A^a_\alpha\bar c^n}(-q,k_2)-gf^{amn}k_{1\alpha}\nonumber \\
&=&-gf^{amn}(k_1-k_2)_\alpha,
\end{eqnarray}
(in this case there is no difference between using the complete or reduced functional).

\section{Feynman rules\label{Frules}}

\renewcommand{\theequation}{F.\arabic{equation}}

\subsection{$R_\xi$ and BFM gauges \label{Appendix:RxiBFMFR}}

\begin{figure}[!t]
\includegraphics{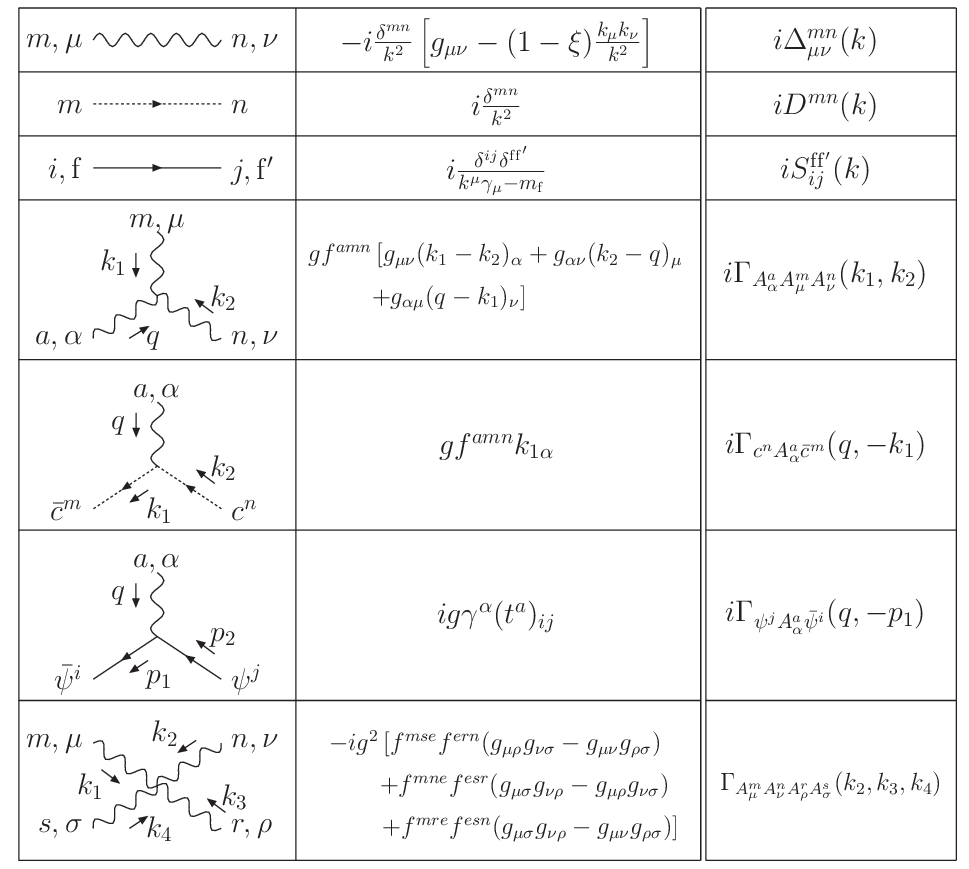}
\caption{Feynman rules for QCD in the $R_\xi$ gauges. The first two columns show the lowest order Feynman diagrams and rule respectively, while the last one shows the corresponding all-order Green's function according to the conventions of Eq.~(\ref{greenfunc}).}
\label{fig:Frules_Rxi}
\end{figure}

The Feynman rules for QCD in $R_\xi$ gauges are given in Fig.~\ref{fig:Frules_Rxi}. 
In the case of the BFM gauge, since the gauge fixing
Lagrangian is quadratic in the quantum fields, apart from vertices involving ghost fields
only vertices containing exactly two quantum fields might differ from the conventional ones. Thus, 
the vertices $\Gamma_{\widehat A\psi\bar\psi}$ and $\Gamma_{\widehat AAAA}$ have to 
lowest order the same expression as the corresponding $R_\xi$ 
ones $\Gamma_{A\psi\bar\psi}$ and $\Gamma_{AAAA}$ 
(to higher order their relation is described by the corresponding BQIs).

\begin{figure}[!t]
\includegraphics{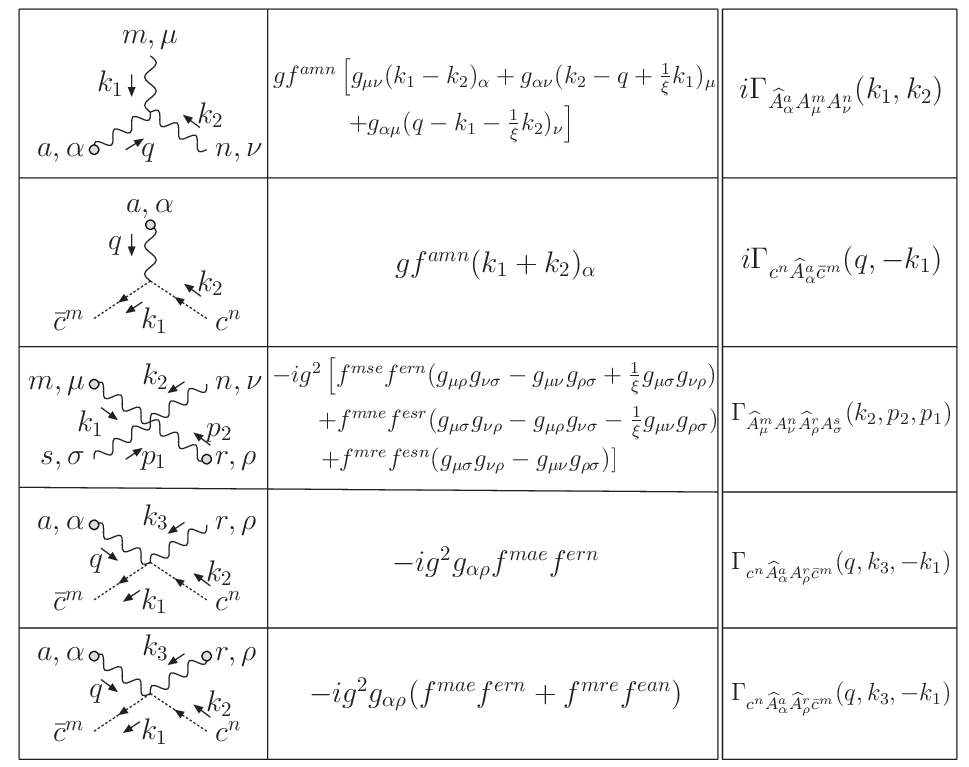}
\caption{Feynman rules for QCD in the BFM gauge. We only include those rules which are different from the $R_\xi$ ones to lowest order. As usual, gray circle on a gluon line indicates a background field.}
\label{fig:Frules_BFM}
\end{figure}

\subsection{Anti-fields \label{Appendix:afFR}}

\begin{figure}[!t]
\includegraphics[scale=1.25]{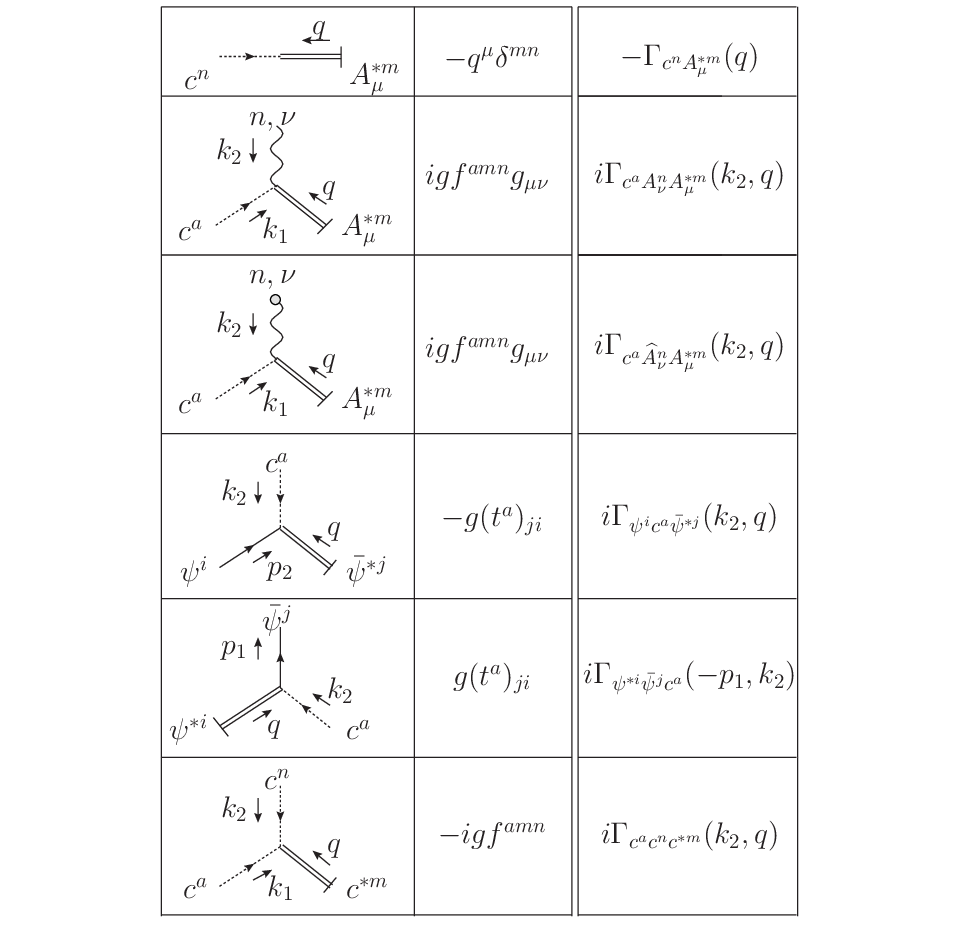}
\caption{Feynman rules for QCD anti-fields.}
\label{fig:Frules_anti}
\end{figure}

The couplings of anti-fields with fields is entirely encoded in the BRST Lagrangian of Eq.~(\ref{BRST_Lag}). 
When choosing the BFM gauge the additional coupling $gf^{amn}A^{*m}_\mu \widehat{A}^n_\nu c^a$ will arise in the BRST Lagrangian ${\cal L}_\mathrm{BRST}$ as a consequence of the BFM splitting $A\to\widehat{A}+A$.
One then gets the Feynman rules given in Fig.~\ref{fig:Frules_anti}.

\subsection{BFM sources \label{Appendix:BFMsFR}}

\begin{figure}[!h]
\includegraphics{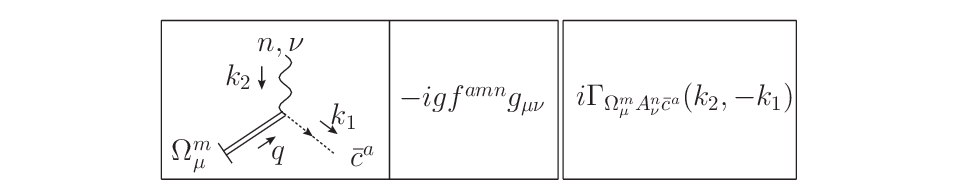}
\caption{Feynman rule for the BFM gluon source $\Omega^m_\mu$.}
\label{fig:Frules_BFMsources}
\end{figure}

The couplings of BFM sources $\Omega^m_\mu$ with fields can be derived from the Faddeev-Popov ghost Lagrangian, since making use of the extended BRST transformation of Eq.~(\ref{extBRST}) we get
\begin{equation}
{\cal L}_\mathrm{FPG}=-\bar c^a s{\cal F}^a_\mathrm{BFM}
\supset-\bar c^a g f^{amn}(s\widehat{A}^m_\mu) A^\mu_n
=-g f^{amn}\bar c^a\Omega^m_\mu A^\mu_n.
\end{equation}
The corresponding Feynman rule is finally given in Fig.~\ref{fig:Frules_BFMsources}.

\end{document}